%% file: seyboth14b.tex
\documentclass[letterpaper, 10 pt, journal]{IEEEtran} 

\pdfoutput=1

\usepackage{cite}
\usepackage{graphicx}
\usepackage{subfigure}
\usepackage{amsmath}
\usepackage{amssymb}
\usepackage{amsthm}
\usepackage{mathptmx}
\usepackage[mathcal]{euscript}
\usepackage{array}

\usepackage{url}
\usepackage{tikz}
\usepackage{tkz-graph}
\usetikzlibrary{shapes,decorations,calc}

\usepackage{pgfplots}

\newlength\figureheight%
\newlength\figurewidth%
\usepackage{hyperref}
\usepackage{xcolor}
\definecolor{dark-red}{rgb}{0.4,0.15,0.15}
\definecolor{dark-blue}{rgb}{0.15,0.15,0.4}
\definecolor{medium-blue}{rgb}{0,0,0.5}
\hypersetup{
    colorlinks, linkcolor={dark-red},
    citecolor={dark-blue}, urlcolor={medium-blue}
}

\usepackage{enumerate}
\usepackage{refcount}

\newtheorem{asmpt}{Assumption}

\newtheorem{lem}{Lemma}
\newtheorem{thm}{Theorem}
\newtheorem{rmk}{Remark}

\newtheorem{exmp}{Example}
\newtheorem{prob}{Problem}

\newcounter{asmptprimecounter}
\newtheorem{asmptprime}[asmptprimecounter]{Assumption}

\newenvironment{myasmptprime}[1]{
\renewcommand{\theasmptprimecounter}{%
  \setcounter{asmptprimecounter}{\getrefnumber{#1}}%
  \arabic{asmptprimecounter}'}
\begin{asmptprime}}%
{
\end{asmptprime}
}

\newcommand{\diag}{{\rm diag}}
\newcommand{\stack}{{\rm stack}}
\renewcommand{\Re}{{\rm \bf Re}}
\renewcommand{\Im}{{\rm \bf Im}}

\usepackage{cite}
\begin{document}

\title{\LARGE \bf Cooperative Control of Linear Multi-Agent Systems\\
  via Distributed Output Regulation and Transient Synchronization}

\author{Georg S. Seyboth$^*$, Wei Ren, Frank Allg\"{o}wer
  \thanks{$^*$Corresponding author. G.S.S and F.A. are with the
    Institute for Systems Theory and Automatic Control, University of
    Stuttgart, Pfaffenwaldring 9, 70550 Stuttgart, Germany.
    ({\tt\scriptsize \{georg.seyboth,
      allgower\}@ist.uni-stuttgart.de}). W.R. is with the Department
    of Electrical Engineering, University of California Riverside,
    Riverside, CA 92521, USA. ({\tt\scriptsize ren@ee.ucr.edu}). The
    authors G.S.S. and F.A. would like to thank the German Research
    Foundation (DFG) for financial support of the project within the
    Cluster of Excellence in Simulation Technology (EXC 310/1) at the
    University of Stuttgart and within the Priority Programme 1305
    ``Control Theory of Digitally Networked Dynamical Systems''.}}

\maketitle
\thispagestyle{empty}
\pagestyle{empty}

\begin{abstract}
  A wide range of multi-agent coordination problems including
  reference tracking and disturbance rejection requirements can be
  formulated as a cooperative output regulation problem.  The general
  framework captures typical problems such as output synchronization,
  leader-follower synchronization, and many more.  In the present
  paper, we propose a novel distributed regulator for groups of
  identical and non-identical linear agents.  We consider global
  external signals affecting all agents and local external signals
  affecting only individual agents in the group.  Both signal types
  may contain references and disturbances.  Our main contribution is a
  novel coupling among the agents based on their transient state
  components or estimates thereof in the output feedback case.  This
  coupling achieves transient synchronization in order to improve the
  cooperative behavior of the group in transient phases and guarantee
  a desired decay rate of the synchronization error.  This leads to a
  cooperative reaction of the group on local disturbances acting on
  individual agents.  The effectiveness of the proposed distributed
  regulator is illustrated by a vehicle platooning example and a
  coordination example for a group of four non-identical 3-DoF
  helicopter models.%
\end{abstract}


\section{Introduction}
\label{sec:intro}

In a variety of modern man-made systems, it is desirable to synthesize
a cooperative behavior among individual dynamical agents, similarly to
bird flocks and fish schools observed in nature.  Examples include
multi-vehicle coordination and formation flight problems, robot
cooperation in production lines, power balancing in micro-grids, and
many more.  Of particular interest are distributed control laws which
require only local information exchange between neighboring subsystems
and no centralized data collection or processing entity.  The main
advantages of distributed control algorithms are scalability for large
networks of dynamical systems, flexibility with respect to addition
and removal of subsystems, and robustness with respect to failure of
individual subsystems.  The following overview of selected
publications on distributed control methods for multi-agent systems
serves as introduction and motivation for the methods developed in the
present paper.

\subsection{Related Work}

A fundamental cooperative control problem is the consensus or
synchronization problem for groups of linear dynamical agents.  The
consensus problem has been studied extensively over the past decade,
starting from single-integrator agents \cite{Olfati-Saber2004,
  Ren2005}, to double-integrator agents \cite{Ren2007}, to identical
general linear agents \cite{Fax2004, Tuna2008, Scardovi2009, Li2010,
  Wieland2011b}, and to non-identical general linear agents
\cite{Wieland2011, Kim2011, Lunze2012b, Grip2012a}.  All these studies
focus on consensus and synchronization of the states or outputs of the
autonomous agents in the group.  Under the distributed control law,
the closed-loop system as a whole is an autonomous system, i.e., it
has no external inputs signals.

From a practical point of view, it is desirable to influence the
behavior of the group via external reference signals.  A solution to
this problem is the leader-follower setup \cite{Hong2006, Ni2010,
  Li2010, Zhang2011b}.  The idea is to select a particular agent as
leader for the group or introduce a virtual leader and design the
distributed control law such that all agents synchronize to this
leader.  The motion of the group is then controlled through the motion
of the active leader.

Moreover, it is important to consider external disturbances acting on
the multi-agent system, to analyze the performance of the closed-loop
system, and to incorporate disturbance rejection or attenuation
requirements in the design procedure.  Rejection of constant
disturbances is addressed in
\cite{Yucelen2012,Andreasson2012,Seyboth2014a}. Disturbance
attenuation and $\mathcal{H}_\infty$ performance criteria are
addressed in \cite{Li2011a, Zhao2012, Wang2013}.

Reference tracking and disturbance rejection problems can be
formulated in the general framework of output regulation which was
developed in the 1970s, \cite{Wonham1973,Francis1975}.  The basic
setup consists of a so-called \emph{exosystem}, an autonomous system
that generates all external signals (references as well as
disturbances) acting on the plant, and a description of the plant.
The signal generated by the exosystem is referred to as the
\emph{generalized disturbance}.  The tracking and regulation
requirements are formulated in terms of a regulation error depending
on the plant state and the external signals.  The objective is to find
a control law, also called \emph{regulator}, which ensures internal
stability of the plant and asymptotic convergence to zero of the
regulation error for all initial conditions.  For the details on the
output regulation theory, the reader is referred to the books
\cite{Knobloch1993}, \cite{Trentelman2001}, and \cite{Huang2004}.

It has been proposed in \cite{Xiang2009, Huang2011} to formulate
multi-agent coordination problems with external reference and
disturbance signals as a \emph{synchronized output regulation}
problem. Since then, a \emph{cooperative output regulation} theory for
a large class of practically relevant cooperative control problems is
under development.  The problem setup in \cite{Xiang2009, Huang2011}
consists of an autonomous exosystem and a group of identical linear
agents, which are affected by the signal generated by the exosystem,
and a tracking error for each agent which shall converge to zero.  The
problem becomes a cooperative control problem since, by assumption,
not all agents have access to the external signal.  In particular, the
group is divided into a group of informed agents which are able to
reconstruct the external signal, and uninformed agents which are
dependent on information exchange with informed agents in order to
solve their regulation task.  The proposed distributed regulator
consists of local feedback laws designed according to the classical
output regulation theory and a distributed estimator for the external
signal.  It has been shown in \cite{Hong2013a} that the cooperative
output regulation problem generalizes existing solutions for the
leader-follower problem.  In \cite{Wang2010}, agents with
non-identical and uncertain dynamics are considered.  The main
limitation of the proposed solution is that the underlying
communication graph is assumed to have no loops.  This assumption is
relaxed in \cite{Su2011}.  Instead, the follower agents are assumed to
have identical nominal dynamics.  The papers \cite{Su2012c} and
\cite{Su2012b} extend the results of \cite{Xiang2009, Huang2011} and
present a solution to the cooperative output regulation problem with
non-identical agents based on state feedback and output feedback,
respectively.  Each agent is described by a generalized plant in which
all matrices may be different for different agents.  Cooperation among
the agents is again required since the uninformed agents are not able
to reconstruct the external signal locally.  The solution proposed in
\cite{Su2012b} consists of three components: local feedback laws which
are constructed based on the classical output regulation theory for
each agent; local observers for the state of each agent; and a
distributed observer for the generalized disturbance.  Further
developments in this area focus on robust cooperative output
regulation for uncertain agent dynamics \cite{Su2013},
\cite{Persis2012}, \cite{Yu2013a}.  All studies mentioned above
consider a single autonomous exosystem generating all reference and
disturbance signals acting on the group. In \cite{Meng2013}, each
agent has an additional local exosystem that generates local reference
signals.  However, only reference signals and no disturbance signals
are considered.  Multiple exosystems are also considered in
\cite{Liu2014}.

\subsection{Contribution}

In the cooperative output regulation framework, the group objective
for the multi-agent system is formulated in terms of a common
reference signal for all agents and local regulation errors for each
agent with respect to the common reference.  Cooperation among the
agents is only required in order to reconstruct the reference signal
at all agents.  The reconstruction is achieved through cooperative
estimation.  Besides that the setup is in fact decentralized in the
sense that there are no couplings based on the agent states or
outputs.  Cooperation is achieved since all agents asymptotically
track a common reference and disturbance rejection is achieved locally
at each agent.  This approach may not lead to a satisfactory behavior
of the group in transient phases.  In particular, a state or output
synchronization objective is not taken into account explicitly.
In the present paper, we propose a novel distributed regulator which
solves the cooperative output regulation problem and additionally
guarantees exponential stability of a suitably defined synchronization
error of the group.  This allows to improve the cooperative behavior
of the group in transient phases significantly and offers more
flexibility in the control design.  Our contributions are the
following.

First, we introduce one global exosystem generating signals affecting
all agents and a local exosystem for each agent generating local
references and disturbances.  This exosystem structure leads to
distributed estimators of lower dimension compared to previous works,
where only one single exosystem (of possibly high dimension) is
considered which generates all external signals acting on the group.

Second, we formulate the overall cooperative control problem as a
single centralized overall output regulation problem.  The solvability
conditions for the overall output regulation problem and its
particular structure allow us to derive necessary and sufficient
solvability conditions for the distributed regulation problem.

Third, we present a distributed regulator based on output feedback
which solves the cooperative output regulation problem.  Our main
contribution is to introduce a novel coupling in the distributed
regulator based on the transient state components of the agents.  We
present a design method for these couplings which allows to impose
performance specifications such as a minimum decay rate of the
synchronization error of the group.  The novel distributed regulator
ensures a cooperative reaction of the group on external disturbances.

Moreover, a detailed numerical example of four 3-DoF helicopters with
non-identical model parameters is presented in order to illustrate the
design procedure for groups of non-identical agents and demonstrate
the performance of the novel distributed regulator with transient
synchronization.

As an auxiliary result, we present a design method based on LMIs for
the coupling gain in networks of identical linear agents such that the
poles corresponding to the synchronization error are placed within a
specified region.  This procedure is the key to incorporate
performance specifications such as a minimum decay rate in the
cooperative control design.

\subsection{Outline}
Section~\ref{sec:preliminaries} contains mathematical preliminaries.
Section~\ref{sec:problem} presents the problem setup and some
auxiliary results.  The distributed regulator for general
non-identical linear agents is presented in
Section~\ref{sec:regulator}.  In Section~\ref{sec:identical}, an
extension of the distributed regulator is derived which guarantees
exponential stability of the synchronization error with a desired
decay rate.  The derivation is based on the assumption that the agents
have identical dynamics.  A vehicle platooning example illustrates the
results. In Section~\ref{sec:nonidentical}, the assumption of
identical agent dynamics is relaxed.  It is shown how the coupling
can be designed in case of non-identical agents based on robust
control methods and the procedure is demonstrated in an example with
four non-identical 3-DoF helicopters.  Section~\ref{sec:conclusion}
concludes the paper.

\section{Preliminaries}\label{sec:preliminaries}

The sets of real and complex numbers are denoted by $\mathbb{R}$ and
$\mathbb{C}$, respectively. The open left half plane, imaginary axis,
and open right half plane of $\mathbb{C}$ are denoted by
$\mathbb{C}^-$, $\mathbb{C}^0$, and $\mathbb{C}^+$, respectively.  For
$z \in \mathbb{C}$, $\bar{z}$ is the complex conjugate, $\Re(z)$ is
the real part and $\Im(z)$ is the imaginary part.  The spectrum of a
matrix $A \in \mathbb{C}^{n \times n}$ is denoted by $\sigma(A)
\subset \mathbb{C}$ and $A$ is called stable if $\sigma(A) \subset
\mathbb{C}^-$, i.e., all its eigenvalues have negative real part.  If
$A$ is real and $\sigma(A) \subset \mathbb{C}^-$, it is also called
Hurwitz.  $\diag(M_k)$ and $\stack(M_k)$ denote a block diagonal
matrix and a vertical stack of matrices with blocks $M_k$, $k =
1,...,N$, respectively.  For a set of vectors $v_k \in \mathbb{R}^n$,
$k=1,...,N$, $v \in \mathbb{R}^{Nn}$ denotes the stack vector $v =
[v_1^\mathsf{T} \; \cdots \; v_N^\mathsf{T} ]^\mathsf{T}$.  For a
vector $v \in \mathbb{R}^n$, $\diag(v)$ is a diagonal matrix with the
entries of $v$ on the diagonal.  The identity matrix of dimension $N$
is $I_N$ and the vector of ones is $\mathbf{1}$.  For a transfer
function matrix $G$, $\|G\|_\infty$ denotes its $\mathcal{H}_\infty$
norm.  The symbol $\otimes$ denotes the Kronecker product.

\section{Problem Setup}\label{sec:problem}

\subsection{Agent Models} 

The dynamics of the non-identical agents are described by linear
state-space models.  The index set of the agents is defined as
$\mathcal{N} = \{1,...,N\}$, where $N$ is the number of agents in the
group.  The dynamics of the undisturbed agents are described by
\begin{equation}
  \label{eq:agent_undisturbed}
    \dot{x}_k = A_k x_k + B_k u_k
\end{equation}
where $x_k(t) \in \mathbb{R}^{n^x_k}$ is the state and $u_k(t) \in
\mathbb{R}^{n^u_k}$ is the control input of agent $k \in \mathcal{N}$.
The coordination problem is formulated in terms of the generalized
plant
\begin{subequations}
  \label{eq:agent}
  \begin{align}
    \dot{x}_k &= A_k x_k + B_k u_k + B^{d^g}_k d^g
    + B^{d^{\ell}}_k d^{\ell}_k \label{eq:agent_x}\\
    y_k &= C_k x_k + D_k u_k + D^{d^g}_k d^g
    + D^{d^{\ell}}_k d^{\ell}_k \label{eq:agent_y}\\
    e_k &= C^e_k x_k + D^e_k u_k + D^{ed^g}_k d^g + D^{ed^{\ell}}_k
    d^{\ell}_k \label{eq:agent_e}
  \end{align}
\end{subequations}
where $y_k(t) \in \mathbb{R}^{n^y_k}$ is the measurement output of
agent $k$ and $d^g(t) \in \mathbb{R}^{n^{d^g}}$, $d^\ell_k(t) \in
\mathbb{R}^{n^{d^\ell}_k}$ are external signals specified next.  The
regulation error $e_k(t) \in \mathbb{R}^{n^e_k}$ is defined such that
asymptotic tracking and disturbance rejection is equivalent to $e_k(t)
\to 0$ as $t \to \infty$ for all initial conditions.

\subsection{External Signals}

We consider two types of external input signals which affect the
group: a global signal that affects all agents and local signals that
affect individual agents in the group.  Each of these signals
represents a generalized disturbance which may consist of reference
signals and disturbances.  While the problem formulation captures this
general case, we may think of the global signal as a pure reference
signal and include all disturbances into the local signals.  We assume
that all signals belong to a known family of signals.  The global
signal $d^g(t) \in \mathbb{R}^{n^{d^g}}$ is a solution of the
autonomous linear system
\begin{equation}\label{eq:global_exosystem}
  \dot{d}^g = S^g d^g,
\end{equation}
where $\sigma(S^g) \subset \mathbb{C}^0 \cup \mathbb{C}^+$.  The local
signal $d^{\ell}_k(t) \in \mathbb{R}^{n^{d^{\ell}}_k}$ acting on agent
$k$ is a solution of the autonomous linear system
\begin{equation}\label{eq:local_exosystem}
  \dot{d}^{\ell}_k = S^\ell_k d^{\ell}_k,
\end{equation}
where $\sigma(S^{\ell}_k) \subset \mathbb{C}^0 \cup \mathbb{C}^+$ for all
$k \in \mathcal{N}$.

\subsection{Information Structure}

All agents have communication capabilities.  The communication
topology is described by a directed graph $\mathcal{G} =
(\mathcal{N},\mathcal{E})$, with node set $\mathcal{N} = \{1,...,N\}$
corresponding to the agent index set and edge set $\mathcal{E} \subset
\mathcal{N} \times \mathcal{N}$.  A directed edge $(j,k) \in
\mathcal{E}$ corresponds to possible information flow from agent $j
\in \mathcal{N}$ to agent $k \in \mathcal{N}$.  We think of these
edges as communication channels.  Hence, each agent can transmit local
measurements or controller states.  There is no forwarding, in the
sense that each agent exchanges information only with its direct
neighbors and not with its two-hop or multi-hop neighbors.  The
neighbor set of agent $k$ is the subset $\mathcal{N}_k \subset
\mathcal{N}$ from which $k$ can receive information, i.e.,
$\mathcal{N}_k = \{j\in\mathcal{N} : (j,k) \in \mathcal{E} \}$.  We
call a directed graph $\mathcal{G}$ \emph{connected} if it contains a
directed spanning tree (which is sometimes called quasi strongly
connected in the literature), and \emph{strongly connected} if there
is a directed path from every node to every other node, i.e., every
node is the root of a spanning tree.  The adjacency matrix
$A_{\mathcal{G}}$ of $\mathcal{G}$ is defined element-wise by $a_{kj}
= 1 \Leftrightarrow (j,k) \in \mathcal{E}$ and $a_{kj} = 0$ otherwise.
In case $\mathcal{G}$ is connected, its Laplacian matrix
$L_{\mathcal{G}} = \diag(A_{\mathcal{G}}\mathbf{1}) - A_{\mathcal{G}}$
has exactly one zero eigenvalue $\lambda_1(L_{\mathcal{G}}) = 0$ and
all other eigenvalues $\lambda_k(L_{\mathcal{G}})$, $k \in \mathcal{N}
\backslash \{1\}$, have positive real parts
$\Re(\lambda_k(L_{\mathcal{G}})) > 0$, cf., \cite{Ren2005}.  In the
following, the eigenvalues of $L_{\mathcal{G}}$ are shortly denoted by
$\lambda_k$, $k \in \mathcal{N}$, and the numbering is always chosen
such that $\lambda_1 = 0$. For further details on graph theory, see
\cite{Ren2005,Wieland2011b,Wieland2011} and the references therein.

\subsection{The Distributed Output Regulation Problem}

The group objective under consideration consists of two parts:
asymptotic tracking of reference signals and asymptotic disturbance
rejection.  Typically, we are interested in synchronization problems
which can be formulated as tracking problem of a common, global
reference signal.  Moreover, our focus will be on the cooperative
behavior of the group in transient phases.  The cooperative control
problem is formally stated as follows.

\begin{prob}\label{prob:regulation}
  For each $k \in \mathcal{N}$, find a distributed regulator
  \begin{subequations}
    \label{eq:controller}
    \begin{align}
      \dot{z}_k &= A^c_{kk} z_k + B^c_{kk} y_k + \sum_{j \in
        \mathcal{N}_k} \left( A^c_{kj} z_j +
        B^c_{kj}y_j \right)\\
      u_k &= C^c_{kk} z_k + D^c_{kk} y_k + \sum_{j \in \mathcal{N}_k}
      \left( C^c_{kj} z_j + D^c_{kj}y_j \right)
    \end{align}
  \end{subequations}
  such that the following two conditions are satisfied:
  \begin{enumerate}[{\bf P1)}]
  \item If $d^g(0) = 0$ and $d^{\ell}_k(0) = 0$, $k \in \mathcal{N}$, then,
    for all initial conditions $x_k(0) = x_{k,0}$ and $z_k(0) =
    z_{k,0}$,
    \begin{equation*}
      \lim_{t \to \infty} x_k(t) = 0 \qquad \text{and} \qquad
      \lim_{t \to \infty} z_k(t) = 0.
    \end{equation*}
  \item For all initial conditions $d^g(0)=d^g_0$,
    $d^{\ell}_k(0)=d^{\ell}_{k,0}$, $x_k(0)=x_{x,0}$ and $z_k(0) = z_{k,0}$, $k
    \in \mathcal{N}$,
    \begin{equation*}
      \lim_{t \to \infty} e_k(t) = 0.
    \end{equation*}
  \end{enumerate}
\end{prob}

Problem~\ref{prob:regulation} for the generalized plants
\eqref{eq:agent} is a very general problem formulation that captures
many particular practically relevant problems such as reference
tracking and disturbance rejection, cf., \cite{Knobloch1993,
  Huang2004}.  Due to the distributed structure of the controller
\eqref{eq:controller}, this problem formulation captures distributed
cooperative control problems.

\subsection{The Overall Output Regulation Problem}

Problem~\ref{prob:regulation} is a distributed output regulation
problem due to the structure imposed on the regulator
\eqref{eq:controller}.  In case of a complete graph, i.e., all-to-all
communication, \eqref{eq:controller} becomes a centralized dynamic
output feedback controller of the form
\begin{subequations}
  \label{eq:centralized_regulator}
  \begin{align}
    \dot{z} &= A^c z + B^c y \\
    u &= C^c z + D^c y,
  \end{align}
\end{subequations}
where $z$, $y$, $u$ are the stack vectors of $z_k$, $y_k$, $u_k$, $k
\in \mathcal{N}$, respectively.  Let $x$, $e$ be the stack vectors of
$x_k$, $e_k$, $k \in \mathcal{N}$, and $d = [ {d^g}^\mathsf{T} \;
{d^{\ell}_1}^\mathsf{T} \: \cdots \; {d^{\ell}_N}^\mathsf{T}
]^\mathsf{T}$.  Then, the overall cooperative control problem can be
formulated as a single classical output regulation problem by
combining all agents \eqref{eq:agent} into one large generalized plant
and one large exosystem as follows:
\begin{subequations}
  \label{eq:overall_plant}
  \begin{align}
    \dot{x} &= \mathcal{A} x + \mathcal{B} u + \mathcal{B}_d d\\
    y &= \mathcal{C} x + \mathcal{D} u + \mathcal{D}_d d \\
    e &= \mathcal{C}_e x + \mathcal{D}_e u + \mathcal{D}_{ed} d
  \end{align}
\end{subequations}
and
\begin{equation}
  \label{eq:overall_exosystem}
  \dot{d} = \mathcal{S} d,
\end{equation}
with the matrices given by $\mathcal{A} = \diag(A_k)$, $\mathcal{B} =
\diag(B_k)$, $\mathcal{B}_d = [ \stack(B^{d^g}_k) \;\;
\diag(B^{d^{\ell}}_k) ]$, $\mathcal{C} = \diag(C_k)$, $\mathcal{D} =
\diag(D_k)$, $\mathcal{D}_d = [ \stack(D^{d^g}_k) \;\;
\diag(D^{d^{\ell}}_k) ]$,~$\mathcal{C}_e =
\diag(C^e_{k})$,~$\mathcal{D}_e~=~\diag(D^e_{k})$, $\mathcal{D}_{ed} =
[\stack(D^{ed^g}_k) \;\; \diag(D^{ed^{\ell}}_k)]$
and
\begin{equation*}
  \mathcal{S} = \begin{bmatrix} S^g & 0 \\ 0 & \diag(S^{\ell}_k) \end{bmatrix}.
\end{equation*}
The distributed output regulation problem can only be solved if the
overall output regulation problem has a centralized solution of the
form \eqref{eq:centralized_regulator}.  Hence, we study the necessary
conditions for the solvability of the overall output regulation
problem and exploit the structure in order to derive necessary
conditions for the local output regulation problems.

\begin{thm}[\hspace{-0.1ex}{\cite[Theorem~1.14]{Huang2004}}]
  \label{thm:centralized_regulator}
  Let the pair $(\mathcal{A},\mathcal{B})$ be stabilizable and the
  pair
  \begin{equation}
    \label{eq:detectable_pair_overall}
    \left( \begin{bmatrix} \mathcal{A} & \mathcal{B}_d \\ 
        0 & \mathcal{S} \end{bmatrix}, \begin{bmatrix} \mathcal{C} &
        \mathcal{D}_d 
      \end{bmatrix} \right)
  \end{equation}
  be detectable and suppose that $\sigma(\mathcal{S}) \subset
  \mathbb{C}^0 \cup \mathbb{C}^+$.  Then,
  Problem~\ref{prob:regulation} has a centralized solution
  \eqref{eq:centralized_regulator}, if and only if the regulator
  equation
  \begin{equation}\label{eq:overall_regulatoreqn}
    \begin{bmatrix} \mathcal{A} & \mathcal{B} \\ \mathcal{C}_e &
      \mathcal{D}_e \end{bmatrix} \begin{bmatrix} \Pi \\
      \Gamma \end{bmatrix} -  \begin{bmatrix} \Pi \\ 0 \end{bmatrix} 
    \mathcal{S} + \begin{bmatrix} \mathcal{B}_d \\ \mathcal{D}_{ed} \end{bmatrix} = 0
  \end{equation}
  is solvable with a solution $\Pi$, $\Gamma$.
\end{thm}

In case all conditions in Theorem~\ref{thm:centralized_regulator} are
fulfilled, a centralized regulator can be constructed as
follows. Choose $\mathbf{F}$ and $\mathbf{L}$ such that
$\mathcal{A}-\mathcal{B}\mathbf{F}$ and
\begin{equation*}
  \begin{bmatrix} \mathcal{A} & \mathcal{B}_d \\ 
    0 & \mathcal{S} \end{bmatrix} -\mathbf{L} \begin{bmatrix} \mathcal{C} &
    \mathcal{D}_d 
  \end{bmatrix}
\end{equation*}
are stable.  Define $\mathbf{G} = \Gamma + \mathbf{F}\Pi$ where
$\Pi$, $\Gamma$ solve \eqref{eq:overall_regulatoreqn}.  Then, the
controller is given by
\begin{align*}
  \begin{bmatrix} \dot{\hat{x}} \\
    \dot{\hat{d}} \end{bmatrix} &= \begin{bmatrix} \mathcal{A} & \mathcal{B}_d \\
    0 & \mathcal{S} \end{bmatrix} \begin{bmatrix} \hat{x} \\
    \hat{d} \end{bmatrix} + \begin{bmatrix} \mathcal{B} \\
    0 \end{bmatrix} u + \mathbf{L} \left( y - \hat{y} \right) \\
  u &= \begin{bmatrix} -\mathbf{F} &
    \mathbf{G} \end{bmatrix} \begin{bmatrix} \hat{x} \\
    \hat{d} \end{bmatrix},
\end{align*}
where $\hat{y} = \mathcal{C}\hat{x} + \mathcal{D}_d \hat{d} +
\mathcal{D} u$.  The control law $u = -\mathbf{F}x + \mathbf{G}d$ is
referred to as the full information control law.  Since $x$ and $d$
are not directly accessible, the observer is constructed to obtain
estimates $\hat{x}$ and $\hat{d}$.  This controller is of the form
\eqref{eq:centralized_regulator} with $z = [ \hat{x}^\mathsf{T} \;\;
\hat{d}^\mathsf{T} ]^\mathsf{T}$.

\begin{lem}\label{lem:regulatoreqn_k}
  The regulator equation \eqref{eq:overall_regulatoreqn} for the
  overall output regulation problem is solvable, if and only if the
  local regulator equations
  \begin{equation}\label{eq:local_regulatoreqn_g_compact}
    \begin{bmatrix} A_k & B_k \\ C^e_k & D^e_k \end{bmatrix} 
    \begin{bmatrix} \Pi^g_k \\ \Gamma^g_k \end{bmatrix} -
    \begin{bmatrix} \Pi^g_k \\ 0 \end{bmatrix} 
    S^g + \begin{bmatrix} B^{d^g}_k \\ D^{ed^g}_k \end{bmatrix} = 0
  \end{equation}
  and
  \begin{equation}\label{eq:local_regulatoreqn_l_compact}
    \begin{bmatrix} A_k & B_k \\ C^e_k & D^e_k \end{bmatrix} 
    \begin{bmatrix} \Pi^{\ell}_k \\ \Gamma^{\ell}_k \end{bmatrix} -
    \begin{bmatrix} \Pi^{\ell}_k \\ 0 \end{bmatrix} 
    S^{\ell}_k + \begin{bmatrix} B^{d^{\ell}}_k \\ D^{ed^{\ell}}_k \end{bmatrix} = 0
  \end{equation}
  are solvable with a solution $\Pi^g_k$, $\Gamma^g_k$,
  $\Pi^{\ell}_k$, and $\Gamma^{\ell}_k$ for all $k \in \mathcal{N}$.
\end{lem}

\begin{IEEEproof}
  We prove the statement by showing that every solution $\Pi$,
  $\Gamma$ of \eqref{eq:overall_regulatoreqn} yields a solution
  $\Pi^g_k$, $\Gamma^g_k$, $\Pi^{\ell}_k$, $\Gamma^{\ell}_k$, $k \in
  \mathcal{N}$ of \eqref{eq:local_regulatoreqn_g_compact},
  \eqref{eq:local_regulatoreqn_l_compact}, and vice versa.  Consider
  \eqref{eq:overall_regulatoreqn} and partition $\Pi$, $\Gamma$
  according to
  \begin{align}
    \Pi &= 
    \begin{bmatrix} \Pi^g_{1} & \Pi^{\ell}_{11} & \cdots & \Pi^{\ell}_{1N} \\
      \vdots & \vdots & & \vdots \\ \Pi^g_N & \Pi^{\ell}_{N1} & \cdots &
      \Pi^{\ell}_{NN} \end{bmatrix}, \label{eq:overall_Pi}\\ %
    \Gamma &=
    \begin{bmatrix} \Gamma^g_1 & \Gamma^{\ell}_{11} & \cdots & \Gamma^{\ell}_{1N} \\
      \vdots & \vdots & & \vdots \\ \Gamma^g_N & \Gamma^{\ell}_{N1} &
      \cdots & \Gamma^{\ell}_{NN} \end{bmatrix}. \label{eq:overall_Gamma}
  \end{align}
  The first equation $\mathcal{A} \Pi + \mathcal{B} \Gamma - \Pi
  \mathcal{S} + \mathcal{B}_d = 0$ of \eqref{eq:overall_regulatoreqn}
  yields
  \begin{align}
    0 &= \begin{bmatrix} A_1 \Pi^g_{1} & A_1 \Pi^{\ell}_{11} & \cdots
      & A_1 \Pi^{\ell}_{1N} \\ \vdots & \vdots & & \vdots \\ A_N
      \Pi^g_{N} & A_N \Pi^{\ell}_{N1} & \cdots & A_N
      \Pi^{\ell}_{NN} \end{bmatrix} \nonumber \\ %
    &\qquad + \begin{bmatrix} B_1 \Gamma^g_{1} & B_1
      \Gamma^{\ell}_{11} & \cdots & B_1 \Gamma^{\ell}_{1N} \\ \vdots &
      \vdots & & \vdots \\ B_N \Gamma^g_{N} & B_N \Gamma^{\ell}_{N1} &
      \cdots & B_N \Gamma^{\ell}_{NN}\end{bmatrix} \nonumber \\ %
    &\qquad - \begin{bmatrix} \Pi^g_{1}S^g & \Pi^{\ell}_{11}S^{\ell}_1
      & \cdots & \Pi^{\ell}_{1N}S^{\ell}_N \\ \vdots & \vdots & &
      \vdots \\ \Pi^g_{N}S^g & \Pi^{\ell}_{N1}S^{\ell}_1 & \cdots &
      \Pi^{\ell}_{NN}S^{\ell}_N \end{bmatrix} \nonumber \\ %
    &\qquad + \begin{bmatrix} B^{d^g}_1 & B^{d^{\ell}}_1 & & 0 \\
      \vdots & & \ddots & \\ B^{d^g}_1 & 0 & &
      B^{d^{\ell}}_{N} \end{bmatrix} \label{eq:eqn1}
  \end{align}
  Eq. \eqref{eq:eqn1} can be decomposed into the set of equations
  \begin{equation}\label{eq:k_g_1}
    A_k\Pi^g_{k} + B_k\Gamma^g_{k} - \Pi^g_{k}S^g + B^{d^g}_k = 0
  \end{equation}
  and 
  \begin{equation}\label{eq:k_l_1}
    A_k\Pi^{\ell}_{kk} + B_k\Gamma^{\ell}_{kk} - \Pi^{\ell}_{kk}S^{\ell}_k + B^{d^{\ell}}_k = 0 
  \end{equation}
  and for $j \neq k$,
  \begin{equation}\label{eq:kj_l_1}
    A_k \Pi^{\ell}_{kj} + B_k \Gamma^{\ell}_{kj} - \Pi^{\ell}_{kj}S^{\ell}_j = 0.
  \end{equation}
  The second equation $\mathcal{C}_e \Pi + \mathcal{D}_e \Gamma +
  \mathcal{D}_{ed} = 0$ of \eqref{eq:overall_regulatoreqn} yields
  \begin{align}
    0 &= 
    \begin{bmatrix} C^e_{1} \Pi^g_{1} & C^e_{1} \Pi^{\ell}_{11} & \cdots &
      C^e_{1} \Pi^{\ell}_{1N} \\ \vdots & \vdots & & \vdots \\ C^e_{N}
      \Pi^g_{N} & C^e_{N} \Pi^{\ell}_{N1} & \cdots & C^e_{N}
      \Pi^{\ell}_{NN} \end{bmatrix} \nonumber \\ %
    &\qquad + \begin{bmatrix} D^e_{1} \Gamma^g_{1} & D^e_{1}
      \Gamma^{\ell}_{11} & \cdots & D^e_{1} \Gamma^{\ell}_{1N} \\ \vdots &
      \vdots & & \vdots \\ D^e_{N} \Gamma^g_{N} & D^e_{N}
      \Gamma^{\ell}_{N1} & \cdots & D^e_{N}
      \Gamma^{\ell}_{NN}\end{bmatrix} \nonumber \\ %
    &\qquad + \begin{bmatrix} D^{ed^g}_{1} & D^{ed^{\ell}}_{1} & & 0 \\
      \vdots & & \ddots & \\ D^{ed^g}_{N} & 0 & &
      D^{ed^{\ell}}_{N}\end{bmatrix} \label{eq:eqn2}
  \end{align}
  Eq. \eqref{eq:eqn2} can be decomposed into the set of equations
  \begin{equation}\label{eq:k_g_2}
    C^e_{k} \Pi^g_{k} + D^e_{k} \Gamma^g_{k} + D^{ed^g}_k = 0
  \end{equation}
  and
  \begin{equation}\label{eq:k_l_2}
    C^e_{k} \Pi^{\ell}_{kk} + D^e_{k} \Gamma^{\ell}_{kk} + D^{ed^{\ell}}_{k} = 0
  \end{equation}
  and for $j \neq k$,
  \begin{equation}\label{eq:kj_l_2}
    C^e_{k} \Pi^{\ell}_{kj} + D^e_{k} \Gamma^{\ell}_{kj} = 0.
  \end{equation}

  (\emph{``Only if'':}) Suppose that \eqref{eq:overall_Pi},
  \eqref{eq:overall_Gamma} solve \eqref{eq:overall_regulatoreqn}.
  Then, according to \eqref{eq:k_g_1}, \eqref{eq:k_g_2} and
  \eqref{eq:k_l_1}, \eqref{eq:k_l_2}, the regulator equations
  \eqref{eq:local_regulatoreqn_g_compact} and
  \eqref{eq:local_regulatoreqn_l_compact} are solved by $\Pi^g_k$,
  $\Gamma^g_k$ and $\Pi^{\ell}_{k} = \Pi^{\ell}_{kk}$, $\Gamma^{\ell}_k =
  \Gamma^{\ell}_{kk}$. 

  (\emph{``If'':}) Suppose that $\Pi^g_k$, $\Gamma^g_k$ and
  $\Pi^{\ell}_{kk}$, $\Gamma^{\ell}_{kk}$ solve
  \eqref{eq:local_regulatoreqn_g_compact} and
  \eqref{eq:local_regulatoreqn_l_compact}.  It is easy to see that
  \eqref{eq:kj_l_1} and \eqref{eq:kj_l_2} can be satisfied by the
  choice $\Pi^{\ell}_{kj} = 0$ and $\Gamma^{\ell}_{kj} = 0$ for $k,j \in
  \mathcal{N}$ with $k \neq j$.  Consequently, \eqref{eq:overall_Pi},
  \eqref{eq:overall_Gamma} with $\Pi^{\ell}_{kj} = 0$ and $\Gamma^{\ell}_{kj} =
  0$ for $k,j \in \mathcal{N}$, $k \neq j$, solve
  \eqref{eq:overall_regulatoreqn}. 
\end{IEEEproof}

\subsection{List of Assumptions}

\begin{asmpt}\label{asmpt:stabilizable} 
  The pair $(A_k,B_k)$ is stabilizable for all $k \in \mathcal{N}$.
\end{asmpt}

\begin{asmpt}\label{asmpt:detectable_k}
  The pair
  \begin{equation*}
    \left( \begin{bmatrix} A_k & B^{d^{\ell}}_k \\ 0 & S^{\ell}_k \end{bmatrix},
      \begin{bmatrix} C_k & D^{d^{\ell}}_k \end{bmatrix} \right)
  \end{equation*}
  is detectable for all $k \in \mathcal{N}$.
\end{asmpt}

\begin{asmpt}\label{asmpt:detectable_1}  
  Agent $1$ has direct access to the signal $d^g$.
\end{asmpt}

\begin{asmpt}\label{asmpt:regulatoreqn_k}
  The regulator equations
  \begin{subequations}
    \label{eq:local_regulatoreqn_g}
    \begin{align}
      &A_k\Pi^g_{k} + B_k\Gamma^g_{k} - \Pi^g_{k}S^g + B^{d^g}_k = 0
      \label{eq:local_regulatoreqn_g_1} \\
      &C^e_{k} \Pi^g_{k} + D^e_{k} \Gamma^g_{k} + D^{ed^g}_k = 0
      \label{eq:local_regulatoreqn_g_2}
    \end{align}
  \end{subequations}
  and
  \begin{subequations}
    \label{eq:local_regulatoreqn_l}
    \begin{align}
      &A_k\Pi^{\ell}_{k} + B_k\Gamma^{\ell}_{k} -
      \Pi^{\ell}_{k}S^{\ell}_k + B^{d^{\ell}}_{k} =
      0 \label{eq:local_regulatoreqn_l_1} \\
      &C^e_{k} \Pi^{\ell}_{k} + D^e_{k} \Gamma^{\ell}_{k} +
      D^{ed^{\ell}}_{k} = 0
      \label{eq:local_regulatoreqn_l_2}
    \end{align}
  \end{subequations}
  have a solution $\Pi^g_{k}$, $\Gamma^g_{k}$, $\Pi^{\ell}_{k}$,
  $\Gamma^{\ell}_{k}$ for all $k \in \mathcal{N}$.
\end{asmpt}

\begin{asmpt}\label{asmpt:spanningtree}
  The communication topology is described by a directed connected
  graph $\mathcal{G}$ and node $1$ is the root of a spanning tree.
\end{asmpt}

In the following, we discuss each assumption separately in order to
point out the importance.

\emph{Assumption~\ref{asmpt:stabilizable}} is equivalent to
stabilizability of the pair $(\mathcal{A},\mathcal{B})$ due to the
structure of \eqref{eq:overall_plant}.  Moreover, it is required for
{\bf P1)} since the plant \eqref{eq:agent_undisturbed} may be
unstable.  

\emph{Assumption~\ref{asmpt:detectable_k}} is necessarily
satisfied if the pair \eqref{eq:detectable_pair_overall} is detectable
due to the structure of \eqref{eq:overall_plant},
\eqref{eq:overall_exosystem}.  The assumption that
\eqref{eq:detectable_pair_overall} is detectable causes no loss of
generality, as discussed in \cite{Knobloch1993}.  Note that
detectability of \eqref{eq:detectable_pair_overall} does {\bf not}
imply detectability of all pairs
\begin{equation}\label{eq:detectable_pair_big_k}
  \left( \begin{bmatrix} A_k & B^{d^g}_k & B^{d^{\ell}}_k \\ 0 & S^g & 0
      \\ 0 & 0 & S^{\ell}_k\end{bmatrix},
    \begin{bmatrix} C_k & D^{d^g}_k & D^{d^{\ell}}_k \end{bmatrix} \right).
\end{equation}
The pair \eqref{eq:detectable_pair_overall} is detectable if
\eqref{eq:detectable_pair_big_k} is detectable for at least one agent
$k \in \mathcal{N}$.  This agent must be the root of a spanning tree
in order to solve the distributed regulation problem.  The agents have
to cooperate in order to obtain an estimate of $d^g$.

\emph{Assumption~\ref{asmpt:detectable_1}} along with
Assumption~\ref{asmpt:detectable_k} guarantees detectability of
\eqref{eq:detectable_pair_overall}.  Following \cite{Su2012b}, we
call agent 1 the \emph{informed agent}.  Note that
Assumption~\ref{asmpt:detectable_1} can easily be relaxed to
detectability of the pair \eqref{eq:detectable_pair_big_k} for agent
$k=1$.  We work with the stricter assumption for ease of presentation.

\emph{Assumption~\ref{asmpt:regulatoreqn_k}} is necessarily satisfied
if there exists a centralized regulator solving
Problem~\ref{prob:regulation}.  It is well known that the solvability
of the regulator equation \eqref{eq:overall_regulatoreqn} is a
necessary and sufficient condition for the solvability of the output
regulation problem, as stated in
Theorem~\ref{thm:centralized_regulator}.  Due to the structure of the
overall generalized plant \eqref{eq:overall_plant} and exosystem
\eqref{eq:overall_exosystem}, the existence of a solution to
\eqref{eq:overall_regulatoreqn} is equivalent to the existence of
solutions to the local regulator equations
\eqref{eq:local_regulatoreqn_g}, \eqref{eq:local_regulatoreqn_l}, as
shown in Lemma~\ref{lem:regulatoreqn_k}.  Hence, there is no loss of
generality in this assumption.

\emph{Assumption~\ref{asmpt:spanningtree}} is required for the
  construction of a distributed regulator, cf., \cite{Su2012b}.

\subsection{Synchronization with Pole Placement Constraints}

Before we present our solution to Problem~\ref{prob:regulation} in the
following Section~\ref{sec:regulator}, we briefly discuss the state
synchronization problem for a group of identical linear agents via
static diffusive couplings since we will encounter two such state
synchronization problems in the remainder of this paper.  We are
interested in a guaranteed performance of such networks in terms of a
desired decay rate of the synchronization error.  For this purpose, we
propose a novel design procedure for suitable coupling gains by pole
placement constraints.
\begin{lem}[\hspace{-0.01ex}\cite{Wieland2011b}]\label{lem:wieland}
  Consider a group of $N$ identical linear agents $\dot{x}_k = A x_k +
  B u_k$, where $x_k \in \mathbb{R}^{n_x}$, $A$ is not Hurwitz, and
  $(A,B)$ is stabilizable.  There exists a coupling gain matrix $K$
  such that the interconnected closed-loop system with static
  diffusive couplings $u_k = K \sum_{j=1}^N a_{kj}(x_j - x_k)$, where
  $a_{kj}$ are the entries of $A_{\mathcal{G}}$, satisfies $x_k(t) -
  x_j(t) \to 0$ as $t \to \infty$ for all $k,j \in \mathcal{N}$ and
  all initial conditions, if and only if the directed graph
  $\mathcal{G}$ describing the communication topology is connected.
\end{lem}
Design procedures for a suitable coupling gain $K$ can be found in
\cite{Wieland2011b}, \cite{Li2010}.  These procedures always lead to a
suitable $K$ if all assumptions of Lemma~\ref{lem:wieland} are
satisfied.  Here, we present a novel design procedure based on LMIs
which takes into account performance specifications in terms of pole
placement constraints.  For this purpose, we make use of the LMI
region introduced in \cite{Chilali1996}, which is defined as
\begin{equation*}
  \mathcal{R} = \{ z \in \mathbb{C} : L + zM + \bar{z}M^\mathsf{T} < 0\},
\end{equation*}
for some fixed real matrices $L = L^\mathsf{T}$ and $M$.  The region
$\mathcal{R}$ is a convex subset of $\mathbb{C}$.

\begin{thm}\label{thm:coupling_pp}
  Suppose that all assumptions of Lemma~\ref{lem:wieland} are
  satisfied and let $\mathcal{R} \subset \mathbb{C}^-$ be an LMI
  region defined by $L = L^\mathsf{T}$ and $M$.  If there exist real
  matrices $Y > 0$ and $Z$ such that for all $k \in \mathcal{N}
  \backslash \{1\}$, the LMIs
  \begin{equation}\label{eq:lmi_pp}
    L \otimes Y + M \otimes (AY - \lambda_k BZ) + M^\mathsf{T} \otimes (AY -
    \bar{\lambda}_k BZ)^\mathsf{T} < 0
  \end{equation}
  are satisfied, then the coupling gain $K = ZY^{-1}$ ensures state
  synchronization.  Moreover, the poles of all modes corresponding to
  the synchronization error $\tilde{x}''$ (defined in the proof) are
  contained in $\mathcal{R}$.
\end{thm}
\begin{IEEEproof}
  The closed-loop system consisting of $\dot{x}_k = A x_k + B u_k$ and
  $u_k = K \sum_{j=1}^N a_{kj}(x_j - x_k)$ is given by
  \begin{equation*}
    \dot{x} = ( I_N \otimes A - L_{\mathcal{G}} \otimes BK ) x,
  \end{equation*}
  We apply the state transformation $\tilde{x} = (T^{-1} \otimes
  I_{n^x}) x$ as introduced by \cite{Fax2004}, where the matrix $T$ is
  chosen such that
  \begin{enumerate}[\it i)]
  \item $\Lambda = T^{-1}L_{\mathcal{G}}T$ is upper-triangular,
  \item the first column of $T$ is the vector of ones $\mathbf{1}$,
  \item the first row of $T^{-1}$ is $p^\mathsf{T}$, with
    $p^\mathsf{T} L_{\mathcal{G}}=0^\mathsf{T}$ and
    $p^\mathsf{T}\mathbf{1}=1$.
  \end{enumerate}
  Note that $\mathbf{1}$ is the right eigenvector and $p^\mathsf{T}$
  is the normalized left eigenvector of $L_{\mathcal{G}}$
  corresponding to the zero eigenvalue.  This state transformation
  yields
  \begin{equation*}
    \dot{\tilde{x}} = ( I_N \otimes A - \Lambda \otimes BK ) \tilde{x}.
  \end{equation*}
  The matrix $(I_N \otimes A - \Lambda \otimes BK)$ is upper block
  triangular with blocks $A - \lambda_kBK$ on the diagonal and
  $\lambda_1 = 0$.  Let $\tilde{x}$ be partitioned into $\tilde{x}'$
  and $\tilde{x}''$.  Then, it follows that
  \begin{align*}
    \dot{\tilde{x}}' &= A \tilde{x}', &
      \dot{\tilde{x}}''&=
    \begin{bmatrix}
      A-\lambda_2BK & \star & \star \\ & \ddots & \star \\ 0 & &
      A-\lambda_NBK
    \end{bmatrix} \tilde{x}''.
  \end{align*}
  Let $T$ be partitioned as $T = [\mathbf{1} \;\; T'']$.  Then, it is
  easy to see that $x = (T \otimes I_{n^x})\tilde{x} = (\mathbf{1}
  \otimes I_{n^x})\tilde{x}' + (T'' \otimes I_{n^x})\tilde{x}''$.
  Since $\mathbf{1}$ is linearly independent of the columns of $T''$,
  state synchronization, i.e., $x_k(t) - x_j(t) \to 0$ as $t \to
  \infty$ for all $k,j \in \mathcal{N}$, is equivalent to
  $\tilde{x}''(t) \to 0$ as $t \to \infty$.  Consequently, the state
  component $\tilde{x}''$ can be seen as the synchronization error of
  the network.  Due to the block-triangular structure of the system
  matrix, the problem under consideration reduces to a simultaneous
  stabilization problem for the $N-1$ systems on the diagonal
  \begin{equation*}
    \dot{\tilde{x}}_k = (A - \lambda_k BK) \tilde{x}_k
  \end{equation*}
  for $k=2,...,N$ via $K$ and with the pole placement constraint
  expressed by the LMI region $\mathcal{R}$.  Note that the
  eigenvalues $\lambda_k$ of $L_{\mathcal{G}}$ are in general complex
  since $\mathcal{G}$ is directed.  If there exists a real symmetric
  matrix $Y > 0$ such that
  \begin{equation}\label{eq:lmi_pp_analysis}
    L \otimes Y + M \otimes (A - \lambda_k BK)Y + M^\mathsf{T} \otimes Y(A -
    \bar{\lambda}_k BK)^\mathsf{T} < 0,
  \end{equation}
  then all poles of $A - \lambda_k BK$ are contained in $\mathcal{R}$
  \cite{Chilali1996}.  The change of variable $Z = KY$ leads to the
  LMIs \eqref{eq:lmi_pp}, which proves the statement. 
\end{IEEEproof}

\begin{rmk}
  For real eigenvalues $\lambda_k$, the existence of a real matrix $Y
  > 0$ such that \eqref{eq:lmi_pp_analysis} is satisfied, is necessary
  and sufficient for all poles of $A - \lambda_k BK$ to be contained
  in $\mathcal{R}$ \cite{Chilali1996}.  The requirement that $Y$ be
  real instead of complex Hermitian \cite{Horn1991} may cause
  conservatism in the design procedure in
  Theorem~\ref{thm:coupling_pp}.  Moreover, conservatism is caused by
  the fact that we search for a common Lyapunov function, i.e., a
  common $Y$, for all $k \in \mathcal{N} \backslash \{1\}$.
\end{rmk}

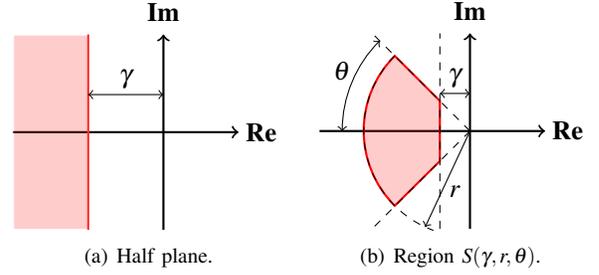
\begin{figure}[t]
  \centering%
  \subfigure[Half plane.]{%
    \begin{tikzpicture}[thick]
      \draw[->] (-1.5,0) -- (1.5,0); \node at (1.8,0) {\Re};%
      \draw[->] (0.5,-1.3) -- (0.5,1.3); \node at (0.5,1.6) {\Im};%
      \draw[color=red] (-0.5,-1.3) -- (-0.5,1.3);%
      \draw[color=white, fill=red, opacity=0.2] (-1.5,-1.3) rectangle
      (-0.5,1.3);%
      \draw[<->,thin] (-0.5,0.5) -- (0.5,0.5);%
      \node at (0,0.7) {$\gamma$};%
    \end{tikzpicture}%
    \label{fig:halfplane}%
  }%
  \hspace{2ex}%
  \subfigure[Region $S(\gamma,r,\theta)$.]{
    \begin{tikzpicture}[thick]
      \draw[->] (-1.5,0) -- (1.5,0); \node at (1.8,0) {\Re};%
      \draw[->] (0.5,-1.3) -- (0.5,1.3); \node at (0.5,1.6) {\Im};%
      \filldraw[fill=red, fill opacity=0.2, draw=red] (0.1,0.4) --
      (-0.5,1) arc (135:225:1.41) -- (0.1,-0.4) -- cycle;%
      \draw[dashed,thin] (0.5,0) -- (-0.8,1.3);%
      \draw[dashed,thin] (0.5,0) -- (-0.8,-1.3);%
      \draw[dashed,thin] (0.1,-1.3) -- (0.1,1.3);%
      \draw[<->,thin] (0.1,0.5) -- (0.5,0.5);%
      \node at (0.3,0.7) {$\gamma$};%
      \draw[<->,thin] (-0.7,1.2) arc (135:180:1.7);%
      \node at (-1.2,0.8) {$\theta$};%
      \draw[dashed,thin] (-0.5,1) arc (135:250:1.41);%
      \draw[->,thin,anchor=west,rotate around={245:(0.5,0)}] (0.5,0) -- (1.91,0);%
      \node at (0.3,-0.8) {$r$};%
    \end{tikzpicture}%
 }%
  \caption{Two exemplary LMI regions.}
  \label{fig:regions}
\end{figure}

Theorem~\ref{thm:coupling_pp} allows to include performance
specifications such as a desired decay rate or damping ratio for the
synchronization error into the design of the coupling gain.  A variety
of convex sets such as vertical strips, horizontal strips, disks,
conic sectors as well as intersections thereof can be expressed as LMI
region, cf., \cite{Chilali1996}.  Two practical examples are
illustrated in Fig.~\ref{fig:regions}.
The half plane in Fig.~\ref{fig:halfplane} is expressed by the scalars
$L = 2\gamma$ and $M = 1$.  For this special case, the synthesis LMIs
in Theorem~\ref{thm:coupling_pp} reduce to those in
\cite[Theorem~4.1]{Wieland2011b}.  The region $S(\gamma,r,\theta)$ is
expressed by $L = \diag(L_1,L_2,L_3)$ and $M = \diag(M_1,M_2,M_3)$
which describes an intersection of the following three regions:
\begin{align*}
 &\text{Half plane:} &  L_1 &= 2\gamma, & M_1 &= 1. \\
 &\text{Disc:} & L_2 &= \begin{bmatrix} -r & 0 \\ 0 & -r\end{bmatrix},
  & M_2 &= \begin{bmatrix} 0 & 1 \\ 0 & 0 \end{bmatrix}.\\
 &\text{Conical sector:} & L_3 &= 0, & M_3 &= \begin{bmatrix} \sin \theta &
    \cos \theta \\ - \cos \theta & \sin \theta \end{bmatrix}.
\end{align*}
Pole placement in the region $S(\gamma,r,\theta)$ guarantees a minimum
decay rate $\gamma$, minimum damping ratio $\zeta = \cos(\theta)$, and
a maximum undamped frequency $\omega_d = r \sin \theta$.

\section{The Distributed Regulator}\label{sec:regulator}

\begin{thm}\label{thm:nominal}
  Consider a group of agents~\eqref{eq:agent} with
  exosystems~\eqref{eq:global_exosystem}, \eqref{eq:local_exosystem}.
  Suppose that Assumptions~\ref{asmpt:stabilizable}--
  \ref{asmpt:spanningtree} are satisfied.  Then, a distributed
  regulator that solves Problem~\ref{prob:regulation} can be
  constructed as follows:
  \begin{itemize}
  \item For all $k \in \mathcal{N}$, choose $F_k$ such that $A_k - B_k
    F_k$ is Hurwitz.
  \item For all $k \in \mathcal{N}$, find a solution for
    \eqref{eq:local_regulatoreqn_g} and
    \eqref{eq:local_regulatoreqn_l} and set
    \begin{subequations}
      \label{eq:gains_G}
      \begin{align}
        G^g_k &= \Gamma^g_k + F_k \Pi^g_k \\
        G^{\ell}_k &= \Gamma^{\ell}_k + F_k \Pi^{\ell}_k
      \end{align}
    \end{subequations}
  \item Set $\hat{d}^g_1 = d^g$, and for all $k \in \mathcal{N}
    \backslash \{1\}$,
    \begin{equation}
      \dot{\hat{d}}^g_k = S^g \hat{d}^g_k + K \sum_{j\in
        \mathcal{N}_k} (\hat{d}^g_j - \hat{d}^g_k) \label{eq:dist_observer},
    \end{equation}
    where $K$ is chosen such that $S^g - \lambda_k K$ is stable for
    the non-zero eigenvalues $\lambda_k$, $k = 2,...,N$, of the
    Laplacian $L_{\mathcal{G}}$.
  \item For all $k \in \mathcal{N}$, choose $L_k$ such that
    \begin{equation}\label{eq:obsv_error_dynamics}
      \begin{bmatrix} A_k & B^{d^{\ell}}_k \\ 0 & S^{\ell}_k \end{bmatrix} 
      - L_k \begin{bmatrix} C_k & D^{d^{\ell}}_k \end{bmatrix}
    \end{equation}
    is Hurwitz and construct the observers
    \begin{align}\label{eq:local_observer}
      \begin{bmatrix} \dot{\hat{x}}_k \\
        \dot{\hat{d}}^{\ell}_k \end{bmatrix} &= \begin{bmatrix} A_k &
        B^{d^{\ell}}_k \\ 0 &
        S^{\ell}_k \end{bmatrix} \begin{bmatrix} \hat{x}_k \\
        \hat{d}^{\ell}_k \end{bmatrix} + \begin{bmatrix} B_k & B^{d^g}_k \\
        0 & 0 \end{bmatrix} \begin{bmatrix} u_k \\
        \hat{d}^g_k \end{bmatrix} + L_k \left(y_k - \hat{y}_k \right)
    \end{align}
    where $\hat{y}_k = C_k \hat{x}_k + D^{d^g}_k \hat{d}^g_k +
    D^{d^{\ell}}_k \hat{d}^{\ell}_k + D_k u_k$.
  \end{itemize}
  Finally, the control law for each agent $k \in \mathcal{N}$ is given
  by
  \begin{equation}\label{eq:nominal_control_law}
    u_k = -F_k \hat{x}_k + G^g_k \hat{d}^g + G^{\ell}_k \hat{d}^{\ell}_k.
  \end{equation}
\end{thm}

\begin{IEEEproof}
  Define the observer errors
  \begin{align}
    \label{eq:observer_errors}
    \epsilon^x_k &= x_k - \hat{x}_k, & \epsilon^{d^{\ell}}_k &=
    d^{\ell}_k - \hat{d}^{\ell}_k, & \epsilon^{d^g}_k &= d^g -
    \hat{d}^g_k.
  \end{align}
  The observer errors $\epsilon^x_k$ and $\epsilon^{d^{\ell}}_k$ satisfy
  \begin{align*}
    \begin{bmatrix} \dot{\epsilon}^x_k \\
      \dot{\epsilon}^{d^{\ell}}_k \end{bmatrix} &= \begin{bmatrix}
      \dot{x}_k \\ \dot{d}^{\ell}_k \end{bmatrix}
    - \begin{bmatrix} \dot{\hat{x}}_k \\ \dot{\hat{d}}^{\ell}_k \end{bmatrix} \\
    &= \begin{bmatrix} A_k & B^{d^{\ell}}_k \\ 0 &
      S^{\ell}_k \end{bmatrix} \begin{bmatrix} x_k \\
      d^{\ell}_k \end{bmatrix} + \begin{bmatrix} B_k & B^{d^g}_k \\
      0 & 0 \end{bmatrix} \begin{bmatrix} u_k \\
      d^g \end{bmatrix} \nonumber \\
    &\quad - \begin{bmatrix} A_k & B^{d^{\ell}}_k \\ 0 &
      S^{\ell}_k \end{bmatrix} \begin{bmatrix} \hat{x}_k \\
      \hat{d}^{\ell}_k \end{bmatrix} - \begin{bmatrix} B_k & B^{d^g}_k \\
      0 & 0 \end{bmatrix} \begin{bmatrix} u_k \\
      \hat{d}^g_k \end{bmatrix} - L_k \left(y_k - \hat{y}_k \right) \\
    &= \begin{bmatrix} A_k & B^{d^{\ell}}_k \\ 0 &
      S^{\ell}_k \end{bmatrix}
    \begin{bmatrix} \epsilon^x_k \\ \epsilon^{d^{\ell}}_k \end{bmatrix} +
    \begin{bmatrix} B^{d^g}_k \\ 0 \end{bmatrix} \epsilon^{d^g}_k
    \nonumber \\
    &\quad - L_k \begin{bmatrix} C_k & D^{d^g}_k &
      D^{d^{\ell}}_k \end{bmatrix} \left( \begin{bmatrix} x_k \\ d^g
        \\ d^{\ell}_k \end{bmatrix}
      - \begin{bmatrix} \hat{x}_k \\ \hat{d}^g_k \\
        \hat{d}^{\ell}_k \end{bmatrix} \right) \\
    &= \left( \begin{bmatrix} A_k & B^{d^{\ell}}_k \\ 0 &
        S^{\ell}_k \end{bmatrix} - L_k \begin{bmatrix} C_k &
        D^{d^{\ell}}_k \end{bmatrix} \right)
    \begin{bmatrix} \epsilon^x_k \\
      \epsilon^{d^{\ell}}_k \end{bmatrix} \nonumber \\
    &\quad + \left( \begin{bmatrix} B^{d^g}_k \\ 0 \end{bmatrix} -
      L_k D^{d^g}_k \right) \epsilon^{d^g}.
  \end{align*}
  The observer error $\epsilon^{d^g}_k$ satisfies $\epsilon^{d^g}_1 =
  0$ and for $k \in \mathcal{N} \backslash \{1\}$,
  \begin{align*}
    \dot{\epsilon}^{d^g}_k &= \dot{d}^g - \dot{\hat{d}}^g_k \\
    &= S^g \epsilon^{d^g}_k - K \sum_{j\in
        \mathcal{N}_k} (\hat{d}^g_j - \hat{d}^g_k) \\
    &= S^g \epsilon^{d^g}_k + K \sum_{j\in
        \mathcal{N}_k} (\epsilon^{d^g}_j - \epsilon^{d^g}_k).
  \end{align*}
  By Assumption~\ref{asmpt:spanningtree}, $\mathcal{G}$ contains a
  spanning tree rooted at $1$.  Hence, a suitable gain matrix $K$ in
  \eqref{eq:dist_observer} exists by Lemma~\ref{lem:wieland}.  A
  design procedure is given in Theorem~\ref{thm:coupling_pp}.
  Consequently, for all $k\in \mathcal{N}$, $\epsilon^{d^g}_k(t) =
  d^g(t) - \hat{d}^g_k(t) \to 0$ as $t \to \infty$. Hence, it follows
  that for all $k \in \mathcal{N}$, $\epsilon^x_k(t) \to 0$ and
  $\epsilon^{d^{\ell}}_k(t) \to 0$ as $t \to \infty$ since
  \eqref{eq:obsv_error_dynamics} is Hurwitz by construction.

  Next, we define the error variable
  \begin{equation}\label{eq:transient_component}
    \epsilon_k = x_k - \Pi_k^g d^g - \Pi_k^{\ell} d^{\ell}_k.
  \end{equation}
  Note that $\epsilon_k$ is the transient state component of agent
  $k$.  In particular, it is the state component in the complement of
  the subspace
  \begin{equation*}
    \mathcal{V}_k^+ = \{(x_k,d^g,d^{\ell}_k) : x_k = \Pi_k^g d^g + \Pi_k^{\ell} d^{\ell}_k\}.
  \end{equation*}
  As discussed in \cite{Knobloch1993}, the equations
  \eqref{eq:local_regulatoreqn_g_1}, \eqref{eq:local_regulatoreqn_l_1}
  express the fact that the subspace $\mathcal{V}_k^+$ is a controlled
  invariant subspace of the system
  \begin{equation*}
    \begin{bmatrix} \dot{x}_k \\ \dot{d}^g \\
      \dot{d}^{\ell}_k \end{bmatrix} = \begin{bmatrix} A_k & B^{d^g}_k &
      B^{d^{\ell}}_k \\ 0 & S^g & 0 \\ 0 & 0 & S^{\ell}_k \end{bmatrix}
    \begin{bmatrix} x_k \\ d^g \\ d^{\ell}_k \end{bmatrix}
    + \begin{bmatrix} B_k \\ 0 \\ 0 \end{bmatrix} u_k
  \end{equation*}
  and $\mathcal{V}_k^+$ is rendered invariant by $u_k = \Gamma^g_k d^g +
  \Gamma^{\ell}_k d^{\ell}_k$. Moreover, this subspace is annihilated by the
  regulation error map \eqref{eq:agent_e} due to
  \eqref{eq:local_regulatoreqn_g_2},
  \eqref{eq:local_regulatoreqn_l_2}.  As we will show next,
  $\mathcal{V}_k^+$ is rendered asymptotically stable by the proposed
  distributed regulator, and $e_k(t)$ indeed converges to zero for all
  initial conditions.  Using \eqref{eq:transient_component},
  \eqref{eq:agent_x}, \eqref{eq:global_exosystem},
  \eqref{eq:local_exosystem}, \eqref{eq:nominal_control_law},
  \eqref{eq:observer_errors}, \eqref{eq:local_regulatoreqn_g_1},
  \eqref{eq:local_regulatoreqn_l_1}, and \eqref{eq:gains_G}, we can
  compute
  \begin{align}
    \dot{\epsilon}_k &= \dot{x}_k - \Pi_k^g \dot{d}^g - \Pi_k^{\ell}
    \dot{d}^{\ell}_k \nonumber \\
    &= A_k x_k + B_k u_k + B^{d^g}_k d^g + B^{d^{\ell}}_k d^{\ell}_k -
    \Pi_k^g S^g d^g - \Pi_k^{\ell}
    S^{\ell}_k d^{\ell}_k \nonumber \\
    &= A_k x_k + B_k \left(-F_k \hat{x}_k + G^g_k \hat{d}^g_k +
      G^{\ell}_k \hat{d}^{\ell}_k \right)  \nonumber \\
    &\qquad + B^{d^g}_k d^g + B^{d^{\ell}}_k d^{\ell}_k - \Pi_k^g S^g
    d^g - \Pi_k^{\ell} S^{\ell}_k d^{\ell}_k \nonumber \\
    &= A_k x_k - B_k F_k \hat{x}_k + B_kG^g_k \hat{d}^g_k +
    B_kG^{\ell}_k  \hat{d}^{\ell}_k \nonumber \\
    &\qquad + B^{d^g}_k d^g + B^{d^{\ell}}_k d^{\ell}_k - \Pi_k^g S^g
    d^g - \Pi_k^{\ell} S^{\ell}_k d^{\ell}_k \nonumber \\
    &= A_k x_k - B_k F_k (x_k - \epsilon^x_k) \nonumber \\
    &\qquad + B_kG^g_k (d^g -
    \epsilon^{d^g}_k) + B_kG^{\ell}_k (d^{\ell}_k - \epsilon^{d^{\ell}}_k) \nonumber \\
    &\qquad + B^{d^g}_k d^g + B^{d^{\ell}}_k d^{\ell}_k - \Pi_k^g S^g
    d^g - \Pi_k^{\ell} S^{\ell}_k d^{\ell}_k \nonumber\\
    &= (A_k - B_k F_k) (\epsilon_k + \Pi_k^g d^g + \Pi_k^{\ell}
    d^{\ell}_k) +  B_k F_k \epsilon^x_k \nonumber \\
    &\qquad + B_kG^g_k d^g - B_kG^g_k \epsilon^{d^g}_k + B_kG^{\ell}_k
    d^{\ell}_k - B_kG^{\ell}_k \epsilon^{d^{\ell}}_k \nonumber \\
    &\qquad + B^{d^g}_k d^g + B^{d^{\ell}}_k d^{\ell}_k - \Pi_k^g S^g
    d^g - \Pi_k^{\ell} S^{\ell}_k d^{\ell}_k \nonumber \\
    &= (A_k - B_k F_k) (\epsilon_k + \Pi_k^g d^g + \Pi_k^{\ell}
    d^{\ell}_k) +  B_k F_k \epsilon^x_k \nonumber \\
    &\qquad+ B_k(\Gamma^g_k + F_k\Pi^g_k) d^g - B_kG^g_k \epsilon^{d^g}_k \nonumber \\
    &\qquad+ B_k(\Gamma^{\ell}_k + F_k\Pi^{\ell}_k) d^{\ell}_k
    - B_kG^{\ell}_k \epsilon^{d^{\ell}}_k \nonumber \\
    &\qquad + B^{d^g}_k d^g + B^{d^{\ell}}_k d^{\ell}_k - \Pi_k^g S^g
    d^g - \Pi_k^{\ell} S^{\ell}_k d^{\ell}_k \nonumber \\
    &= (A_k - B_k F_k) \epsilon_k + B_k F_k \epsilon^x_k- B_kG^g_k
    \epsilon^{d^g}_k - B_kG^{\ell}_k
    \epsilon^{d^{\ell}}_k \label{eq:epsilon_dot}
  \end{align}
  Since the observer errors $\epsilon^x_k$, $\epsilon^{d^g}_k$,
  $\epsilon^{d^{\ell}}_k$ vanish asymptotically and $A_k-B_kF_k$ is Hurwitz
  by construction, we can conclude that $\epsilon_k(t) \to 0$ as $t
  \to \infty$ for all $k \in \mathcal{N}$.  If $d^g(0) = 0$
  and $d^{\ell}_k(0)=0$ for all $k \in \mathcal{N}$,  then $x_k(t) =
  \epsilon_k(t) \to 0$ as $t \to \infty$, and since the observer
  errors vanish asymptotically, also $\hat{x}_k(t) \to 0$,
  $\hat{d}^g_k(t) \to 0$, and $\hat{d}^{\ell}_k(t) \to 0$ as $t \to
  \infty$.  Consequently, {\bf P1)} is fulfilled.

  It remains to show that the regulation errors $e_k$ converge to zero
  for all initial conditions.  Using \eqref{eq:agent_e},
  \eqref{eq:nominal_control_law}, \eqref{eq:observer_errors},
  \eqref{eq:local_regulatoreqn_g_2},
  \eqref{eq:local_regulatoreqn_l_2}, and \eqref{eq:gains_G}, we can
  compute
  \begin{align}
    e_k &= C^e_k x_k + D^{e}_k u_k + D^{ed^g}_k d^g + D^{ed^{\ell}}_k
    d^{\ell}_k \nonumber \\
    &= C^e_k (\epsilon_k + \Pi_k^g d^g + \Pi_k^{\ell} d^{\ell}_k)
    + D^{e}_k u_k + D^{ed^g}_k d^g + D^{ed^{\ell}}_k d^{\ell}_k
    \nonumber \\
    &= C^e_k \epsilon_k +(C^e_k \Pi_k^g + D^{ed^g}_k) d^g + (C^e_k
    \Pi_k^{\ell} + D^{ed^{\ell}}_k) d^{\ell}_k + D^{e}_k u_k \nonumber
    \\
    &= C^e_k \epsilon_k +(C^e_k \Pi_k^g + D^{ed^g}_k) d^g + (C^e_k
    \Pi_k^{\ell} + D^{ed^{\ell}}_k) d^{\ell}_k \nonumber \\
    &\qquad + D^{e}_k (- F_k x_k + G^g_k d^g + G^{\ell}_k d^{\ell}_k) \nonumber \\
    &\qquad - D^{e}_k (- F_k \epsilon^x_k + G^g_k \epsilon^{d^g}_k +
    G^{\ell}_k \epsilon^{d^{\ell}}_k) \nonumber \\
    &= C^e_k \epsilon_k+(C^e_k \Pi_k^g + D^{ed^g}_k) d^g + (C^e_k
    \Pi_k^{\ell} + D^{ed^{\ell}}_k) d^{\ell}_k \nonumber \\
    &\qquad + D^{e}_k (- F_k (\epsilon_k + \Pi^g_k
    d^g + \Pi_k^{\ell} d^{\ell}_k) + G^g_k d^g + G^{\ell}_k d^{\ell}_k) \nonumber \\
    &\qquad - D^{e}_k (- F_k \epsilon^x_k + G^g_k \epsilon^{d^g}_k +
    G^{\ell}_k \epsilon^{d^{\ell}}_k) \nonumber \\
    &= (C^e_k - D^{e}_kF_k) \epsilon_k+ (C^e_k \Pi_k^g +
    D^e_k\Gamma^g_k + D^{ed^g}_k) d^g \nonumber \\
    &\qquad + (C^e_k \Pi_k^{\ell} + D^e_k \Gamma^{\ell}_k +
    D^{ed^{\ell}}_k) d^{\ell}_k \nonumber \\
    &\qquad - D^{e}_k (- F_k \epsilon^x_k + G^g_k \epsilon^{d^g}_k +
    G^{\ell}_k \epsilon^{d^{\ell}}_k) \nonumber \\
    &= (C^e_k - D^{e}_kF_k) \epsilon_k - D^{e}_k (- F_k \epsilon^x_k +
    G^g_k \epsilon^{d^g}_k + G^{\ell}_k
    \epsilon^{d^{\ell}}_k). \label{eq:ek}
  \end{align}
  Both the state component $\epsilon_k$ and the observer errors
  $\epsilon^x_k$, $\epsilon^{d^g}_k$, $\epsilon^{d^{\ell}}_k$ vanish
  asymptotically.  Hence, $e_k(t) \to 0$ as $t \to \infty$ for
  all $k \in \mathcal{N}$, i.e., {\bf P2)} is fulfilled and
  Problem~\ref{prob:regulation} is solved. 
\end{IEEEproof}

\begin{rmk}
  Problem~\ref{prob:regulation} can be solved by the distributed
  regulator proposed in \cite{Su2012b}, when
  \eqref{eq:overall_exosystem} is regarded as a single large
  exosystem.  However, according to the formulation in \cite{Su2012b},
  each agent reconstructs the full vector $d$ of all reference and
  disturbance signals acting on the group via a distributed estimation
  protocol, which leads to dynamic controllers of high order and is
  unnecessary.  Our distributed regulator takes the structure of
  \eqref{eq:overall_exosystem} into account: each agent estimates only
  the global generalized disturbance as well as the local generalized
  disturbance acting on itself.
\end{rmk}

The distributed regulator constructed in Theorem~\ref{thm:nominal}
solves Problem~\ref{prob:regulation}.  The following example shows
that this result allows to solve a platooning problem and illustrates
the distributed regulator.  Moreover, this examples points out a
limitation of the control scheme and motivates the extension of the
distributed regulator, which will be presented in the next section.

\begin{exmp}\label{exmp:platoon}
  Consider a group of $N=5$ vehicles in a platoon, each modeled as a
  double-integrator system of the form
  \begin{equation*}
    \dot{x}_k = \underbrace{\begin{bmatrix} 0 & 1 \\ 0 & 0 \end{bmatrix}}_{A} x_k
    + \underbrace{\begin{bmatrix} 0 \\ 1 \end{bmatrix}}_{B} u_k + 
    \underbrace{\begin{bmatrix} 0 \\ 1 \end{bmatrix}}_{B_w} w_k,
  \end{equation*}
  for all $k \in \mathcal{N}$.  Moreover, there is a virtual leader
  generating a reference signal for the group, given by $\dot{d^g} =
  S^g d^g$ with $S^g = A$.  The state $x_k \in \mathbb{R}^2$ consists
  of the vehicle position $s_k$ and velocity $v_k$, i.e., $x_k = [s_k
  \; v_k]^\mathsf{T}$.  The input $w_k$ is a local disturbance acting
  on vehicle $k$, e.g., due to a mass change, gear shift, or other
  external influence.  We assume a sinusoidal disturbance $w_2$ for
  agent $2$ and constant disturbances for all other agents.

  The task is to find a control law $u_k$ for each vehicle such that
  the following requirements are met for all initial conditions:
  \begin{enumerate}
  \item The velocities $v_k$ of all vehicles converge to the velocity
    commanded by the virtual leader:
    \begin{equation*}
      \forall k \in \mathcal{N}: \quad \lim_{t \to \infty} v_k(t)-[ 0 \;\; 1 ]d^g(t) = 0.
    \end{equation*}
  \item The relative distance of each vehicle with respect to the
    virtual leader converges to a desired constant value $r_k$:
    \begin{equation*}
      \forall k \in \mathcal{N}: \quad \lim_{t \to \infty}
      s_k(t) - [ 1 \;\; 0 ]d^g(t) = r_k.
    \end{equation*}
  \end{enumerate}
  The external local disturbance $w_k$ and local reference $r_k$ are
  combined into the local generalized disturbance signals $d^{\ell}_k$.
  For agent $2$, we have
  \begin{align*}
    \dot{d}^{\ell}_2 &= \underbrace{\begin{bmatrix} 0 & 0 & 0 \\ 0 & 0 & 1
        \\ 0 & -1 & 0 \end{bmatrix}}_{S^{\ell}_2} d^{\ell}_2, & \begin{bmatrix}
      r_2 \\ w_2 \end{bmatrix} &= \begin{bmatrix} 1 & 0 & 0 \\ 0 & 1 &
      0 \end{bmatrix} d^{\ell}_2.
  \end{align*}
  For all other agents, we have $d^{\ell}_k = [ r_k \;\; w_k
  ]^\mathsf{T}$ and
  \begin{equation*}
    \dot{d}^{\ell}_k = \underbrace{\begin{bmatrix} 0 & 0 \\ 0 &
        0 \end{bmatrix}}_{S^{\ell}_k} d^{\ell}_k.
  \end{equation*}
  This leads to $B^{d^{\ell}}_2 = B_w [ 0 \;\; 1 \;\; 0 ]$ and
  $B^{d^{\ell}}_k = B_w [ 0 \;\; 1 ]$ for all other agents.  We assume
  that each vehicle can communicate with its follower and predecessor
  in the platoon as illustrated by the graph in
  Fig.~\ref{fig:platoon_graph}.
  \begin{figure}[t]
    \centering
    \input{figures/platoon.tex}
    \caption{Communication graph and external signals of the platoon.}
    \label{fig:platoon_graph}
  \end{figure}
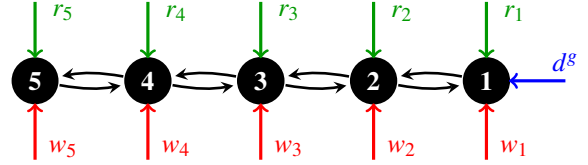  
  The platooning problem can be formulated as a distributed output
  regulation problem with regulation errors defined as
  \begin{equation*}
    e_k = \underbrace{\begin{bmatrix} 1 & 0 \\ 0 & 1 \end{bmatrix}}_{C^e_k} x_k
    + \underbrace{\begin{bmatrix} -1 & 0 \\ 0 &
        -1 \end{bmatrix}}_{D^{ed^g}_k} d^g + 
    \underbrace{\begin{bmatrix} -1 & 0 \\ 0 & 0 \end{bmatrix}}_{D^{ed^{\ell}}_k} d^{\ell}_k.
  \end{equation*}
  Note that for $k=2$, $D^{ed^{\ell}}_k$ has a third column of zeros.
  We assume that only agent $1$ has access to the global reference
  signal $d^g$ and all agents have access to their local reference
  signal $r_k$ and to their state $x_k$.  Local disturbance estimators
  as in \eqref{eq:local_observer} are used in order to reconstruct
  $w_k$.

  Simulation results for two different choices of the control gain
  $F_k$ and with local exogenous input signals according to
  Fig.~\ref{fig:platooning_signals} are shown in
  Fig.~\ref{fig:platooning}.  The distributed regulation problem is
  solved in both cases, i.e., the reference signals are tracked and
  the disturbance signals are rejected asymptotically.
  \begin{figure}[t]
    \setlength\figurewidth{0.7\linewidth}%
    \setlength\figureheight{0.25\figurewidth}%
    \centering
    \begin{tabular}{r}
      \subfigure[Local reference
      signals.]{\input{figures/sim_platoon_r.tikz}%
      } \\
      \subfigure[Local
      disturbances.]{\input{figures/sim_platoon_w.tikz}%
      }
    \end{tabular}
    \caption{Local exogenous input signals for Example~\ref{exmp:platoon}.}
  \label{fig:platooning_signals}
  \end{figure}
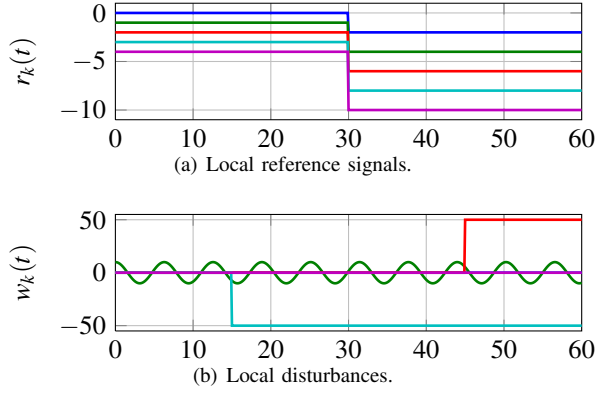
  \begin{figure}[t]
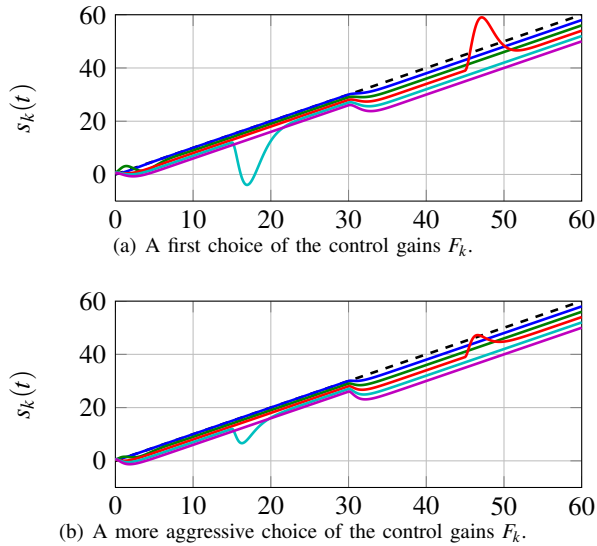

    \setlength\figurewidth{0.7\linewidth}%
    \setlength\figureheight{0.4\figurewidth}%
    \centering
    \begin{tabular}{r}
      \subfigure[A first choice of the control gains
      $F_k$.]{\input{figures/sim_platoon_x.tikz}%
      } \\
      \subfigure[A more aggressive choice of the control gains
      $F_k$.]{\input{figures/sim_platoon_x_loc.tikz}%
      }
    \end{tabular}
    \caption{Simulation results for Example~\ref{exmp:platoon}.}
  \label{fig:platooning}
  \end{figure}

  This examples illustrates an inherent limitation of the control
  scheme: The group does not react cooperatively on local disturbances
  acting on individual agents.  The steps in the constant disturbances
  $w_4$ and $w_3$ at $t = 15$ and $t = 45$, respectively, are rejected
  by the local controllers of these agents.  As can be seen in
  Fig.~\ref{fig:platooning}, a more aggressive choice of the control
  gains $F_k$ leads to a faster disturbance rejection.  But the other
  agents in the group have no information about the disturbance and
  cannot adjust their actions to this situation.  A more desirable
  and cooperative reaction on the local disturbances would be that the
  other vehicles in the platoon slow down or accelerate in order to
  maintain the desired relative distances during the transient phase.
  Maintaining the desired inter-vehicle distances is of higher
  importance than maintaining the desired velocity for a vehicle
  platoon.  Such performance requirements cannot be taken into
  account explicitly with the distributed regulator described in
  Theorem~\ref{thm:nominal}.
\end{exmp}

Up to now, the proposed control strategy is a hierarchical control
strategy.  Cooperation happens only on the network level in order to
spread the reference signal $d^g$ all over the network.  Such a
hierarchical approach has many benefits but also has its drawbacks,
cf., \cite{Seyboth2014a}.  In particular, there is no cooperative
reaction on the individual external disturbances.

\section{Improving the Transient Behavior:\\ Identical Agents}\label{sec:identical}

The distributed regulator presented in Section~\ref{sec:regulator}
solves the distributed output regulation problem and realizes a
cooperative behavior of the group.  In this section, we show that the
cooperative behavior can be improved significantly by a suitable
extension of the distributed regulator.  As motivated in
Example~\ref{exmp:platoon}, it is desirable to establish a
\emph{cooperative reaction on local disturbances}, meaning that the
group disagreement or synchronization error is kept small in transient
phases.  For this purpose, we present a novel distributed regulator
with additional couplings among the agents which stabilize the
synchronization error and guarantee a desired performance.  This is
our main contribution and improvement of the distributed regulator
compared to previous works \cite{Su2012b}.

In a first step, we consider groups of agents with identical dynamics.
In particular, we impose the following assumption.
\begin{asmpt}\label{asmpt:identical_agents}
  Assume that for all $k \in \mathcal{N}$,
  \begin{align*}
    A_k &= A, & B_k &= B, & C^e_k &= C^e, & D^e_k &= D^e, & F_k &= F.
  \end{align*}
\end{asmpt}
Note that not all matrices in \eqref{eq:agent} are required to be
identical for the individual agents.  In particular, the measurement
output maps \eqref{eq:agent_y} are allowed to be non-identical, and
the local exosystems \eqref{eq:local_exosystem} as well as the
generalized disturbance input matrices in \eqref{eq:agent_x},
\eqref{eq:agent_e}, can be non-identical.  This leaves great
flexibility in the problem formulation despite
Assumption~\ref{asmpt:identical_agents}.  Moreover,
Assumption~\ref{asmpt:identical_agents} will be relaxed in
Section~\ref{sec:nonidentical}.

Our goal is to improve the cooperative behavior of the group in
transient phases.  The disagreement of the group can be quantified
based on the transient state components $\epsilon_k$ defined in
\eqref{eq:transient_component}.  For this purpose, we consider the
transient synchronization errors defined by
\begin{equation}\label{eq:sync_error}
  \epsilon^s_k = \epsilon_k - \frac{1}{N} \sum_{j=1}^N \epsilon_j.
\end{equation}
From \eqref{eq:ek}, we know that $e_k = (C^e - D^eF)\epsilon_k$ under
Assumption~\ref{asmpt:identical_agents} and without observer errors.
Since $C^e - D^eF$ is identical for all $k \in \mathcal{N}$, agreement
of $\epsilon_k$ corresponds to agreement of the regulation errors $e_k$.
In the following, we propose a distributed regulator which solves
Problem~\ref{prob:regulation} and, at the same time, exponentially
stabilizes the synchronization errors $\epsilon^s_k$ with a certain
desired decay rate $\gamma > 0$.  In order to guarantee that this
decay rate can be achieved, we refine
Assumption~\ref{asmpt:stabilizable} as follows.
\begin{myasmptprime}{asmpt:stabilizable}
  \label{asmpt:stabilizableprime} 
  The pair $(A_k+\gamma I_{n^x_k},B_k)$, where $\gamma > 0$, is
  stabilizable for all $k \in \mathcal{N}$.
\end{myasmptprime}

We start with the full information case.  This allows to present the
main idea in a clear way and will be instrumental in the proof of the
output feedback case.

\begin{lem}[Full information regulator]\label{lem:couplings_fi}
  Consider a group of agents~\eqref{eq:agent} with
  exosystems~\eqref{eq:global_exosystem}, \eqref{eq:local_exosystem}.
  Suppose that Assumptions~\ref{asmpt:stabilizableprime},
  \ref{asmpt:regulatoreqn_k}, \ref{asmpt:spanningtree} and
  \ref{asmpt:identical_agents} are satisfied.  Then, a distributed
  full-information regulator that solves Problem~\ref{prob:regulation}
  and additionally achieves exponential stability of synchronization
  errors $\epsilon^s_k$ with decay rate $\gamma > 0$ can be
  constructed as follows:
  \begin{itemize}
  \item Choose $F$ such that $A-BF$ is Hurwitz.
  \item For all $k \in \mathcal{N}$, find a solution for
    \eqref{eq:local_regulatoreqn_g} and
    \eqref{eq:local_regulatoreqn_l} and set
    \begin{align*}
      G^g_k &= \Gamma^g_k + F \Pi^g_k \\
      G^{\ell}_k &= \Gamma^{\ell}_k + F \Pi^{\ell}_k 
    \end{align*}
  \item Choose $H$ such that for $k = 2,...,N$,
    \begin{equation}\label{eq:real_smaller_gamma}
      \max\{ \Re(\mu), \mu \in \sigma(A-BF-\lambda_k BH) \} < -\gamma,
    \end{equation}
    i.e., the maximal real part of all eigenvalues of the matrices
    $A-BF-\lambda_k BH$ is smaller than $-\gamma$.
  \end{itemize}
  Finally, the control law for each agent $k \in \mathcal{N}$ is given by
  \begin{equation}\label{eq:couplings_control_law_fi}
    u_k = -F x_k + G^g_k d^g_k + G^{\ell}_k d^{\ell}_k + H
    \sum_{j \in \mathcal{N}_k} \left(\epsilon_j - \epsilon_k \right).
  \end{equation}
\end{lem}

\begin{IEEEproof}
  Analogously to \eqref{eq:epsilon_dot}, the dynamics of $\epsilon_k$
  can be computed as
  \begin{equation}\label{eq:error_network}
    \dot{\epsilon}_k = (A - B F) \epsilon_k + BH \sum_{j \in \mathcal{N}_k}
    \left(\epsilon_j - \epsilon_k \right).
  \end{equation}
  The novel term in the control law
  \eqref{eq:couplings_control_law_fi} couples the transient state
  components of the agents.  Eq. \eqref{eq:error_network} is a
  diffusively coupled network of $N$ identical stable linear systems.
  The same transformation as in the proof of
  Theorem~\ref{thm:coupling_pp}, i.e., $\tilde{\epsilon} = (T^{-1}
  \otimes I_{n^x}) \epsilon$, leads to
  \begin{equation*}
    \dot{\tilde{\epsilon}} = \big(I_N \otimes (A-BF) - \Lambda \otimes
    BH \big) \tilde{\epsilon}.
  \end{equation*}
  Let $\tilde{\epsilon}$ be partitioned into $\tilde{\epsilon}'$ and
  $\tilde{\epsilon}''$.  Then, $\dot{\tilde{\epsilon}}' = (A-BF)
  \tilde{\epsilon}'$ and
  \begin{align*}
    \dot{\tilde{\epsilon}}''&=
    \begin{bmatrix}
      A-BF-\lambda_2BH & \star & \star \\ & \ddots & \star \\ 0 & &
      A-BF-\lambda_NBH 
    \end{bmatrix} \tilde{\epsilon}''.
  \end{align*}
  By \eqref{eq:real_smaller_gamma} the maximal real part of the latter
  matrix is smaller than $-\gamma$.  Note that such a gain $H$ exists
  due to Assumptions~\ref{asmpt:stabilizableprime} and
  Lemma~\ref{lem:wieland}.  Hence, there exists a constant
  $\tilde{c} > 0$ such that
  \begin{equation*}
    \|\tilde{\epsilon}''(t)\| \leq \tilde{c} \|\tilde{\epsilon}''(0)\|
    e^{-\gamma t},
  \end{equation*}
  for all $t \geq 0$, cf., \cite{Bernstein2009}.  We define the
  projection matrix $P = I_N -
  \frac{1}{N}\mathbf{1}\mathbf{1}^\mathsf{T}$ and partition $T = [
  \mathbf{1} \;\; T'' ]$.  Then, the stack vector of synchronization
  errors \eqref{eq:sync_error} is given by $\epsilon^s = (P \otimes
  I_{n^x}) \epsilon = ( PT \otimes I_{n^x}) \tilde{\epsilon} = ( PT''
  \otimes I_{n^x}) \tilde{\epsilon}''$ since $P\mathbf{1} = 0$.
  Hence, it follows that there exists a constant $c > 0$ such that
  \begin{equation*}
    \|\epsilon^s(t)\| \leq c \|( P \otimes I_{n^x}) \epsilon(0) \| e^{-\gamma t}.
  \end{equation*}
  Consequently, the synchronization error is exponentially stable with
  a decay rate of at least $\gamma$.

  The matrix $A-BF$ is Hurwitz by the choice of $F$, which guarantees
  that $\tilde{\epsilon}'(t) \to 0$ as $t \to \infty$.  Therefore, it
  holds that $\epsilon(t) = (T \otimes I_{n^x})\tilde{\epsilon}(t) \to
  0$ as $t \to \infty$.  Since $e_k = (C^e - D^eF)\epsilon_k$, this
  shows that Problem~\ref{prob:regulation} is solved.
\end{IEEEproof}

The novel control law \eqref{eq:couplings_control_law_fi} solves
Problem~\ref{prob:regulation} and additionally enforces
synchronization of the regulation errors with a desired decay rate and
therefore has the desired effect, that is, a cooperative reaction of
the group on disturbances acting on individual agents.  The gain
matrices $F$ and $H$ allow to tune separately the local disturbance
rejection of each agent and the synchronization of the group.
Theorem~\ref{thm:coupling_pp} serves as a design method for $H$.

The full information control law \eqref{eq:couplings_control_law_fi}
is impractical since the agents do not have direct access to their
state and generalized disturbance signals.  The following theorem
shows that \eqref{eq:couplings_control_law_fi} can be implemented
based on observers, analogously to \eqref{eq:nominal_control_law} in
Theorem~\ref{thm:nominal}.  In order to guarantee that the desired
decay rate can be achieved, we refine
Assumption~\ref{asmpt:detectable_k}.
\begin{myasmptprime}{asmpt:detectable_k}
  \label{asmpt:detectableprime} 
  The pair 
  \begin{equation*}
    \left( \begin{bmatrix} A_k & B^{d^{\ell}}_k \\ 0 &
        S^{\ell}_k \end{bmatrix} + \eta I_{(n^x_k + n^{d^{\ell}}_k)},
      \begin{bmatrix} C_k & D^{d^{\ell}}_k \end{bmatrix} \right),
  \end{equation*}
  where $\eta > 0$, is detectable for all $k \in \mathcal{N}$.
\end{myasmptprime}

\begin{thm}[Distributed output feedback
  regulator]\label{thm:couplings}
  Consider a group of agents~\eqref{eq:agent} with
  exosystems~\eqref{eq:global_exosystem}, \eqref{eq:local_exosystem}.
  Suppose that Assumptions~\ref{asmpt:stabilizableprime},
  \ref{asmpt:detectableprime}, \ref{asmpt:detectable_1},
  \ref{asmpt:regulatoreqn_k}, \ref{asmpt:spanningtree} and
  \ref{asmpt:identical_agents} are satisfied.  Construct a distributed
  regulator as in Theorem~\ref{thm:nominal} and choose $L_k$ and $K$
  such that the corresponding systems are exponentially stable with
  decay rate $\eta > \gamma > 0$.  Choose $H$ as in
  \eqref{eq:real_smaller_gamma}.  Then, the control law
  \begin{equation}\label{eq:couplings_control_law}
    u_k = -F \hat{x}_k + G^g_k \hat{d}^g_k + G^{\ell}_k \hat{d}^{\ell}_k + H
    \sum_{j \in \mathcal{N}_k} \left(\hat{\epsilon}_j - \hat{\epsilon}_k\right),
  \end{equation}
  solves Problem~\ref{prob:regulation} and additionally achieves
  exponential stability of the synchronization errors $\epsilon^s_k$
  with decay rate $\gamma > 0$.
\end{thm}

\begin{IEEEproof}
  The novel term in \eqref{eq:couplings_control_law} has no influence
  on the dynamics of the observer errors $\epsilon^x_k$,
  $\epsilon^{d^{\ell}}_k$, $\epsilon^{d^g}_k$.  They obey the same dynamics
  as in the proof of Theorem~\ref{thm:nominal}.  Moreover, by
  assumption, there exist constants $c_k > 0$ such that for all $t
  \geq 0$,
  \begin{equation}\label{eq:exp_observer_errors}
    \left\| \begin{bmatrix} \epsilon^x_k(t) \\ \epsilon^{d^g}_k(t) \\
        \epsilon^{d^{\ell}}_k(t) \end{bmatrix} \right\| \leq c_k \left\| 
      \begin{bmatrix} \epsilon^x_k(0) \\ \epsilon^{d^g}_k(0) \\ 
        \epsilon^{d^{\ell}}_k(0) \end{bmatrix} \right\| e^{-\eta t}.
  \end{equation}
  Note that
  \begin{align*}
    \hat{\epsilon}_k &= \hat{x}_k - \Pi^g_k \hat{d}^g_k - \Pi_k^{\ell} \hat{d}^{\ell}_k \\
    &= x_k - \epsilon^x_k - \Pi^g_k (d^g - \epsilon^{d^g}_k) - \Pi_k^{\ell}
    (d^{\ell}_k - \epsilon^{d^{\ell}}_k) \\
    &= \epsilon_k - \epsilon^x_k + \Pi_k^g \epsilon^{d^g}_k + \Pi_k^{\ell}
    \epsilon^{d^{\ell}}_k.
  \end{align*}
  Analogously to the proof of Theorem~\ref{thm:nominal}, the dynamics
  of $\epsilon_k$ can be computed as
  \begin{align*}
    \dot{\epsilon}_k 
    &= (A - B F) \epsilon_k + BH \sum_{j \in \mathcal{N}_k}
    \left(\epsilon_j - \epsilon_k\right) + \delta_k(t),
  \end{align*}
  where $\delta_k(t)$ captures the influence of the observer errors
  \begin{align*}
    \delta_k(t) &= B F \epsilon^x_k - B G^g_k
    \epsilon^{d^g}_k - B G^{\ell}_k \epsilon^{d^{\ell}}_k  \\ 
    &\; - BH \sum_{j \in \mathcal{N}_k} \left(\epsilon^x_j -
      \Pi_j^g \epsilon^{d^g}_j - \Pi_j^{\ell} \epsilon^{d^{\ell}}_j -
      \epsilon^x_k + \Pi_k^g \epsilon^{d^g}_k + \Pi_k^{\ell}
      \epsilon^{d^{\ell}}_k\right).
  \end{align*}
  We define the stack vector $\delta = [ \delta_1^\mathsf{T} \; \cdots
  \; \delta_N^\mathsf{T} ]^\mathsf{T}$.  From
  \eqref{eq:exp_observer_errors} it follows that there exists a
  constant $c_\delta$ such that for all $t \geq 0$,
  \begin{equation}\label{eq:exp_delta}
    \|\delta(t)\| \leq c_\delta \|\delta(0)\| e^{-\eta t}.
  \end{equation}
  Analogously to the proof of Lemma~\ref{lem:couplings_fi}, we obtain
  \begin{equation*}
    \dot{\tilde{\epsilon}} = \big( I_N \otimes (A-BF) - \Lambda \otimes
    BH \big) \tilde{\epsilon} + (T^{-1} \otimes I_{n^x}) \delta.
  \end{equation*}
  With \eqref{eq:exp_delta} and since $\eta > \gamma$, it follows that
  $\tilde{\epsilon}''(t) \to 0$ exponentially as $t \to \infty$ with
  decay rate $\gamma$.  

  Moreover, we also have $\tilde{\epsilon}'(t) \to 0$ as $t \to
  \infty$.  Since $\epsilon_k(t) \to 0$ as $t \to \infty$ for all $k
  \in \mathcal{N}$, the coupling term in
  \eqref{eq:couplings_control_law} vanishes asymptotically.
  Consequently, analogously to the proof of Theorem~\ref{thm:nominal},
  it holds that $e_k(t) \to 0$ as $t \to \infty$ and
  Problem~\ref{prob:regulation} is solved.
\end{IEEEproof}

\begin{exmp}\label{exmp:platoon_couplings}
  We consider the same setup as in Example~\ref{exmp:platoon}. Now, we
  use the novel distributed regulator as described in
  Theorem~\ref{thm:couplings}.  Simulation results with local
  exogenous input signals according to
  Fig.~\ref{fig:platooning_signals} are shown in
  Fig.~\ref{fig:platooning_novel}.  The effect of the disturbances on
  the inter-vehicle distances is indeed rejected much more
  efficiently, compared to Example~\ref{exmp:platoon}.  The platoon
  reacts cooperatively on the local disturbances and maintains small
  synchronization errors.  The two different simulations in
  Fig.~\ref{fig:platooning_novel} with different choices of the
  control gains $F$ and $H$ illustrate the flexibility of the control
  design.  Depending on the requirements, it is possible to put more
  emphasis on the disturbance rejection or on the synchronization.
  \begin{figure}[t]
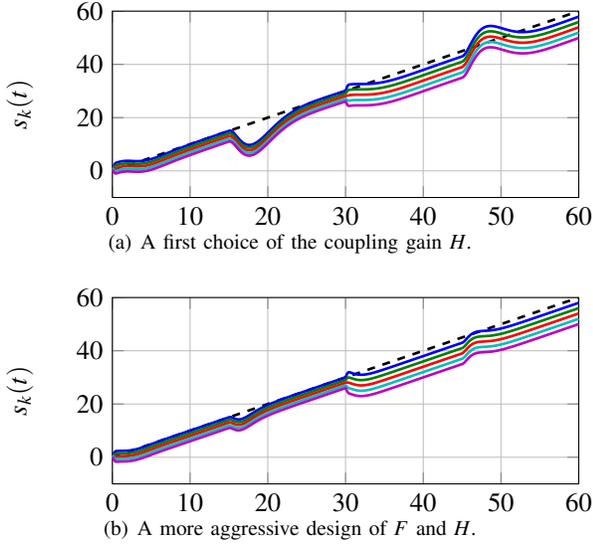

    \setlength\figurewidth{0.7\linewidth}%
    \setlength\figureheight{0.4\figurewidth}%
    \centering
    \begin{tabular}{r}
      \subfigure[A first choice of the coupling gain
      $H$.]{\input{figures/sim_platoon_x_net.tikz}%
      } \\
      \subfigure[A more aggressive design of $F$ and
      $H$.]{\input{figures/sim_platoon_x_loc_net.tikz}%
      }
    \end{tabular}
    \caption{Simulation results for Example~\ref{exmp:platoon_couplings}.}
  \label{fig:platooning_novel}
  \end{figure}
\end{exmp}

\section{Improving the Transient Behavior:\\ Non-identical
  Agents}\label{sec:nonidentical}

In this section, we aim at relaxing
Assumption~\ref{asmpt:identical_agents}.  In case of non-identical
agents, an analogue coupling term as in
\eqref{eq:couplings_control_law_fi} leads to a diffusively coupled
network of non-identical stable linear systems in
\eqref{eq:error_network}.  In this case, it is hard to find a suitable
coupling gain $H$.  In case of non-identical state dimensions $n^x_k$
such couplings cannot be realized at all.  We exclude this case and
treat the non-identical agents as perturbed versions of a nominal
system.  We present a design method for $H$ based on this nominal
system and use robustness arguments in order to prove exponential
stability of the heterogeneous network.  For ease of presentation, we
discuss only the full information case in the following.  The output
feedback implementation can be carried out analogously to
Theorem~\ref{thm:couplings}.

\begin{figure}[b]
  \centering
  \begin{tikzpicture}[scale=0.8,thick]
    \draw (-0.6,-0.6) rectangle (0.6,0.6); \node at (0,0) {$P$};
    \draw[->] (-1.6,-0.2) -- (-0.6,-0.2); \node at (-2,-0.2)
    {$\nu_k$}; \draw[->] (0.6,-0.2) -- (1.6,-0.2); \node at (2,-0.2)
    {$e_k$}; \draw (-0.6,1) rectangle (0.6,2.2); \node at (0,1.6)
    {$\Delta_k$}; \draw [->] (-0.6,1.6) -- (-1.2,1.6) -- (-1.2,0.2) --
    (-0.6,0.2); \node at (-1.6,0.8) {$\omega_k$}; \draw [<-] (0.6,1.6)
    -- (1.2,1.6) -- (1.2,0.2) -- (0.6,0.2); \node at (1.6,0.8)
    {$\zeta_k$};
  \end{tikzpicture}
  \caption{Nominal agent dynamics $P$ with individual perturbation
    $\Delta_k$.}
  \label{fig:uncertain_agent}
\end{figure}
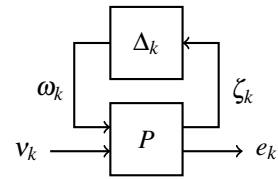

Suppose that we implement a control law like
\eqref{eq:couplings_control_law} in a group of non-identical agents,
i.e.,
\begin{equation}\label{eq:couplings_control_law_heterog}
  u_k = -F_k x_k + G^g_k d^g_k + G^{\ell}_k d^{\ell}_k 
  + H \sum_{j \in \mathcal{N}_k} \left(\epsilon_j - \epsilon_k \right)
\end{equation}
where $F_k$, $G^g_k$, $G^{\ell}_k$ are designed as in
Theorem~\ref{thm:nominal} and $H$ is a coupling gain to be specified
in the following.  Then, analogously to \eqref{eq:error_network}, the
error variables $\epsilon_k$ obey the dynamics
\begin{equation*}
  \dot{\epsilon}_k = (A_k - B_k F_k) \epsilon_k + B_k H \sum_{j \in \mathcal{N}_k}
  \left(\epsilon_j - \epsilon_k \right),
\end{equation*}
where $A_k-B_kF_k$ is Hurwitz.  The key assumption in the following is
that the systems
\begin{equation}\label{eq:nonidentical_agent}
  \dot{\epsilon}_k = (A_k - B_k F_k)\epsilon_k + B_k\nu_k
\end{equation}
with artificial inputs $\nu_k$ have similar dynamics.  In particular,
we express each system as perturbed version of the same nominal system
$P$ with uncertainty $\Delta_k$ according to
\begin{subequations}
  \label{eq:uncertain_agent}
  \begin{align}
    P: \qquad \dot{\epsilon}_k &= \tilde{A} \epsilon_k + \tilde{B}
    \nu_k + B^\omega \omega_k \\
    \zeta_k &= C^\zeta \epsilon_k + D^\zeta \nu_k \\
    \Delta_k: \qquad \omega_k &= \Delta_k \zeta_k,
  \end{align}
\end{subequations}
as illustrated in Fig.~\ref{fig:uncertain_agent}.  Note that the
regulation error $e_k = (C^e - FD^e)\epsilon_k$ can be interpreted as
output of the system $P$.
The fact that the systems \eqref{eq:nonidentical_agent} are similar is
formally expressed in the following assumption.
\begin{asmpt}\label{asmpt:similar}
  The matrix $\tilde{A}$ is Hurwitz.  The uncertainties $\Delta_k$ are
  proper real-rational stable transfer matrices and satisfy
  $\|\Delta_k\|_\infty < \eta^\Delta$, where $\eta^\Delta > 0$, for
  all $k \in \mathcal{N}$.
\end{asmpt}
The following procedure yields a description of the systems
\eqref{eq:nonidentical_agent} as in \eqref{eq:uncertain_agent}.  We
define the nominal system $G$ based on the average matrices
\begin{align*}
  \tilde{A} &= \frac{1}{N}\sum_{k=1}^N (A_k - B_k F_k), & 
  \tilde{B} &= \frac{1}{N}\sum_{k=1}^N B_k,
\end{align*}
i.e., the transfer function matrix from $\nu_k$ to $\epsilon_k$ is
given by
\begin{equation*}
  G(s) = (sI_{n^x}-\tilde{A})^{-1}\tilde{B}.
\end{equation*}
Note that we choose $\epsilon_k$ as output since we want to construct
a state feedback controller for the nominal system later on.  At this
point, we have to check whether $\tilde{A}$ is Hurwitz.  If it is not
Hurwitz, the description \eqref{eq:uncertain_agent} must be
constructed in a different way such that $\tilde{A}$ is Hurwitz.  If
it is Hurwitz, we proceed as follows.  We define the individual
transfer function matrices
\begin{equation*}
  G_k(s) = \big(sI_{n^x}-(A_k-B_kF_k)\big)^{-1}B_k.
\end{equation*}
Then, we express $G_k(s)$ as perturbed version of the nominal system
$G(s)$ with dynamical additive uncertainty $\Delta_k$, i.e.,
\begin{equation*}
  G_k(s) = G(s) + \Delta_k(s),
\end{equation*}
where $\Delta_k$ is simply obtained from $\Delta_k(s) = G_k(s) -
G(s)$.  Since $G(s)$ and $G_k(s)$ are proper real-rational and stable,
$\Delta_k$ is proper real-rational and stable as well.  The bound
$\eta^\Delta$ in Assumption~\ref{asmpt:similar} is computed as
\begin{equation*}
  \eta^\Delta = \max_{k \in \mathcal{N}} \|\Delta_k\|_\infty.
\end{equation*}
With the matrices $B^\omega = \tilde{B}$, $C^\zeta = I_{n^x}$, and
$D^\zeta = 0$, and the uncertainties $\Delta_k$ obtained from the
procedure above, the systems \eqref{eq:nonidentical_agent} are
represented as in \eqref{eq:uncertain_agent} and
Assumption~\ref{asmpt:similar} is satisfied.

\begin{figure}[b]
  \centering
  \begin{tikzpicture}[scale=1.3,thick]
    \draw (-0.6,-0.6) rectangle (0.6,0.6); %
    \node at (0,0) {$T^{\omega \zeta}$}; %
    \draw (-0.6,1) rectangle (0.6,2.2); %
    \node at (0,1.6) {\tiny$\begin{bmatrix} \Delta_1 & & \\ & \ddots &
        \\ & & \Delta_N\end{bmatrix}$}; %
    \draw [->] (-0.6,1.6) -- (-1.2,1.6) -- (-1.2,0) -- (-0.6,0); %
    \node at (-1.6,0.8) {$\omega$}; %
    \draw [<-] (0.6,1.6) -- (1.2,1.6) -- (1.2,0) -- (0.6,0); %
    \node at (1.6,0.8) {$\zeta$}; %
  \end{tikzpicture}
  \caption{Nominal network with uncertainties.}
  \label{fig:uncertain_network}
\end{figure}
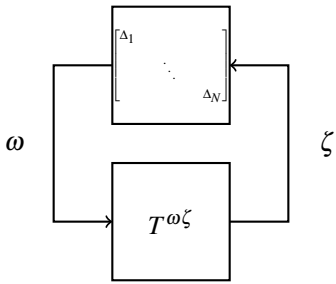

With couplings
\begin{equation*}
  \nu_k = H \sum_{j \in \mathcal{N}_k} (\epsilon_j - \epsilon_k),
\end{equation*}
between the systems and the representation \eqref{eq:uncertain_agent}
for each agent, we obtain
\begin{subequations}
  \label{eq:uncertain_network}
  \begin{align}
    T^{\omega \zeta}: \qquad \dot{\epsilon} &= (I_N \otimes
    \tilde{A} - L_{\mathcal{G}} \otimes \tilde{B}H) \epsilon %
    + (I_N \otimes B^\omega)  \omega \\
    \zeta &= (I_N \otimes C^\zeta - L_{\mathcal{G}} \otimes
    D^\zeta H) \epsilon \\
    \Delta: \qquad \omega &= \diag(\Delta_k)\zeta.
  \end{align}
\end{subequations}
This is a diffusively coupled network of uncertain stable linear
systems, which is illustrated in Fig.~\ref{fig:uncertain_network}.
The next step is to design $H$ such that the $\mathcal{H}_\infty$ norm
of the network with input $\omega$ and output $\zeta$ is minimized.
For this purpose, we follow the ideas of \cite{Li2011a} and extend the
synthesis procedure such that a desired decay rate $\gamma > 0$ is
achieved for the synchronization error.  Note that a lower bound on
the achievable $\mathcal{H}_\infty$ performance is given by the
autonomous system since we consider only diffusive couplings and
$\lambda_1 = 0$.
\begin{lem}\label{lem:decomposition}
  Let the graph $\mathcal{G}$ be undirected and connected.  Then, the
  network $T^{\omega \zeta}$ is exponentially stable and satisfies
  $\|T^{\omega\zeta}\|_\infty < \eta$ for a given $\eta > 0$ and
  achieves state synchronization exponentially with a decay rate
  $\gamma > 0$, if and only if the $N$ systems
  \begin{align*}
    T^{\tilde{\omega} \tilde{\zeta}}_k: \qquad
    \dot{\tilde{\epsilon}}_k &= (\tilde{A} - \lambda_k\tilde{B}H)
    \tilde{\epsilon}_k + B^\omega \tilde{\omega}_k \\
    \tilde{\zeta}_k &= (C^\zeta - \lambda_k D^\zeta H)
    \tilde{\epsilon}_k
  \end{align*}
  are exponentially stable and satisfy $\|T^{\tilde{\omega}
    \tilde{\zeta}}_k\|_\infty < \eta$ for all $k \in \mathcal{N}$, and
  additionally for $k=2,...,N$, have a decay rate $\gamma$.
\end{lem}
\begin{IEEEproof}
  Since $\mathcal{G}$ is undirected and connected there exists an
  orthogonal matrix $U$ such that $U^\mathsf{T} L_{\mathcal{G}} U =
  \Lambda = \diag(\lambda_k)$ and $\lambda_1 = 0$ is the top left
  diagonal element of $\Lambda$.  With the coordinate transformation
  $\tilde{\epsilon} = (U^\mathsf{T} \otimes I_N) \epsilon$,
  $\tilde{\omega} = (U^\mathsf{T} \otimes I_N) \omega$, $\tilde{\zeta}
  = (U^\mathsf{T} \otimes I_N) \zeta$, we obtain
  \begin{align*}
    \dot{\tilde{\epsilon}} &= (I_N \otimes \tilde{A} - \Lambda \otimes
    \tilde{B}H) \tilde{\epsilon} + (I_N \otimes B^\omega) \tilde{\omega} \\
    \tilde{\zeta} &= (I_N \otimes C^\zeta - \Lambda \otimes D^\zeta H)
    \tilde{\epsilon}.
  \end{align*}
  The latter system is decomposed into the decoupled systems
  \begin{align*}
    \dot{\tilde{\epsilon}}_k &= (\tilde{A} - \lambda_k\tilde{B}H) %
    \tilde{\epsilon}_k + B^\omega \tilde{\omega}_k \\
    \tilde{\zeta}_k &= (C^\zeta - \lambda_k D^\zeta H)
    \tilde{\epsilon}_k.
  \end{align*}
  Hence, exponential stability of $T^{\omega \zeta}$ corresponds to
  exponential stability of these systems for all $k\in \mathcal{N}$.
  Analogously to the proof of Lemma~\ref{lem:couplings_fi}, the state
  components $k=2,...,N$ correspond to the synchronization error of
  the network.  Exponential stability with decay rate $\gamma$ of
  these systems is equivalent to state synchronization with decay rate
  $\gamma$, as stated in the proof of Lemma~\ref{lem:couplings_fi}.
  Moreover, since $\|U\|_\infty = \|U^\mathsf{T}\|_\infty = 1$ and the maximal
  singular value of a block diagonal matrix equals the maximum of the
  maximal singular values of each block, it follows that $\|T^{\omega
    \zeta}\|_\infty = \max_k \|T^{\tilde{\omega}
    \tilde{\zeta}}_k\|_\infty < \eta$.
\end{IEEEproof}

\begin{thm}\label{thm:H_nonidentical}
  Let the graph $\mathcal{G}$ be undirected and connected.  Suppose
  that Assumption~\ref{asmpt:similar} is satisfied and let $\eta < 1 /
  \eta^\Delta$ and $\gamma > 0$.  Suppose that there exist $X > 0$, $Y
  > 0$, $Z$ such that the LMIs
  \begin{equation}\label{eq:lmi_1}
    \begin{bmatrix} \tilde{A}^\mathsf{T} X + X\tilde{A} & XB^\omega &
      (C^\zeta)^\mathsf{T} \\ (B^\omega)^\mathsf{T} X & -\eta I & 0 \\
      C^\zeta & 0 & -\eta I \end{bmatrix} < 0
  \end{equation}
  and, for $k = 2,...,N$,
  \begin{equation}\label{eq:lmi_k}
    \begin{bmatrix} \Xi & B^\omega &
      (C^\zeta Y -\lambda_k D^\zeta Z)^\mathsf{T} \\ (B^\omega)^\mathsf{T} & -\eta I & 0 \\
      C^\zeta Y -\lambda_k D^\zeta Z & 0 & -\eta I \end{bmatrix} < 0
  \end{equation}
  are satisfied, where $\Xi = (\tilde{A}Y - \lambda_k \tilde{B}
  Z)^\mathsf{T} + (\tilde{A}Y - \lambda_k \tilde{B} Z) + 2\gamma Y$.  Then,
  the network \eqref{eq:uncertain_network} with $H = ZY^{-1}$ is
  exponentially stable.
\end{thm}

\begin{IEEEproof}
  The $N$ subsystems $T^{\tilde{\omega} \tilde{\zeta}}_k$ are composed
  of the transformed systems
  \begin{align*}
    \tilde{P}: \qquad \dot{\tilde{\epsilon}}_k &= \tilde{A}
    \tilde{\epsilon}_k
    - \lambda_k \tilde{B} \nu_k + B^\omega \tilde{\omega}_k \\
    \tilde{\zeta}_k &= C^\zeta \tilde{\epsilon}_k - \lambda_k D^\zeta
    \nu_k
  \end{align*}
  and state feedback law $\nu_k = H \tilde{\epsilon}_k$.  The key idea
  is to design one single controller $\nu_k = H \tilde{\epsilon}_k$
  guaranteeing the desired $\mathcal{H}_\infty$ norm $\eta$ for all
  $N$ subsystems $\tilde{P}$ as well as the desired decay rate $\eta$
  for $k=2,...,N$.  According to the Bounded Real Lemma
  \cite{Zhou1998}, the closed loop of $\tilde{P}$ with control law
  $\nu_k = H \tilde{\epsilon}_k$ is asymptotically stable and has
  $\mathcal{H}_\infty$ norm less than or equal $\eta$, if and only if
  there exists $X > 0$ such that
  \begin{equation*}
    \begin{bmatrix} \Theta & XB^\omega &
      (C^\zeta-\lambda_k D^\zeta H)^\mathsf{T} \\ (B^\omega)^\mathsf{T} X & -\eta I & 0 \\
      C^\zeta-\lambda_k D^\zeta H & 0 & -\eta I \end{bmatrix} < 0,
  \end{equation*}
  where $ \Theta = (\tilde{A}-\lambda_k \tilde{B} H)^\mathsf{T} X +
  X(\tilde{A}-\lambda_k \tilde{B} H)$. Since $\lambda_1 = 0$, the best
  achievable $\mathcal{H}_\infty$ norm is limited by the uncontrolled
  system $\tilde{P}$.  For $k=1$, the matrix inequality above
  simplifies to the analysis LMI \eqref{eq:lmi_1}, which allows to
  find a lower bound on the achievable $\mathcal{H}_\infty$ norm
  $\eta$.  For $k = 2,...,N$, synthesis LMIs for $H$ are derived as
  follows.  In order to guarantee the desired decay rate $\gamma$, we
  add $2\gamma X$ to the upper left block $\Theta$.  This guarantees
  that the real parts of all corresponding eigenvalues are no larger
  than $-\gamma$.  Then, we multiply the matrix inequality from both
  sides with $\diag(X^{-1},I,I)$ and perform the change of variables
  $Y = X^{-1}$ and $Z = KY$.  This yields the LMIs \eqref{eq:lmi_k}
  for $k = 2,...,N$.  The corresponding control gain is obtained from
  $H = Z Y^{-1}$.

  By Lemma~\ref{lem:decomposition}, it follows that $T^{\omega \zeta}$
  is exponentially stable and $\|T^{\omega \zeta}\|_\infty < \eta$.
  Since $\eta < 1 / \eta^\Delta$ by assumption, exponential stability
  of \eqref{eq:uncertain_network} is a direct consequence of the Small
  Gain Theorem, cf., \cite{Trentelman2001}. %
\end{IEEEproof}

The design procedure for a suitable coupling gain $H$ for a group of
non-identical agents is summarized as follows.  First, express the
non-identical systems \eqref{eq:nonidentical_agent} as perturbed
versions of a common nominal plant $P$ with uncertainties $\Delta_k$
according to \eqref{eq:uncertain_agent} such that
Assumption~\ref{asmpt:similar} is satisfied with $\eta^\Delta$ as
small as possible.  Second, design a gain matrix $H$ according to
Theorem~\ref{thm:H_nonidentical} for some $\eta < 1 / \eta^\Delta$ and
a desired $\gamma > 0$.  Finally, examine the design and evaluate the
behavior of the closed-loop system.

\begin{rmk}
  The nominal network has $n^x$ poles which correspond to the
  synchronous motion of the variables $\epsilon_k$ and $(N-1)n^x$
  poles which correspond to the synchronization error and which, by
  design of the coupling gain $H$, have a decay rate of $\gamma$.  In
  presence of the perturbations, the network
  \eqref{eq:uncertain_network} is still guaranteed to be exponentially
  stable by Theorem~\ref{thm:H_nonidentical}.  However, the
  perturbations introduce a coupling between the $n^x$ modes of the
  synchronous motion and the $(N-1)n^x$ modes of the synchronization
  error.  The procedure in Theorem~\ref{thm:H_nonidentical} does not
  guarantee robust pole placement of the $(N-1)n^x$ poles such that
  their real parts are smaller than $-\gamma$ in presence of the
  perturbations.  It rather guarantees robust stability and nominal
  performance of the network.  The influence from the $n^x$ modes with
  lower decay rate to the synchronization error may shift poles and
  result in slower convergence of the synchronization error.
  Nevertheless, the additional coupling term can be expected to
  improve the convergence speed of the synchronization error
  significantly.  For details on the robust pole placement problem,
  the reader is referred to \cite{Chilali1999}.
\end{rmk}

\begin{rmk}
  Note that the design procedure in Theorem~\ref{thm:H_nonidentical}
  can be extended to pole placement constraints in terms of general
  LMI regions analogously to Theorem~\ref{thm:coupling_pp}, based on
  the results of \cite{Chilali1999}.  In this case, the LMIs
  \eqref{eq:lmi_k} have to be replaced by %
  \small%
  \begin{equation}\label{eq:lmi_k_pp}
    \begin{bmatrix} \Xi_{\mathcal{R}}  & M_1^\mathsf{T} \otimes B^\omega &
      M_2^\mathsf{T} \otimes (C^\zeta Y -\lambda_k D^\zeta Z)^\mathsf{T} \\ M_1
      \otimes (B^\omega)^\mathsf{T} & -\eta I & 0 \\
      M_2 \otimes (C^\zeta Y -\lambda_k D^\zeta Z) & 0 & -\eta I \end{bmatrix} < 0,
  \end{equation}
  \normalsize%
  where $\Xi_{\mathcal{R}} = L \otimes Y + M \otimes (\tilde{A}Y -
  \lambda_k \tilde{B} Z) + M^\mathsf{T} \otimes (\tilde{A}Y - \lambda_k
  \tilde{B} Z)^\mathsf{T}$ and $M_1^\mathsf{T} M_2 = M$ is a decomposition such
  that $M_1$ and $M_2$ have full column rank.  Such a decomposition
  can be obtained from the singular value decomposition of $M$.
\end{rmk}

\begin{exmp}\label{exmp:heli}
  Here, our main results are applied to a coordination problem of four
  Quanser 3-DoF Lab Helicopters\footnote{
    \url{http://www.quanser.com/products/3dof_helicopter} }.  Suppose
  we have a group of helicopters modeled by %
  \footnotesize%
  \begin{align*}
    \dot{x}_k &= \underbrace{\begin{bmatrix} 0 & 1 & 0 & 0 & 0 & 0 \\
        0 & 0 & -p^1_k & 0 & 0 & 0 \\ 0 & 0 & 0 & 1 & 0 & 0 \\ 0 & 0 &
        0 & 0 & 0 & 0 \\ 0 & 0 & 0 & 0 & 0 & 1 \\ 0 & 0 & 0 & 0 &
        -p^2_k & 0 \end{bmatrix}}_{A_k} x_k +
    \underbrace{\begin{bmatrix}
        0 & 0 \\ 0 & 0 \\ 0 & 0 \\ p^3_k & -p^3_k \\ 0 & 0 \\
        p^4_k & p^4_k \end{bmatrix}}_{B_k} u_k +
    \underbrace{\begin{bmatrix}
        0 & 0 \\ 0 & 0 \\ 0 & 0 \\ 1 & -1 \\ 0 & 0 \\
        1 & 1 \end{bmatrix}}_{B_k^{d^{\ell}}} d^{\ell}_k, \\
    y_k &= \underbrace{\begin{bmatrix} 1 & 0 & 0 & 0 & 0 &
        0 \\ 0 & 0 & 1 & 0 & 0 & 0 \\ 0 & 0 & 0 & 0 & 1 & 0
        \\ \end{bmatrix}}_{C_k} x_k
  \end{align*}
  \normalsize%
  for $k \in \mathcal{N}$.  The state $x_k \in \mathbb{R}^6$ consists
  of the travel angle $\alpha_k$, pitch angle $\beta_k$, and elevation
  angle $\gamma_k$ and the respective angular velocities, i.e., $x_k =
  [ \alpha_k \;\; \dot{\alpha}_k \;\; \beta_k \;\; \dot{\beta}_k \;\;
  \gamma_k \;\; \dot{\gamma}_k ]^\mathsf{T}$.  The measurement output
  $y_k$ consists of the three absolute angles.  The parameters of the
  four helicopters are given in vector form by $p_1 = [ 0.7 \;\; 1.17
  \;\; 6.0 \;\; 0.58 ]^\mathsf{T}$, $p_2 = [ 0.5 \;\; 1.05 \;\; 5.8
  \;\; 0.50 ]^\mathsf{T}$, $p_3 = [ 0.6 \;\; 1.10 \;\; 6.1 \;\; 0.63
  ]^\mathsf{T}$, $p_4 = [ 0.7 \;\; 1.25 \;\; 5.6 \;\; 0.48
  ]^\mathsf{T}$.  Note that the helicopters are non-identical due to
  the different parameter values assumed for each system.

  The objective is that all helicopters track a reference signal for
  travel and elevation angles while asymptotically rejecting the
  external disturbances.  The reference is generated by
  \begin{equation*}
    \dot{d^g} = \underbrace{\begin{bmatrix} 0 & 1 & 0 \\ 0 & 0 & 0 \\
        0 & 0 & 0 \end{bmatrix}}_{S^g} d^g,
  \end{equation*}
  where $r_\alpha = [1 \;\; 0 \;\; 0 ]d^g$ is a ramp signal and
  reference for the travel angels and where $r_\gamma = [ 0 \;\; 0
  \;\; 1 ]d^g$ is a constant signal and reference for the elevation
  angles.  Consequently, the regulation errors are defined as
  \begin{equation*}
    e_k = \underbrace{\begin{bmatrix} 1 & 0 & 0 & 0 & 0 & 0 \\ 0 & 0 & 0 & 0 & 1 &
        0 \end{bmatrix}}_{C^e_k} x_k + \underbrace{\begin{bmatrix} -1 & 0 & 0 \\ 0 &
        0 & -1 \end{bmatrix}}_{D^{ed^g}_k} d^g.
  \end{equation*}
  Each helicopter is affected by constant disturbances $d^{\ell}_k$ acting
  as forces on the rotors, which is described by the local exosystems
  with $S^{\ell}_k = \diag(0,0)$ for all $k \in \mathcal{N}$.  The
  communication graph $\mathcal{G}$ is an undirected cycle.
  \begin{figure}[t]
    \setlength\figurewidth{0.7\linewidth}%
    \setlength\figureheight{0.4\figurewidth}%
    \centering
    \input{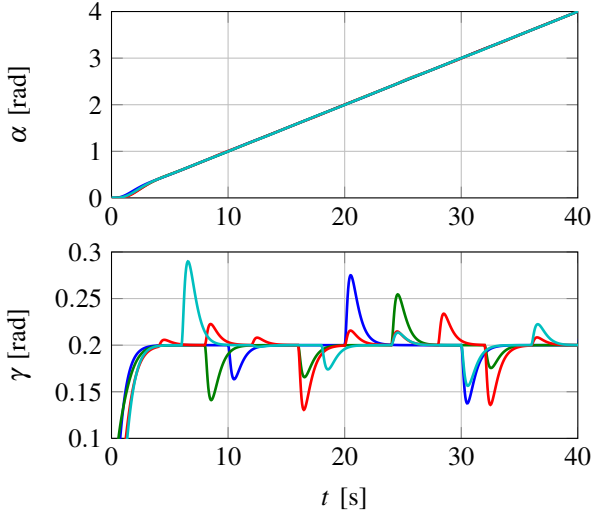}
    \caption{Simulation result with distributed regulator according
      to Theorem~\ref{thm:nominal} (Example~\ref{exmp:heli}).}
    \label{fig:heli_dec}
  \end{figure}

  In a first step, we construct a distributed output regulator
  according to Theorem~\ref{thm:nominal}.  The control gains $F$ are
  designed as LQR with weight matrices $Q = \diag(50,1,1,1,100,50)$
  and $R = \diag(2,2)$.  The observer poles are placed between $-10$
  and $-12$ on the real axis.  We simulate the closed-loop system and
  apply a random piecewise constant disturbance signal with magnitude
  in the range of $-2.5$N to $2.5$N each helicopter.
  Fig.~\ref{fig:heli_dec} shows the corresponding simulation result.
  Next, we extend the distributed regulator
  \eqref{eq:nominal_control_law} by couplings as in
  \eqref{eq:couplings_control_law_heterog} in order to improve the
  cooperative behavior of the group and synchronize the transient
  state components.  The control gains $F_k$ are now designed with a
  higher weight $R =\diag(4,4)$ on the control signals.  Following the
  procedure described in Section~\ref{sec:nonidentical}, we compute a
  nominal model $G$, individual models $G_k$, and quantify the
  $\mathcal{H}_\infty$ norm of the additive uncertainties $\Delta_k$,
  where $G_k = G + \Delta_k$.  Fig.~\ref{fig:svd} shows the singular
  value plots of the transfer matrices $\Delta_k = G_k - G$.
  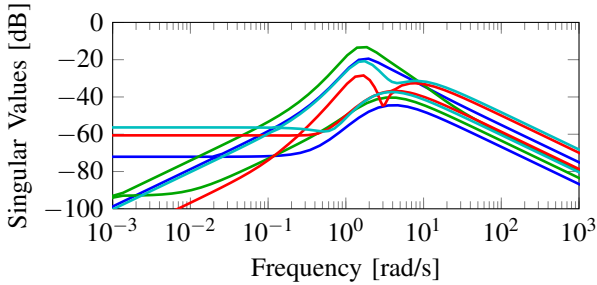
\begin{figure}[t]
    \setlength\figurewidth{0.7\linewidth}%
    \setlength\figureheight{0.4\figurewidth}%
    \centering
    \input{figures/sim_svd_Delta.tikz}
    \caption{Singular value plots of the additive uncertainties
      $\Delta_k$ (Example~\ref{exmp:heli}).}
    \label{fig:svd}
  \end{figure}
  \begin{figure}[t]
    \setlength\figurewidth{0.7\linewidth}%
    \setlength\figureheight{0.4\figurewidth}%
    \centering
    \input{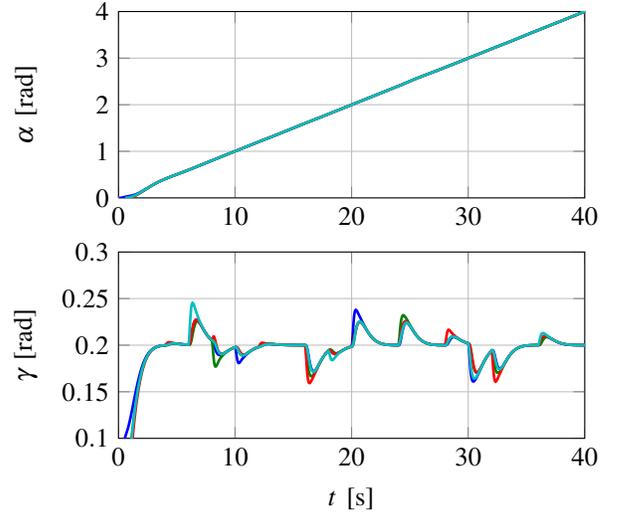}
    \caption{Simulation result with distributed regulator according to
      Theorem~\ref{thm:nominal} and additional coupling term as in
      \eqref{eq:couplings_control_law_heterog}, with coupling gain $H$
      designed according to Theorem~\ref{thm:H_nonidentical}
      and \eqref{eq:lmi_k_pp} (Example~\ref{exmp:heli}).}
    \label{fig:heli_coop}
  \end{figure}
  We obtain $\eta^\Delta = \max_k \|\Delta_k\|_\infty = 0.3129$.  We
  compute a suitable coupling gain $H$ based on the LMIs
  \eqref{eq:lmi_k_pp} where we choose $\eta = 0.95 (\eta^\Delta)^{-1}$
  and the pole placement region $S(3,30,\pi/3)$ ensuring a decay rate
  of $\gamma = 3$.  This leads to a feasible design of $H$.
  Simulation results with the same setup as before and the extended
  distributed regulator are shown in Fig.~\ref{fig:heli_coop}.
  Obviously, the cooperative behavior is improved considerably
  compared to Fig.~\ref{fig:heli_dec}.  The synchronization error of
  the elevation angles remains very small in the transient phases.
  The group of non-identical helicopters reacts cooperatively on the
  local external disturbances, as desired.
\end{exmp}

\section{Conclusion}\label{sec:conclusion}

The cooperative output regulation problem captures a wide range of
practical multi-agent coordination problems.  In this paper, we have
presented a novel distributed regulator which solves the coordination
problem and additionally allows to improve and tune the
synchronization error dynamics of the group.  A novel coupling term
based on the transient state components of each agent allows to impose
a desired exponential decay rate on the synchronization error among
agents, which leads to a significant improvement of the cooperative
behavior of the group in transient phases.  Under the novel control
law, the group is able to react cooperatively on external disturbances
acting on individual agents.  We have discussed a vehicle platooning
example and a coordination example for a group of four 3-DoF
helicopters in order to emphasize the importance of a cooperative
reaction on disturbances and in order to illustrate the design
procedure and effectiveness of the novel distributed regulator with
transient synchronization.

\section*{Acknowledgment}

We gratefully thank Prof. Jie Chen and Dr. Jie Mei for helpful
discussions.

\bibliography{mybibliography}

\end{document}

%% file: figures/platoon.tex
\definecolor{COLOR0}{rgb}{0.0,0.0,0.0}
\definecolor{COLOR1}{rgb}{0.0,0.0,0.0}
\definecolor{grn}{rgb}{0.0,0.6,0.0}
\definecolor{COLOR2}{rgb}{0.0,0.0,0.0}
\pgfdeclarelayer{background}
\pgfdeclarelayer{foreground}
\pgfsetlayers{background,main,foreground}
\begin{tikzpicture}[scale=1.5]

  \coordinate (n0) at (6,0);%
  \coordinate (n1) at (5,0);%
  \coordinate (n2) at (4,0);%
  \coordinate (n3) at (3,0);%
  \coordinate (n4) at (2,0);%
  \coordinate (n5) at (1,0);%

  \node at (n1) [circle, line width=1, fill=COLOR1] (1)
  {\color{white}\bf 1};%
  \node at (n2) [circle, line width=1, fill=COLOR1] (2)
  {\color{white}\bf 2};%
  \node at (n3) [circle, line width=1, fill=COLOR0] (3)
  {\color{white}\bf 3};%
  \node at (n4) [circle, line width=1, fill=COLOR0] (4)
  {\color{white}\bf 4};%
  \node at (n5) [circle, line width=1, fill=COLOR0] (5)
  {\color{white}\bf 5};%

\begin{pgfonlayer}{background}
\tikzset{EdgeStyle/.style = {->, shorten >=1pt, >=stealth, bend
    right=10, line width=1, color=black}}
\Edge [label=](1)(2)
\Edge [label=](2)(3)
\Edge [label=](3)(4)
\Edge [label=](4)(5)
\Edge [label=](5)(4)
\Edge [label=](4)(3)
\Edge [label=](3)(2)
\Edge [label=](2)(1)

\end{pgfonlayer}

\foreach \x in {1,2,3,4,5} {%
  \draw[->,very thick,color=grn] ($(n\x)+(0,0.7)$) --
  ($(n\x)+(0,0.2)$);%
  \node[color=grn] at ($(n\x)+(0.25,0.45)$) [above] {$r_\x$};
  \draw[->,very thick,color=red] ($(n\x)+(0,-0.7)$) --
  ($(n\x)+(0,-0.2)$);%
  \node[color=red] at ($(n\x)+(0.25,-0.45)$) [below] {$w_\x$};
}%

\node[color=blue] at ($(n1)+(0.7,0)$) [above] {$d^g$}; \draw[->,very
thick,color=blue] ($(n1)+(0.7,0)$) -- ($(n1)+(0.2,0)$);%

\end{tikzpicture}

%% file: figures/sim_platoon_r.tikz
%
%
%

\definecolor{mycolor1}{rgb}{0,0.75,0.75}
\definecolor{mycolor2}{rgb}{0.75,0,0.75}

\begin{tikzpicture}

\begin{axis}[%
width=\figurewidth,
height=\figureheight,
scale only axis,
xmin=0, xmax=60,
xmajorgrids,
ymin=-11, ymax=1,
ylabel={$r_k(t)$},
ymajorgrids]
\addplot [
color=blue,
solid,
line width=1.0pt,
forget plot
]
coordinates{
 (0,0)(0.1,0)(0.2,0)(0.3,0)(0.4,0)(0.5,0)(0.6,0)(0.7,0)(0.8,0)(0.9,0)(1,0)(1.1,0)(1.2,0)(1.3,0)(1.4,0)(1.5,0)(1.6,0)(1.7,0)(1.8,0)(1.9,0)(2,0)(2.1,0)(2.2,0)(2.3,0)(2.4,0)(2.5,0)(2.6,0)(2.7,0)(2.8,0)(2.9,0)(3,0)(3.1,0)(3.2,0)(3.3,0)(3.4,0)(3.5,0)(3.6,0)(3.7,0)(3.8,0)(3.9,0)(4,0)(4.1,0)(4.2,0)(4.3,0)(4.4,0)(4.5,0)(4.6,0)(4.7,0)(4.8,0)(4.9,0)(5,0)(5.1,0)(5.2,0)(5.3,0)(5.4,0)(5.5,0)(5.6,0)(5.7,0)(5.8,0)(5.9,0)(6,0)(6.1,0)(6.2,0)(6.3,0)(6.4,0)(6.5,0)(6.6,0)(6.7,0)(6.8,0)(6.9,0)(7,0)(7.1,0)(7.2,0)(7.3,0)(7.4,0)(7.5,0)(7.6,0)(7.7,0)(7.8,0)(7.9,0)(8,0)(8.1,0)(8.2,0)(8.3,0)(8.4,0)(8.5,0)(8.6,0)(8.7,0)(8.8,0)(8.9,0)(9,0)(9.1,0)(9.2,0)(9.3,0)(9.4,0)(9.5,0)(9.6,0)(9.7,0)(9.8,0)(9.9,0)(10,0)(10.1,0)(10.2,0)(10.3,0)(10.4,0)(10.5,0)(10.6,0)(10.7,0)(10.8,0)(10.9,0)(11,0)(11.1,0)(11.2,0)(11.3,0)(11.4,0)(11.5,0)(11.6,0)(11.7,0)(11.8,0)(11.9,0)(12,0)(12.1,0)(12.2,0)(12.3,0)(12.4,0)(12.5,0)(12.6,0)(12.7,0)(12.8,0)(12.9,0)(13,0)(13.1,0)(13.2,0)(13.3,0)(13.4,0)(13.5,0)(13.6,0)(13.7,0)(13.8,0)(13.9,0)(14,0)(14.1,0)(14.2,0)(14.3,0)(14.4,0)(14.5,0)(14.6,0)(14.7,0)(14.8,0)(14.9,0)(15,0)(15.1,0)(15.2,0)(15.3,0)(15.4,0)(15.5,0)(15.6,0)(15.7,0)(15.8,0)(15.9,0)(16,0)(16.1,0)(16.2,0)(16.3,0)(16.4,0)(16.5,0)(16.6,0)(16.7,0)(16.8,0)(16.9,0)(17,0)(17.1,0)(17.2,0)(17.3,0)(17.4,0)(17.5,0)(17.6,0)(17.7,0)(17.8,0)(17.9,0)(18,0)(18.1,0)(18.2,0)(18.3,0)(18.4,0)(18.5,0)(18.6,0)(18.7,0)(18.8,0)(18.9,0)(19,0)(19.1,0)(19.2,0)(19.3,0)(19.4,0)(19.5,0)(19.6,0)(19.7,0)(19.8,0)(19.9,0)(20,0)(20.1,0)(20.2,0)(20.3,0)(20.4,0)(20.5,0)(20.6,0)(20.7,0)(20.8,0)(20.9,0)(21,0)(21.1,0)(21.2,0)(21.3,0)(21.4,0)(21.5,0)(21.6,0)(21.7,0)(21.8,0)(21.9,0)(22,0)(22.1,0)(22.2,0)(22.3,0)(22.4,0)(22.5,0)(22.6,0)(22.7,0)(22.8,0)(22.9,0)(23,0)(23.1,0)(23.2,0)(23.3,0)(23.4,0)(23.5,0)(23.6,0)(23.7,0)(23.8,0)(23.9,0)(24,0)(24.1,0)(24.2,0)(24.3,0)(24.4,0)(24.5,0)(24.6,0)(24.7,0)(24.8,0)(24.9,0)(25,0)(25.1,0)(25.2,0)(25.3,0)(25.4,0)(25.5,0)(25.6,0)(25.7,0)(25.8,0)(25.9,0)(26,0)(26.1,0)(26.2,0)(26.3,0)(26.4,0)(26.5,0)(26.6,0)(26.7,0)(26.8,0)(26.9,0)(27,0)(27.1,0)(27.2,0)(27.3,0)(27.4,0)(27.5,0)(27.6,0)(27.7,0)(27.8,0)(27.9,0)(28,0)(28.1,0)(28.2,0)(28.3,0)(28.4,0)(28.5,0)(28.6,0)(28.7,0)(28.8,0)(28.9,0)(29,0)(29.1,0)(29.2,0)(29.3,0)(29.4,0)(29.5,0)(29.6,0)(29.7,0)(29.8,0)(29.9,0)(30,-2)(30.1,-2)(30.2,-2)(30.3,-2)(30.4,-2)(30.5,-2)(30.6,-2)(30.7,-2)(30.8,-2)(30.9,-2)(31,-2)(31.1,-2)(31.2,-2)(31.3,-2)(31.4,-2)(31.5,-2)(31.6,-2)(31.7,-2)(31.8,-2)(31.9,-2)(32,-2)(32.1,-2)(32.2,-2)(32.3,-2)(32.4,-2)(32.5,-2)(32.6,-2)(32.7,-2)(32.8,-2)(32.9,-2)(33,-2)(33.1,-2)(33.2,-2)(33.3,-2)(33.4,-2)(33.5,-2)(33.6,-2)(33.7,-2)(33.8,-2)(33.9,-2)(34,-2)(34.1,-2)(34.2,-2)(34.3,-2)(34.4,-2)(34.5,-2)(34.6,-2)(34.7,-2)(34.8,-2)(34.9,-2)(35,-2)(35.1,-2)(35.2,-2)(35.3,-2)(35.4,-2)(35.5,-2)(35.6,-2)(35.7,-2)(35.8,-2)(35.9,-2)(36,-2)(36.1,-2)(36.2,-2)(36.3,-2)(36.4,-2)(36.5,-2)(36.6,-2)(36.7,-2)(36.8,-2)(36.9,-2)(37,-2)(37.1,-2)(37.2,-2)(37.3,-2)(37.4,-2)(37.5,-2)(37.6,-2)(37.7,-2)(37.8,-2)(37.9,-2)(38,-2)(38.1,-2)(38.2,-2)(38.3,-2)(38.4,-2)(38.5,-2)(38.6,-2)(38.7,-2)(38.8,-2)(38.9,-2)(39,-2)(39.1,-2)(39.2,-2)(39.3,-2)(39.4,-2)(39.5,-2)(39.6,-2)(39.7,-2)(39.8,-2)(39.9,-2)(40,-2)(40.1,-2)(40.2,-2)(40.3,-2)(40.4,-2)(40.5,-2)(40.6,-2)(40.7,-2)(40.8,-2)(40.9,-2)(41,-2)(41.1,-2)(41.2,-2)(41.3,-2)(41.4,-2)(41.5,-2)(41.6,-2)(41.7,-2)(41.8,-2)(41.9,-2)(42,-2)(42.1,-2)(42.2,-2)(42.3,-2)(42.4,-2)(42.5,-2)(42.6,-2)(42.7,-2)(42.8,-2)(42.9,-2)(43,-2)(43.1,-2)(43.2,-2)(43.3,-2)(43.4,-2)(43.5,-2)(43.6,-2)(43.7,-2)(43.8,-2)(43.9,-2)(44,-2)(44.1,-2)(44.2,-2)(44.3,-2)(44.4,-2)(44.5,-2)(44.6,-2)(44.7,-2)(44.8,-2)(44.9,-2)(45,-2)(45.1,-2)(45.2,-2)(45.3,-2)(45.4,-2)(45.5,-2)(45.6,-2)(45.7,-2)(45.8,-2)(45.9,-2)(46,-2)(46.1,-2)(46.2,-2)(46.3,-2)(46.4,-2)(46.5,-2)(46.6,-2)(46.7,-2)(46.8,-2)(46.9,-2)(47,-2)(47.1,-2)(47.2,-2)(47.3,-2)(47.4,-2)(47.5,-2)(47.6,-2)(47.7,-2)(47.8,-2)(47.9,-2)(48,-2)(48.1,-2)(48.2,-2)(48.3,-2)(48.4,-2)(48.5,-2)(48.6,-2)(48.7,-2)(48.8,-2)(48.9,-2)(49,-2)(49.1,-2)(49.2,-2)(49.3,-2)(49.4,-2)(49.5,-2)(49.6,-2)(49.7,-2)(49.8,-2)(49.9,-2)(50,-2)(50.1,-2)(50.2,-2)(50.3,-2)(50.4,-2)(50.5,-2)(50.6,-2)(50.7,-2)(50.8,-2)(50.9,-2)(51,-2)(51.1,-2)(51.2,-2)(51.3,-2)(51.4,-2)(51.5,-2)(51.6,-2)(51.7,-2)(51.8,-2)(51.9,-2)(52,-2)(52.1,-2)(52.2,-2)(52.3,-2)(52.4,-2)(52.5,-2)(52.6,-2)(52.7,-2)(52.8,-2)(52.9,-2)(53,-2)(53.1,-2)(53.2,-2)(53.3,-2)(53.4,-2)(53.5,-2)(53.6,-2)(53.7,-2)(53.8,-2)(53.9,-2)(54,-2)(54.1,-2)(54.2,-2)(54.3,-2)(54.4,-2)(54.5,-2)(54.6,-2)(54.7,-2)(54.8,-2)(54.9,-2)(55,-2)(55.1,-2)(55.2,-2)(55.3,-2)(55.4,-2)(55.5,-2)(55.6,-2)(55.7,-2)(55.8,-2)(55.9,-2)(56,-2)(56.1,-2)(56.2,-2)(56.3,-2)(56.4,-2)(56.5,-2)(56.6,-2)(56.7,-2)(56.8,-2)(56.9,-2)(57,-2)(57.1,-2)(57.2,-2)(57.3,-2)(57.4,-2)(57.5,-2)(57.6,-2)(57.7,-2)(57.8,-2)(57.9,-2)(58,-2)(58.1,-2)(58.2,-2)(58.3,-2)(58.4,-2)(58.5,-2)(58.6,-2)(58.7,-2)(58.8,-2)(58.9,-2)(59,-2)(59.1,-2)(59.2,-2)(59.3,-2)(59.4,-2)(59.5,-2)(59.6,-2)(59.7,-2)(59.8,-2)(59.9,-2)(60,-2) 
};
\addplot [
color=green!50!black,
solid,
line width=1.0pt,
forget plot
]
coordinates{
 (0,-1)(0.1,-1)(0.2,-1)(0.3,-1)(0.4,-1)(0.5,-1)(0.6,-1)(0.7,-1)(0.8,-1)(0.9,-1)(1,-1)(1.1,-1)(1.2,-1)(1.3,-1)(1.4,-1)(1.5,-1)(1.6,-1)(1.7,-1)(1.8,-1)(1.9,-1)(2,-1)(2.1,-1)(2.2,-1)(2.3,-1)(2.4,-1)(2.5,-1)(2.6,-1)(2.7,-1)(2.8,-1)(2.9,-1)(3,-1)(3.1,-1)(3.2,-1)(3.3,-1)(3.4,-1)(3.5,-1)(3.6,-1)(3.7,-1)(3.8,-1)(3.9,-1)(4,-1)(4.1,-1)(4.2,-1)(4.3,-1)(4.4,-1)(4.5,-1)(4.6,-1)(4.7,-1)(4.8,-1)(4.9,-1)(5,-1)(5.1,-1)(5.2,-1)(5.3,-1)(5.4,-1)(5.5,-1)(5.6,-1)(5.7,-1)(5.8,-1)(5.9,-1)(6,-1)(6.1,-1)(6.2,-1)(6.3,-1)(6.4,-1)(6.5,-1)(6.6,-1)(6.7,-1)(6.8,-1)(6.9,-1)(7,-1)(7.1,-1)(7.2,-1)(7.3,-1)(7.4,-1)(7.5,-1)(7.6,-1)(7.7,-1)(7.8,-1)(7.9,-1)(8,-1)(8.1,-1)(8.2,-1)(8.3,-1)(8.4,-1)(8.5,-1)(8.6,-1)(8.7,-1)(8.8,-1)(8.9,-1)(9,-1)(9.1,-1)(9.2,-1)(9.3,-1)(9.4,-1)(9.5,-1)(9.6,-1)(9.7,-1)(9.8,-1)(9.9,-1)(10,-1)(10.1,-1)(10.2,-1)(10.3,-1)(10.4,-1)(10.5,-1)(10.6,-1)(10.7,-1)(10.8,-1)(10.9,-1)(11,-1)(11.1,-1)(11.2,-1)(11.3,-1)(11.4,-1)(11.5,-1)(11.6,-1)(11.7,-1)(11.8,-1)(11.9,-1)(12,-1)(12.1,-1)(12.2,-1)(12.3,-1)(12.4,-1)(12.5,-1)(12.6,-1)(12.7,-1)(12.8,-1)(12.9,-1)(13,-1)(13.1,-1)(13.2,-1)(13.3,-1)(13.4,-1)(13.5,-1)(13.6,-1)(13.7,-1)(13.8,-1)(13.9,-1)(14,-1)(14.1,-1)(14.2,-1)(14.3,-1)(14.4,-1)(14.5,-1)(14.6,-1)(14.7,-1)(14.8,-1)(14.9,-1)(15,-1)(15.1,-1)(15.2,-1)(15.3,-1)(15.4,-1)(15.5,-1)(15.6,-1)(15.7,-1)(15.8,-1)(15.9,-1)(16,-1)(16.1,-1)(16.2,-1)(16.3,-1)(16.4,-1)(16.5,-1)(16.6,-1)(16.7,-1)(16.8,-1)(16.9,-1)(17,-1)(17.1,-1)(17.2,-1)(17.3,-1)(17.4,-1)(17.5,-1)(17.6,-1)(17.7,-1)(17.8,-1)(17.9,-1)(18,-1)(18.1,-1)(18.2,-1)(18.3,-1)(18.4,-1)(18.5,-1)(18.6,-1)(18.7,-1)(18.8,-1)(18.9,-1)(19,-1)(19.1,-1)(19.2,-1)(19.3,-1)(19.4,-1)(19.5,-1)(19.6,-1)(19.7,-1)(19.8,-1)(19.9,-1)(20,-1)(20.1,-1)(20.2,-1)(20.3,-1)(20.4,-1)(20.5,-1)(20.6,-1)(20.7,-1)(20.8,-1)(20.9,-1)(21,-1)(21.1,-1)(21.2,-1)(21.3,-1)(21.4,-1)(21.5,-1)(21.6,-1)(21.7,-1)(21.8,-1)(21.9,-1)(22,-1)(22.1,-1)(22.2,-1)(22.3,-1)(22.4,-1)(22.5,-1)(22.6,-1)(22.7,-1)(22.8,-1)(22.9,-1)(23,-1)(23.1,-1)(23.2,-1)(23.3,-1)(23.4,-1)(23.5,-1)(23.6,-1)(23.7,-1)(23.8,-1)(23.9,-1)(24,-1)(24.1,-1)(24.2,-1)(24.3,-1)(24.4,-1)(24.5,-1)(24.6,-1)(24.7,-1)(24.8,-1)(24.9,-1)(25,-1)(25.1,-1)(25.2,-1)(25.3,-1)(25.4,-1)(25.5,-1)(25.6,-1)(25.7,-1)(25.8,-1)(25.9,-1)(26,-1)(26.1,-1)(26.2,-1)(26.3,-1)(26.4,-1)(26.5,-1)(26.6,-1)(26.7,-1)(26.8,-1)(26.9,-1)(27,-1)(27.1,-1)(27.2,-1)(27.3,-1)(27.4,-1)(27.5,-1)(27.6,-1)(27.7,-1)(27.8,-1)(27.9,-1)(28,-1)(28.1,-1)(28.2,-1)(28.3,-1)(28.4,-1)(28.5,-1)(28.6,-1)(28.7,-1)(28.8,-1)(28.9,-1)(29,-1)(29.1,-1)(29.2,-1)(29.3,-1)(29.4,-1)(29.5,-1)(29.6,-1)(29.7,-1)(29.8,-1)(29.9,-1)(30,-4)(30.1,-4)(30.2,-4)(30.3,-4)(30.4,-4)(30.5,-4)(30.6,-4)(30.7,-4)(30.8,-4)(30.9,-4)(31,-4)(31.1,-4)(31.2,-4)(31.3,-4)(31.4,-4)(31.5,-4)(31.6,-4)(31.7,-4)(31.8,-4)(31.9,-4)(32,-4)(32.1,-4)(32.2,-4)(32.3,-4)(32.4,-4)(32.5,-4)(32.6,-4)(32.7,-4)(32.8,-4)(32.9,-4)(33,-4)(33.1,-4)(33.2,-4)(33.3,-4)(33.4,-4)(33.5,-4)(33.6,-4)(33.7,-4)(33.8,-4)(33.9,-4)(34,-4)(34.1,-4)(34.2,-4)(34.3,-4)(34.4,-4)(34.5,-4)(34.6,-4)(34.7,-4)(34.8,-4)(34.9,-4)(35,-4)(35.1,-4)(35.2,-4)(35.3,-4)(35.4,-4)(35.5,-4)(35.6,-4)(35.7,-4)(35.8,-4)(35.9,-4)(36,-4)(36.1,-4)(36.2,-4)(36.3,-4)(36.4,-4)(36.5,-4)(36.6,-4)(36.7,-4)(36.8,-4)(36.9,-4)(37,-4)(37.1,-4)(37.2,-4)(37.3,-4)(37.4,-4)(37.5,-4)(37.6,-4)(37.7,-4)(37.8,-4)(37.9,-4)(38,-4)(38.1,-4)(38.2,-4)(38.3,-4)(38.4,-4)(38.5,-4)(38.6,-4)(38.7,-4)(38.8,-4)(38.9,-4)(39,-4)(39.1,-4)(39.2,-4)(39.3,-4)(39.4,-4)(39.5,-4)(39.6,-4)(39.7,-4)(39.8,-4)(39.9,-4)(40,-4)(40.1,-4)(40.2,-4)(40.3,-4)(40.4,-4)(40.5,-4)(40.6,-4)(40.7,-4)(40.8,-4)(40.9,-4)(41,-4)(41.1,-4)(41.2,-4)(41.3,-4)(41.4,-4)(41.5,-4)(41.6,-4)(41.7,-4)(41.8,-4)(41.9,-4)(42,-4)(42.1,-4)(42.2,-4)(42.3,-4)(42.4,-4)(42.5,-4)(42.6,-4)(42.7,-4)(42.8,-4)(42.9,-4)(43,-4)(43.1,-4)(43.2,-4)(43.3,-4)(43.4,-4)(43.5,-4)(43.6,-4)(43.7,-4)(43.8,-4)(43.9,-4)(44,-4)(44.1,-4)(44.2,-4)(44.3,-4)(44.4,-4)(44.5,-4)(44.6,-4)(44.7,-4)(44.8,-4)(44.9,-4)(45,-4)(45.1,-4)(45.2,-4)(45.3,-4)(45.4,-4)(45.5,-4)(45.6,-4)(45.7,-4)(45.8,-4)(45.9,-4)(46,-4)(46.1,-4)(46.2,-4)(46.3,-4)(46.4,-4)(46.5,-4)(46.6,-4)(46.7,-4)(46.8,-4)(46.9,-4)(47,-4)(47.1,-4)(47.2,-4)(47.3,-4)(47.4,-4)(47.5,-4)(47.6,-4)(47.7,-4)(47.8,-4)(47.9,-4)(48,-4)(48.1,-4)(48.2,-4)(48.3,-4)(48.4,-4)(48.5,-4)(48.6,-4)(48.7,-4)(48.8,-4)(48.9,-4)(49,-4)(49.1,-4)(49.2,-4)(49.3,-4)(49.4,-4)(49.5,-4)(49.6,-4)(49.7,-4)(49.8,-4)(49.9,-4)(50,-4)(50.1,-4)(50.2,-4)(50.3,-4)(50.4,-4)(50.5,-4)(50.6,-4)(50.7,-4)(50.8,-4)(50.9,-4)(51,-4)(51.1,-4)(51.2,-4)(51.3,-4)(51.4,-4)(51.5,-4)(51.6,-4)(51.7,-4)(51.8,-4)(51.9,-4)(52,-4)(52.1,-4)(52.2,-4)(52.3,-4)(52.4,-4)(52.5,-4)(52.6,-4)(52.7,-4)(52.8,-4)(52.9,-4)(53,-4)(53.1,-4)(53.2,-4)(53.3,-4)(53.4,-4)(53.5,-4)(53.6,-4)(53.7,-4)(53.8,-4)(53.9,-4)(54,-4)(54.1,-4)(54.2,-4)(54.3,-4)(54.4,-4)(54.5,-4)(54.6,-4)(54.7,-4)(54.8,-4)(54.9,-4)(55,-4)(55.1,-4)(55.2,-4)(55.3,-4)(55.4,-4)(55.5,-4)(55.6,-4)(55.7,-4)(55.8,-4)(55.9,-4)(56,-4)(56.1,-4)(56.2,-4)(56.3,-4)(56.4,-4)(56.5,-4)(56.6,-4)(56.7,-4)(56.8,-4)(56.9,-4)(57,-4)(57.1,-4)(57.2,-4)(57.3,-4)(57.4,-4)(57.5,-4)(57.6,-4)(57.7,-4)(57.8,-4)(57.9,-4)(58,-4)(58.1,-4)(58.2,-4)(58.3,-4)(58.4,-4)(58.5,-4)(58.6,-4)(58.7,-4)(58.8,-4)(58.9,-4)(59,-4)(59.1,-4)(59.2,-4)(59.3,-4)(59.4,-4)(59.5,-4)(59.6,-4)(59.7,-4)(59.8,-4)(59.9,-4)(60,-4) 
};
\addplot [
color=red,
solid,
line width=1.0pt,
forget plot
]
coordinates{
 (0,-2)(0.1,-2)(0.2,-2)(0.3,-2)(0.4,-2)(0.5,-2)(0.6,-2)(0.7,-2)(0.8,-2)(0.9,-2)(1,-2)(1.1,-2)(1.2,-2)(1.3,-2)(1.4,-2)(1.5,-2)(1.6,-2)(1.7,-2)(1.8,-2)(1.9,-2)(2,-2)(2.1,-2)(2.2,-2)(2.3,-2)(2.4,-2)(2.5,-2)(2.6,-2)(2.7,-2)(2.8,-2)(2.9,-2)(3,-2)(3.1,-2)(3.2,-2)(3.3,-2)(3.4,-2)(3.5,-2)(3.6,-2)(3.7,-2)(3.8,-2)(3.9,-2)(4,-2)(4.1,-2)(4.2,-2)(4.3,-2)(4.4,-2)(4.5,-2)(4.6,-2)(4.7,-2)(4.8,-2)(4.9,-2)(5,-2)(5.1,-2)(5.2,-2)(5.3,-2)(5.4,-2)(5.5,-2)(5.6,-2)(5.7,-2)(5.8,-2)(5.9,-2)(6,-2)(6.1,-2)(6.2,-2)(6.3,-2)(6.4,-2)(6.5,-2)(6.6,-2)(6.7,-2)(6.8,-2)(6.9,-2)(7,-2)(7.1,-2)(7.2,-2)(7.3,-2)(7.4,-2)(7.5,-2)(7.6,-2)(7.7,-2)(7.8,-2)(7.9,-2)(8,-2)(8.1,-2)(8.2,-2)(8.3,-2)(8.4,-2)(8.5,-2)(8.6,-2)(8.7,-2)(8.8,-2)(8.9,-2)(9,-2)(9.1,-2)(9.2,-2)(9.3,-2)(9.4,-2)(9.5,-2)(9.6,-2)(9.7,-2)(9.8,-2)(9.9,-2)(10,-2)(10.1,-2)(10.2,-2)(10.3,-2)(10.4,-2)(10.5,-2)(10.6,-2)(10.7,-2)(10.8,-2)(10.9,-2)(11,-2)(11.1,-2)(11.2,-2)(11.3,-2)(11.4,-2)(11.5,-2)(11.6,-2)(11.7,-2)(11.8,-2)(11.9,-2)(12,-2)(12.1,-2)(12.2,-2)(12.3,-2)(12.4,-2)(12.5,-2)(12.6,-2)(12.7,-2)(12.8,-2)(12.9,-2)(13,-2)(13.1,-2)(13.2,-2)(13.3,-2)(13.4,-2)(13.5,-2)(13.6,-2)(13.7,-2)(13.8,-2)(13.9,-2)(14,-2)(14.1,-2)(14.2,-2)(14.3,-2)(14.4,-2)(14.5,-2)(14.6,-2)(14.7,-2)(14.8,-2)(14.9,-2)(15,-2)(15.1,-2)(15.2,-2)(15.3,-2)(15.4,-2)(15.5,-2)(15.6,-2)(15.7,-2)(15.8,-2)(15.9,-2)(16,-2)(16.1,-2)(16.2,-2)(16.3,-2)(16.4,-2)(16.5,-2)(16.6,-2)(16.7,-2)(16.8,-2)(16.9,-2)(17,-2)(17.1,-2)(17.2,-2)(17.3,-2)(17.4,-2)(17.5,-2)(17.6,-2)(17.7,-2)(17.8,-2)(17.9,-2)(18,-2)(18.1,-2)(18.2,-2)(18.3,-2)(18.4,-2)(18.5,-2)(18.6,-2)(18.7,-2)(18.8,-2)(18.9,-2)(19,-2)(19.1,-2)(19.2,-2)(19.3,-2)(19.4,-2)(19.5,-2)(19.6,-2)(19.7,-2)(19.8,-2)(19.9,-2)(20,-2)(20.1,-2)(20.2,-2)(20.3,-2)(20.4,-2)(20.5,-2)(20.6,-2)(20.7,-2)(20.8,-2)(20.9,-2)(21,-2)(21.1,-2)(21.2,-2)(21.3,-2)(21.4,-2)(21.5,-2)(21.6,-2)(21.7,-2)(21.8,-2)(21.9,-2)(22,-2)(22.1,-2)(22.2,-2)(22.3,-2)(22.4,-2)(22.5,-2)(22.6,-2)(22.7,-2)(22.8,-2)(22.9,-2)(23,-2)(23.1,-2)(23.2,-2)(23.3,-2)(23.4,-2)(23.5,-2)(23.6,-2)(23.7,-2)(23.8,-2)(23.9,-2)(24,-2)(24.1,-2)(24.2,-2)(24.3,-2)(24.4,-2)(24.5,-2)(24.6,-2)(24.7,-2)(24.8,-2)(24.9,-2)(25,-2)(25.1,-2)(25.2,-2)(25.3,-2)(25.4,-2)(25.5,-2)(25.6,-2)(25.7,-2)(25.8,-2)(25.9,-2)(26,-2)(26.1,-2)(26.2,-2)(26.3,-2)(26.4,-2)(26.5,-2)(26.6,-2)(26.7,-2)(26.8,-2)(26.9,-2)(27,-2)(27.1,-2)(27.2,-2)(27.3,-2)(27.4,-2)(27.5,-2)(27.6,-2)(27.7,-2)(27.8,-2)(27.9,-2)(28,-2)(28.1,-2)(28.2,-2)(28.3,-2)(28.4,-2)(28.5,-2)(28.6,-2)(28.7,-2)(28.8,-2)(28.9,-2)(29,-2)(29.1,-2)(29.2,-2)(29.3,-2)(29.4,-2)(29.5,-2)(29.6,-2)(29.7,-2)(29.8,-2)(29.9,-2)(30,-6)(30.1,-6)(30.2,-6)(30.3,-6)(30.4,-6)(30.5,-6)(30.6,-6)(30.7,-6)(30.8,-6)(30.9,-6)(31,-6)(31.1,-6)(31.2,-6)(31.3,-6)(31.4,-6)(31.5,-6)(31.6,-6)(31.7,-6)(31.8,-6)(31.9,-6)(32,-6)(32.1,-6)(32.2,-6)(32.3,-6)(32.4,-6)(32.5,-6)(32.6,-6)(32.7,-6)(32.8,-6)(32.9,-6)(33,-6)(33.1,-6)(33.2,-6)(33.3,-6)(33.4,-6)(33.5,-6)(33.6,-6)(33.7,-6)(33.8,-6)(33.9,-6)(34,-6)(34.1,-6)(34.2,-6)(34.3,-6)(34.4,-6)(34.5,-6)(34.6,-6)(34.7,-6)(34.8,-6)(34.9,-6)(35,-6)(35.1,-6)(35.2,-6)(35.3,-6)(35.4,-6)(35.5,-6)(35.6,-6)(35.7,-6)(35.8,-6)(35.9,-6)(36,-6)(36.1,-6)(36.2,-6)(36.3,-6)(36.4,-6)(36.5,-6)(36.6,-6)(36.7,-6)(36.8,-6)(36.9,-6)(37,-6)(37.1,-6)(37.2,-6)(37.3,-6)(37.4,-6)(37.5,-6)(37.6,-6)(37.7,-6)(37.8,-6)(37.9,-6)(38,-6)(38.1,-6)(38.2,-6)(38.3,-6)(38.4,-6)(38.5,-6)(38.6,-6)(38.7,-6)(38.8,-6)(38.9,-6)(39,-6)(39.1,-6)(39.2,-6)(39.3,-6)(39.4,-6)(39.5,-6)(39.6,-6)(39.7,-6)(39.8,-6)(39.9,-6)(40,-6)(40.1,-6)(40.2,-6)(40.3,-6)(40.4,-6)(40.5,-6)(40.6,-6)(40.7,-6)(40.8,-6)(40.9,-6)(41,-6)(41.1,-6)(41.2,-6)(41.3,-6)(41.4,-6)(41.5,-6)(41.6,-6)(41.7,-6)(41.8,-6)(41.9,-6)(42,-6)(42.1,-6)(42.2,-6)(42.3,-6)(42.4,-6)(42.5,-6)(42.6,-6)(42.7,-6)(42.8,-6)(42.9,-6)(43,-6)(43.1,-6)(43.2,-6)(43.3,-6)(43.4,-6)(43.5,-6)(43.6,-6)(43.7,-6)(43.8,-6)(43.9,-6)(44,-6)(44.1,-6)(44.2,-6)(44.3,-6)(44.4,-6)(44.5,-6)(44.6,-6)(44.7,-6)(44.8,-6)(44.9,-6)(45,-6)(45.1,-6)(45.2,-6)(45.3,-6)(45.4,-6)(45.5,-6)(45.6,-6)(45.7,-6)(45.8,-6)(45.9,-6)(46,-6)(46.1,-6)(46.2,-6)(46.3,-6)(46.4,-6)(46.5,-6)(46.6,-6)(46.7,-6)(46.8,-6)(46.9,-6)(47,-6)(47.1,-6)(47.2,-6)(47.3,-6)(47.4,-6)(47.5,-6)(47.6,-6)(47.7,-6)(47.8,-6)(47.9,-6)(48,-6)(48.1,-6)(48.2,-6)(48.3,-6)(48.4,-6)(48.5,-6)(48.6,-6)(48.7,-6)(48.8,-6)(48.9,-6)(49,-6)(49.1,-6)(49.2,-6)(49.3,-6)(49.4,-6)(49.5,-6)(49.6,-6)(49.7,-6)(49.8,-6)(49.9,-6)(50,-6)(50.1,-6)(50.2,-6)(50.3,-6)(50.4,-6)(50.5,-6)(50.6,-6)(50.7,-6)(50.8,-6)(50.9,-6)(51,-6)(51.1,-6)(51.2,-6)(51.3,-6)(51.4,-6)(51.5,-6)(51.6,-6)(51.7,-6)(51.8,-6)(51.9,-6)(52,-6)(52.1,-6)(52.2,-6)(52.3,-6)(52.4,-6)(52.5,-6)(52.6,-6)(52.7,-6)(52.8,-6)(52.9,-6)(53,-6)(53.1,-6)(53.2,-6)(53.3,-6)(53.4,-6)(53.5,-6)(53.6,-6)(53.7,-6)(53.8,-6)(53.9,-6)(54,-6)(54.1,-6)(54.2,-6)(54.3,-6)(54.4,-6)(54.5,-6)(54.6,-6)(54.7,-6)(54.8,-6)(54.9,-6)(55,-6)(55.1,-6)(55.2,-6)(55.3,-6)(55.4,-6)(55.5,-6)(55.6,-6)(55.7,-6)(55.8,-6)(55.9,-6)(56,-6)(56.1,-6)(56.2,-6)(56.3,-6)(56.4,-6)(56.5,-6)(56.6,-6)(56.7,-6)(56.8,-6)(56.9,-6)(57,-6)(57.1,-6)(57.2,-6)(57.3,-6)(57.4,-6)(57.5,-6)(57.6,-6)(57.7,-6)(57.8,-6)(57.9,-6)(58,-6)(58.1,-6)(58.2,-6)(58.3,-6)(58.4,-6)(58.5,-6)(58.6,-6)(58.7,-6)(58.8,-6)(58.9,-6)(59,-6)(59.1,-6)(59.2,-6)(59.3,-6)(59.4,-6)(59.5,-6)(59.6,-6)(59.7,-6)(59.8,-6)(59.9,-6)(60,-6) 
};
\addplot [
color=mycolor1,
solid,
line width=1.0pt,
forget plot
]
coordinates{
 (0,-3)(0.1,-3)(0.2,-3)(0.3,-3)(0.4,-3)(0.5,-3)(0.6,-3)(0.7,-3)(0.8,-3)(0.9,-3)(1,-3)(1.1,-3)(1.2,-3)(1.3,-3)(1.4,-3)(1.5,-3)(1.6,-3)(1.7,-3)(1.8,-3)(1.9,-3)(2,-3)(2.1,-3)(2.2,-3)(2.3,-3)(2.4,-3)(2.5,-3)(2.6,-3)(2.7,-3)(2.8,-3)(2.9,-3)(3,-3)(3.1,-3)(3.2,-3)(3.3,-3)(3.4,-3)(3.5,-3)(3.6,-3)(3.7,-3)(3.8,-3)(3.9,-3)(4,-3)(4.1,-3)(4.2,-3)(4.3,-3)(4.4,-3)(4.5,-3)(4.6,-3)(4.7,-3)(4.8,-3)(4.9,-3)(5,-3)(5.1,-3)(5.2,-3)(5.3,-3)(5.4,-3)(5.5,-3)(5.6,-3)(5.7,-3)(5.8,-3)(5.9,-3)(6,-3)(6.1,-3)(6.2,-3)(6.3,-3)(6.4,-3)(6.5,-3)(6.6,-3)(6.7,-3)(6.8,-3)(6.9,-3)(7,-3)(7.1,-3)(7.2,-3)(7.3,-3)(7.4,-3)(7.5,-3)(7.6,-3)(7.7,-3)(7.8,-3)(7.9,-3)(8,-3)(8.1,-3)(8.2,-3)(8.3,-3)(8.4,-3)(8.5,-3)(8.6,-3)(8.7,-3)(8.8,-3)(8.9,-3)(9,-3)(9.1,-3)(9.2,-3)(9.3,-3)(9.4,-3)(9.5,-3)(9.6,-3)(9.7,-3)(9.8,-3)(9.9,-3)(10,-3)(10.1,-3)(10.2,-3)(10.3,-3)(10.4,-3)(10.5,-3)(10.6,-3)(10.7,-3)(10.8,-3)(10.9,-3)(11,-3)(11.1,-3)(11.2,-3)(11.3,-3)(11.4,-3)(11.5,-3)(11.6,-3)(11.7,-3)(11.8,-3)(11.9,-3)(12,-3)(12.1,-3)(12.2,-3)(12.3,-3)(12.4,-3)(12.5,-3)(12.6,-3)(12.7,-3)(12.8,-3)(12.9,-3)(13,-3)(13.1,-3)(13.2,-3)(13.3,-3)(13.4,-3)(13.5,-3)(13.6,-3)(13.7,-3)(13.8,-3)(13.9,-3)(14,-3)(14.1,-3)(14.2,-3)(14.3,-3)(14.4,-3)(14.5,-3)(14.6,-3)(14.7,-3)(14.8,-3)(14.9,-3)(15,-3)(15.1,-3)(15.2,-3)(15.3,-3)(15.4,-3)(15.5,-3)(15.6,-3)(15.7,-3)(15.8,-3)(15.9,-3)(16,-3)(16.1,-3)(16.2,-3)(16.3,-3)(16.4,-3)(16.5,-3)(16.6,-3)(16.7,-3)(16.8,-3)(16.9,-3)(17,-3)(17.1,-3)(17.2,-3)(17.3,-3)(17.4,-3)(17.5,-3)(17.6,-3)(17.7,-3)(17.8,-3)(17.9,-3)(18,-3)(18.1,-3)(18.2,-3)(18.3,-3)(18.4,-3)(18.5,-3)(18.6,-3)(18.7,-3)(18.8,-3)(18.9,-3)(19,-3)(19.1,-3)(19.2,-3)(19.3,-3)(19.4,-3)(19.5,-3)(19.6,-3)(19.7,-3)(19.8,-3)(19.9,-3)(20,-3)(20.1,-3)(20.2,-3)(20.3,-3)(20.4,-3)(20.5,-3)(20.6,-3)(20.7,-3)(20.8,-3)(20.9,-3)(21,-3)(21.1,-3)(21.2,-3)(21.3,-3)(21.4,-3)(21.5,-3)(21.6,-3)(21.7,-3)(21.8,-3)(21.9,-3)(22,-3)(22.1,-3)(22.2,-3)(22.3,-3)(22.4,-3)(22.5,-3)(22.6,-3)(22.7,-3)(22.8,-3)(22.9,-3)(23,-3)(23.1,-3)(23.2,-3)(23.3,-3)(23.4,-3)(23.5,-3)(23.6,-3)(23.7,-3)(23.8,-3)(23.9,-3)(24,-3)(24.1,-3)(24.2,-3)(24.3,-3)(24.4,-3)(24.5,-3)(24.6,-3)(24.7,-3)(24.8,-3)(24.9,-3)(25,-3)(25.1,-3)(25.2,-3)(25.3,-3)(25.4,-3)(25.5,-3)(25.6,-3)(25.7,-3)(25.8,-3)(25.9,-3)(26,-3)(26.1,-3)(26.2,-3)(26.3,-3)(26.4,-3)(26.5,-3)(26.6,-3)(26.7,-3)(26.8,-3)(26.9,-3)(27,-3)(27.1,-3)(27.2,-3)(27.3,-3)(27.4,-3)(27.5,-3)(27.6,-3)(27.7,-3)(27.8,-3)(27.9,-3)(28,-3)(28.1,-3)(28.2,-3)(28.3,-3)(28.4,-3)(28.5,-3)(28.6,-3)(28.7,-3)(28.8,-3)(28.9,-3)(29,-3)(29.1,-3)(29.2,-3)(29.3,-3)(29.4,-3)(29.5,-3)(29.6,-3)(29.7,-3)(29.8,-3)(29.9,-3)(30,-8)(30.1,-8)(30.2,-8)(30.3,-8)(30.4,-8)(30.5,-8)(30.6,-8)(30.7,-8)(30.8,-8)(30.9,-8)(31,-8)(31.1,-8)(31.2,-8)(31.3,-8)(31.4,-8)(31.5,-8)(31.6,-8)(31.7,-8)(31.8,-8)(31.9,-8)(32,-8)(32.1,-8)(32.2,-8)(32.3,-8)(32.4,-8)(32.5,-8)(32.6,-8)(32.7,-8)(32.8,-8)(32.9,-8)(33,-8)(33.1,-8)(33.2,-8)(33.3,-8)(33.4,-8)(33.5,-8)(33.6,-8)(33.7,-8)(33.8,-8)(33.9,-8)(34,-8)(34.1,-8)(34.2,-8)(34.3,-8)(34.4,-8)(34.5,-8)(34.6,-8)(34.7,-8)(34.8,-8)(34.9,-8)(35,-8)(35.1,-8)(35.2,-8)(35.3,-8)(35.4,-8)(35.5,-8)(35.6,-8)(35.7,-8)(35.8,-8)(35.9,-8)(36,-8)(36.1,-8)(36.2,-8)(36.3,-8)(36.4,-8)(36.5,-8)(36.6,-8)(36.7,-8)(36.8,-8)(36.9,-8)(37,-8)(37.1,-8)(37.2,-8)(37.3,-8)(37.4,-8)(37.5,-8)(37.6,-8)(37.7,-8)(37.8,-8)(37.9,-8)(38,-8)(38.1,-8)(38.2,-8)(38.3,-8)(38.4,-8)(38.5,-8)(38.6,-8)(38.7,-8)(38.8,-8)(38.9,-8)(39,-8)(39.1,-8)(39.2,-8)(39.3,-8)(39.4,-8)(39.5,-8)(39.6,-8)(39.7,-8)(39.8,-8)(39.9,-8)(40,-8)(40.1,-8)(40.2,-8)(40.3,-8)(40.4,-8)(40.5,-8)(40.6,-8)(40.7,-8)(40.8,-8)(40.9,-8)(41,-8)(41.1,-8)(41.2,-8)(41.3,-8)(41.4,-8)(41.5,-8)(41.6,-8)(41.7,-8)(41.8,-8)(41.9,-8)(42,-8)(42.1,-8)(42.2,-8)(42.3,-8)(42.4,-8)(42.5,-8)(42.6,-8)(42.7,-8)(42.8,-8)(42.9,-8)(43,-8)(43.1,-8)(43.2,-8)(43.3,-8)(43.4,-8)(43.5,-8)(43.6,-8)(43.7,-8)(43.8,-8)(43.9,-8)(44,-8)(44.1,-8)(44.2,-8)(44.3,-8)(44.4,-8)(44.5,-8)(44.6,-8)(44.7,-8)(44.8,-8)(44.9,-8)(45,-8)(45.1,-8)(45.2,-8)(45.3,-8)(45.4,-8)(45.5,-8)(45.6,-8)(45.7,-8)(45.8,-8)(45.9,-8)(46,-8)(46.1,-8)(46.2,-8)(46.3,-8)(46.4,-8)(46.5,-8)(46.6,-8)(46.7,-8)(46.8,-8)(46.9,-8)(47,-8)(47.1,-8)(47.2,-8)(47.3,-8)(47.4,-8)(47.5,-8)(47.6,-8)(47.7,-8)(47.8,-8)(47.9,-8)(48,-8)(48.1,-8)(48.2,-8)(48.3,-8)(48.4,-8)(48.5,-8)(48.6,-8)(48.7,-8)(48.8,-8)(48.9,-8)(49,-8)(49.1,-8)(49.2,-8)(49.3,-8)(49.4,-8)(49.5,-8)(49.6,-8)(49.7,-8)(49.8,-8)(49.9,-8)(50,-8)(50.1,-8)(50.2,-8)(50.3,-8)(50.4,-8)(50.5,-8)(50.6,-8)(50.7,-8)(50.8,-8)(50.9,-8)(51,-8)(51.1,-8)(51.2,-8)(51.3,-8)(51.4,-8)(51.5,-8)(51.6,-8)(51.7,-8)(51.8,-8)(51.9,-8)(52,-8)(52.1,-8)(52.2,-8)(52.3,-8)(52.4,-8)(52.5,-8)(52.6,-8)(52.7,-8)(52.8,-8)(52.9,-8)(53,-8)(53.1,-8)(53.2,-8)(53.3,-8)(53.4,-8)(53.5,-8)(53.6,-8)(53.7,-8)(53.8,-8)(53.9,-8)(54,-8)(54.1,-8)(54.2,-8)(54.3,-8)(54.4,-8)(54.5,-8)(54.6,-8)(54.7,-8)(54.8,-8)(54.9,-8)(55,-8)(55.1,-8)(55.2,-8)(55.3,-8)(55.4,-8)(55.5,-8)(55.6,-8)(55.7,-8)(55.8,-8)(55.9,-8)(56,-8)(56.1,-8)(56.2,-8)(56.3,-8)(56.4,-8)(56.5,-8)(56.6,-8)(56.7,-8)(56.8,-8)(56.9,-8)(57,-8)(57.1,-8)(57.2,-8)(57.3,-8)(57.4,-8)(57.5,-8)(57.6,-8)(57.7,-8)(57.8,-8)(57.9,-8)(58,-8)(58.1,-8)(58.2,-8)(58.3,-8)(58.4,-8)(58.5,-8)(58.6,-8)(58.7,-8)(58.8,-8)(58.9,-8)(59,-8)(59.1,-8)(59.2,-8)(59.3,-8)(59.4,-8)(59.5,-8)(59.6,-8)(59.7,-8)(59.8,-8)(59.9,-8)(60,-8) 
};
\addplot [
color=mycolor2,
solid,
line width=1.0pt,
forget plot
]
coordinates{
 (0,-4)(0.1,-4)(0.2,-4)(0.3,-4)(0.4,-4)(0.5,-4)(0.6,-4)(0.7,-4)(0.8,-4)(0.9,-4)(1,-4)(1.1,-4)(1.2,-4)(1.3,-4)(1.4,-4)(1.5,-4)(1.6,-4)(1.7,-4)(1.8,-4)(1.9,-4)(2,-4)(2.1,-4)(2.2,-4)(2.3,-4)(2.4,-4)(2.5,-4)(2.6,-4)(2.7,-4)(2.8,-4)(2.9,-4)(3,-4)(3.1,-4)(3.2,-4)(3.3,-4)(3.4,-4)(3.5,-4)(3.6,-4)(3.7,-4)(3.8,-4)(3.9,-4)(4,-4)(4.1,-4)(4.2,-4)(4.3,-4)(4.4,-4)(4.5,-4)(4.6,-4)(4.7,-4)(4.8,-4)(4.9,-4)(5,-4)(5.1,-4)(5.2,-4)(5.3,-4)(5.4,-4)(5.5,-4)(5.6,-4)(5.7,-4)(5.8,-4)(5.9,-4)(6,-4)(6.1,-4)(6.2,-4)(6.3,-4)(6.4,-4)(6.5,-4)(6.6,-4)(6.7,-4)(6.8,-4)(6.9,-4)(7,-4)(7.1,-4)(7.2,-4)(7.3,-4)(7.4,-4)(7.5,-4)(7.6,-4)(7.7,-4)(7.8,-4)(7.9,-4)(8,-4)(8.1,-4)(8.2,-4)(8.3,-4)(8.4,-4)(8.5,-4)(8.6,-4)(8.7,-4)(8.8,-4)(8.9,-4)(9,-4)(9.1,-4)(9.2,-4)(9.3,-4)(9.4,-4)(9.5,-4)(9.6,-4)(9.7,-4)(9.8,-4)(9.9,-4)(10,-4)(10.1,-4)(10.2,-4)(10.3,-4)(10.4,-4)(10.5,-4)(10.6,-4)(10.7,-4)(10.8,-4)(10.9,-4)(11,-4)(11.1,-4)(11.2,-4)(11.3,-4)(11.4,-4)(11.5,-4)(11.6,-4)(11.7,-4)(11.8,-4)(11.9,-4)(12,-4)(12.1,-4)(12.2,-4)(12.3,-4)(12.4,-4)(12.5,-4)(12.6,-4)(12.7,-4)(12.8,-4)(12.9,-4)(13,-4)(13.1,-4)(13.2,-4)(13.3,-4)(13.4,-4)(13.5,-4)(13.6,-4)(13.7,-4)(13.8,-4)(13.9,-4)(14,-4)(14.1,-4)(14.2,-4)(14.3,-4)(14.4,-4)(14.5,-4)(14.6,-4)(14.7,-4)(14.8,-4)(14.9,-4)(15,-4)(15.1,-4)(15.2,-4)(15.3,-4)(15.4,-4)(15.5,-4)(15.6,-4)(15.7,-4)(15.8,-4)(15.9,-4)(16,-4)(16.1,-4)(16.2,-4)(16.3,-4)(16.4,-4)(16.5,-4)(16.6,-4)(16.7,-4)(16.8,-4)(16.9,-4)(17,-4)(17.1,-4)(17.2,-4)(17.3,-4)(17.4,-4)(17.5,-4)(17.6,-4)(17.7,-4)(17.8,-4)(17.9,-4)(18,-4)(18.1,-4)(18.2,-4)(18.3,-4)(18.4,-4)(18.5,-4)(18.6,-4)(18.7,-4)(18.8,-4)(18.9,-4)(19,-4)(19.1,-4)(19.2,-4)(19.3,-4)(19.4,-4)(19.5,-4)(19.6,-4)(19.7,-4)(19.8,-4)(19.9,-4)(20,-4)(20.1,-4)(20.2,-4)(20.3,-4)(20.4,-4)(20.5,-4)(20.6,-4)(20.7,-4)(20.8,-4)(20.9,-4)(21,-4)(21.1,-4)(21.2,-4)(21.3,-4)(21.4,-4)(21.5,-4)(21.6,-4)(21.7,-4)(21.8,-4)(21.9,-4)(22,-4)(22.1,-4)(22.2,-4)(22.3,-4)(22.4,-4)(22.5,-4)(22.6,-4)(22.7,-4)(22.8,-4)(22.9,-4)(23,-4)(23.1,-4)(23.2,-4)(23.3,-4)(23.4,-4)(23.5,-4)(23.6,-4)(23.7,-4)(23.8,-4)(23.9,-4)(24,-4)(24.1,-4)(24.2,-4)(24.3,-4)(24.4,-4)(24.5,-4)(24.6,-4)(24.7,-4)(24.8,-4)(24.9,-4)(25,-4)(25.1,-4)(25.2,-4)(25.3,-4)(25.4,-4)(25.5,-4)(25.6,-4)(25.7,-4)(25.8,-4)(25.9,-4)(26,-4)(26.1,-4)(26.2,-4)(26.3,-4)(26.4,-4)(26.5,-4)(26.6,-4)(26.7,-4)(26.8,-4)(26.9,-4)(27,-4)(27.1,-4)(27.2,-4)(27.3,-4)(27.4,-4)(27.5,-4)(27.6,-4)(27.7,-4)(27.8,-4)(27.9,-4)(28,-4)(28.1,-4)(28.2,-4)(28.3,-4)(28.4,-4)(28.5,-4)(28.6,-4)(28.7,-4)(28.8,-4)(28.9,-4)(29,-4)(29.1,-4)(29.2,-4)(29.3,-4)(29.4,-4)(29.5,-4)(29.6,-4)(29.7,-4)(29.8,-4)(29.9,-4)(30,-10)(30.1,-10)(30.2,-10)(30.3,-10)(30.4,-10)(30.5,-10)(30.6,-10)(30.7,-10)(30.8,-10)(30.9,-10)(31,-10)(31.1,-10)(31.2,-10)(31.3,-10)(31.4,-10)(31.5,-10)(31.6,-10)(31.7,-10)(31.8,-10)(31.9,-10)(32,-10)(32.1,-10)(32.2,-10)(32.3,-10)(32.4,-10)(32.5,-10)(32.6,-10)(32.7,-10)(32.8,-10)(32.9,-10)(33,-10)(33.1,-10)(33.2,-10)(33.3,-10)(33.4,-10)(33.5,-10)(33.6,-10)(33.7,-10)(33.8,-10)(33.9,-10)(34,-10)(34.1,-10)(34.2,-10)(34.3,-10)(34.4,-10)(34.5,-10)(34.6,-10)(34.7,-10)(34.8,-10)(34.9,-10)(35,-10)(35.1,-10)(35.2,-10)(35.3,-10)(35.4,-10)(35.5,-10)(35.6,-10)(35.7,-10)(35.8,-10)(35.9,-10)(36,-10)(36.1,-10)(36.2,-10)(36.3,-10)(36.4,-10)(36.5,-10)(36.6,-10)(36.7,-10)(36.8,-10)(36.9,-10)(37,-10)(37.1,-10)(37.2,-10)(37.3,-10)(37.4,-10)(37.5,-10)(37.6,-10)(37.7,-10)(37.8,-10)(37.9,-10)(38,-10)(38.1,-10)(38.2,-10)(38.3,-10)(38.4,-10)(38.5,-10)(38.6,-10)(38.7,-10)(38.8,-10)(38.9,-10)(39,-10)(39.1,-10)(39.2,-10)(39.3,-10)(39.4,-10)(39.5,-10)(39.6,-10)(39.7,-10)(39.8,-10)(39.9,-10)(40,-10)(40.1,-10)(40.2,-10)(40.3,-10)(40.4,-10)(40.5,-10)(40.6,-10)(40.7,-10)(40.8,-10)(40.9,-10)(41,-10)(41.1,-10)(41.2,-10)(41.3,-10)(41.4,-10)(41.5,-10)(41.6,-10)(41.7,-10)(41.8,-10)(41.9,-10)(42,-10)(42.1,-10)(42.2,-10)(42.3,-10)(42.4,-10)(42.5,-10)(42.6,-10)(42.7,-10)(42.8,-10)(42.9,-10)(43,-10)(43.1,-10)(43.2,-10)(43.3,-10)(43.4,-10)(43.5,-10)(43.6,-10)(43.7,-10)(43.8,-10)(43.9,-10)(44,-10)(44.1,-10)(44.2,-10)(44.3,-10)(44.4,-10)(44.5,-10)(44.6,-10)(44.7,-10)(44.8,-10)(44.9,-10)(45,-10)(45.1,-10)(45.2,-10)(45.3,-10)(45.4,-10)(45.5,-10)(45.6,-10)(45.7,-10)(45.8,-10)(45.9,-10)(46,-10)(46.1,-10)(46.2,-10)(46.3,-10)(46.4,-10)(46.5,-10)(46.6,-10)(46.7,-10)(46.8,-10)(46.9,-10)(47,-10)(47.1,-10)(47.2,-10)(47.3,-10)(47.4,-10)(47.5,-10)(47.6,-10)(47.7,-10)(47.8,-10)(47.9,-10)(48,-10)(48.1,-10)(48.2,-10)(48.3,-10)(48.4,-10)(48.5,-10)(48.6,-10)(48.7,-10)(48.8,-10)(48.9,-10)(49,-10)(49.1,-10)(49.2,-10)(49.3,-10)(49.4,-10)(49.5,-10)(49.6,-10)(49.7,-10)(49.8,-10)(49.9,-10)(50,-10)(50.1,-10)(50.2,-10)(50.3,-10)(50.4,-10)(50.5,-10)(50.6,-10)(50.7,-10)(50.8,-10)(50.9,-10)(51,-10)(51.1,-10)(51.2,-10)(51.3,-10)(51.4,-10)(51.5,-10)(51.6,-10)(51.7,-10)(51.8,-10)(51.9,-10)(52,-10)(52.1,-10)(52.2,-10)(52.3,-10)(52.4,-10)(52.5,-10)(52.6,-10)(52.7,-10)(52.8,-10)(52.9,-10)(53,-10)(53.1,-10)(53.2,-10)(53.3,-10)(53.4,-10)(53.5,-10)(53.6,-10)(53.7,-10)(53.8,-10)(53.9,-10)(54,-10)(54.1,-10)(54.2,-10)(54.3,-10)(54.4,-10)(54.5,-10)(54.6,-10)(54.7,-10)(54.8,-10)(54.9,-10)(55,-10)(55.1,-10)(55.2,-10)(55.3,-10)(55.4,-10)(55.5,-10)(55.6,-10)(55.7,-10)(55.8,-10)(55.9,-10)(56,-10)(56.1,-10)(56.2,-10)(56.3,-10)(56.4,-10)(56.5,-10)(56.6,-10)(56.7,-10)(56.8,-10)(56.9,-10)(57,-10)(57.1,-10)(57.2,-10)(57.3,-10)(57.4,-10)(57.5,-10)(57.6,-10)(57.7,-10)(57.8,-10)(57.9,-10)(58,-10)(58.1,-10)(58.2,-10)(58.3,-10)(58.4,-10)(58.5,-10)(58.6,-10)(58.7,-10)(58.8,-10)(58.9,-10)(59,-10)(59.1,-10)(59.2,-10)(59.3,-10)(59.4,-10)(59.5,-10)(59.6,-10)(59.7,-10)(59.8,-10)(59.9,-10)(60,-10) 
};
\end{axis}
\end{tikzpicture}%

%% file: figures/sim_platoon_w.tikz
%
%
%

\definecolor{mycolor1}{rgb}{0,0.75,0.75}
\definecolor{mycolor2}{rgb}{0.75,0,0.75}

\begin{tikzpicture}

\begin{axis}[%
width=\figurewidth,
height=\figureheight,
scale only axis,
xmin=0, xmax=60,
xmajorgrids,
ymin=-55, ymax=55,
ylabel={$w_k(t)$},
ymajorgrids]
\addplot [
color=blue,
solid,
line width=1.0pt,
forget plot
]
coordinates{
 (0,0)(0.1,0)(0.2,0)(0.3,0)(0.4,0)(0.5,0)(0.6,0)(0.7,0)(0.8,0)(0.9,0)(1,0)(1.1,0)(1.2,0)(1.3,0)(1.4,0)(1.5,0)(1.6,0)(1.7,0)(1.8,0)(1.9,0)(2,0)(2.1,0)(2.2,0)(2.3,0)(2.4,0)(2.5,0)(2.6,0)(2.7,0)(2.8,0)(2.9,0)(3,0)(3.1,0)(3.2,0)(3.3,0)(3.4,0)(3.5,0)(3.6,0)(3.7,0)(3.8,0)(3.9,0)(4,0)(4.1,0)(4.2,0)(4.3,0)(4.4,0)(4.5,0)(4.6,0)(4.7,0)(4.8,0)(4.9,0)(5,0)(5.1,0)(5.2,0)(5.3,0)(5.4,0)(5.5,0)(5.6,0)(5.7,0)(5.8,0)(5.9,0)(6,0)(6.1,0)(6.2,0)(6.3,0)(6.4,0)(6.5,0)(6.6,0)(6.7,0)(6.8,0)(6.9,0)(7,0)(7.1,0)(7.2,0)(7.3,0)(7.4,0)(7.5,0)(7.6,0)(7.7,0)(7.8,0)(7.9,0)(8,0)(8.1,0)(8.2,0)(8.3,0)(8.4,0)(8.5,0)(8.6,0)(8.7,0)(8.8,0)(8.9,0)(9,0)(9.1,0)(9.2,0)(9.3,0)(9.4,0)(9.5,0)(9.6,0)(9.7,0)(9.8,0)(9.9,0)(10,0)(10.1,0)(10.2,0)(10.3,0)(10.4,0)(10.5,0)(10.6,0)(10.7,0)(10.8,0)(10.9,0)(11,0)(11.1,0)(11.2,0)(11.3,0)(11.4,0)(11.5,0)(11.6,0)(11.7,0)(11.8,0)(11.9,0)(12,0)(12.1,0)(12.2,0)(12.3,0)(12.4,0)(12.5,0)(12.6,0)(12.7,0)(12.8,0)(12.9,0)(13,0)(13.1,0)(13.2,0)(13.3,0)(13.4,0)(13.5,0)(13.6,0)(13.7,0)(13.8,0)(13.9,0)(14,0)(14.1,0)(14.2,0)(14.3,0)(14.4,0)(14.5,0)(14.6,0)(14.7,0)(14.8,0)(14.9,0)(15,0)(15.1,0)(15.2,0)(15.3,0)(15.4,0)(15.5,0)(15.6,0)(15.7,0)(15.8,0)(15.9,0)(16,0)(16.1,0)(16.2,0)(16.3,0)(16.4,0)(16.5,0)(16.6,0)(16.7,0)(16.8,0)(16.9,0)(17,0)(17.1,0)(17.2,0)(17.3,0)(17.4,0)(17.5,0)(17.6,0)(17.7,0)(17.8,0)(17.9,0)(18,0)(18.1,0)(18.2,0)(18.3,0)(18.4,0)(18.5,0)(18.6,0)(18.7,0)(18.8,0)(18.9,0)(19,0)(19.1,0)(19.2,0)(19.3,0)(19.4,0)(19.5,0)(19.6,0)(19.7,0)(19.8,0)(19.9,0)(20,0)(20.1,0)(20.2,0)(20.3,0)(20.4,0)(20.5,0)(20.6,0)(20.7,0)(20.8,0)(20.9,0)(21,0)(21.1,0)(21.2,0)(21.3,0)(21.4,0)(21.5,0)(21.6,0)(21.7,0)(21.8,0)(21.9,0)(22,0)(22.1,0)(22.2,0)(22.3,0)(22.4,0)(22.5,0)(22.6,0)(22.7,0)(22.8,0)(22.9,0)(23,0)(23.1,0)(23.2,0)(23.3,0)(23.4,0)(23.5,0)(23.6,0)(23.7,0)(23.8,0)(23.9,0)(24,0)(24.1,0)(24.2,0)(24.3,0)(24.4,0)(24.5,0)(24.6,0)(24.7,0)(24.8,0)(24.9,0)(25,0)(25.1,0)(25.2,0)(25.3,0)(25.4,0)(25.5,0)(25.6,0)(25.7,0)(25.8,0)(25.9,0)(26,0)(26.1,0)(26.2,0)(26.3,0)(26.4,0)(26.5,0)(26.6,0)(26.7,0)(26.8,0)(26.9,0)(27,0)(27.1,0)(27.2,0)(27.3,0)(27.4,0)(27.5,0)(27.6,0)(27.7,0)(27.8,0)(27.9,0)(28,0)(28.1,0)(28.2,0)(28.3,0)(28.4,0)(28.5,0)(28.6,0)(28.7,0)(28.8,0)(28.9,0)(29,0)(29.1,0)(29.2,0)(29.3,0)(29.4,0)(29.5,0)(29.6,0)(29.7,0)(29.8,0)(29.9,0)(30,0)(30.1,0)(30.2,0)(30.3,0)(30.4,0)(30.5,0)(30.6,0)(30.7,0)(30.8,0)(30.9,0)(31,0)(31.1,0)(31.2,0)(31.3,0)(31.4,0)(31.5,0)(31.6,0)(31.7,0)(31.8,0)(31.9,0)(32,0)(32.1,0)(32.2,0)(32.3,0)(32.4,0)(32.5,0)(32.6,0)(32.7,0)(32.8,0)(32.9,0)(33,0)(33.1,0)(33.2,0)(33.3,0)(33.4,0)(33.5,0)(33.6,0)(33.7,0)(33.8,0)(33.9,0)(34,0)(34.1,0)(34.2,0)(34.3,0)(34.4,0)(34.5,0)(34.6,0)(34.7,0)(34.8,0)(34.9,0)(35,0)(35.1,0)(35.2,0)(35.3,0)(35.4,0)(35.5,0)(35.6,0)(35.7,0)(35.8,0)(35.9,0)(36,0)(36.1,0)(36.2,0)(36.3,0)(36.4,0)(36.5,0)(36.6,0)(36.7,0)(36.8,0)(36.9,0)(37,0)(37.1,0)(37.2,0)(37.3,0)(37.4,0)(37.5,0)(37.6,0)(37.7,0)(37.8,0)(37.9,0)(38,0)(38.1,0)(38.2,0)(38.3,0)(38.4,0)(38.5,0)(38.6,0)(38.7,0)(38.8,0)(38.9,0)(39,0)(39.1,0)(39.2,0)(39.3,0)(39.4,0)(39.5,0)(39.6,0)(39.7,0)(39.8,0)(39.9,0)(40,0)(40.1,0)(40.2,0)(40.3,0)(40.4,0)(40.5,0)(40.6,0)(40.7,0)(40.8,0)(40.9,0)(41,0)(41.1,0)(41.2,0)(41.3,0)(41.4,0)(41.5,0)(41.6,0)(41.7,0)(41.8,0)(41.9,0)(42,0)(42.1,0)(42.2,0)(42.3,0)(42.4,0)(42.5,0)(42.6,0)(42.7,0)(42.8,0)(42.9,0)(43,0)(43.1,0)(43.2,0)(43.3,0)(43.4,0)(43.5,0)(43.6,0)(43.7,0)(43.8,0)(43.9,0)(44,0)(44.1,0)(44.2,0)(44.3,0)(44.4,0)(44.5,0)(44.6,0)(44.7,0)(44.8,0)(44.9,0)(45,0)(45.1,0)(45.2,0)(45.3,0)(45.4,0)(45.5,0)(45.6,0)(45.7,0)(45.8,0)(45.9,0)(46,0)(46.1,0)(46.2,0)(46.3,0)(46.4,0)(46.5,0)(46.6,0)(46.7,0)(46.8,0)(46.9,0)(47,0)(47.1,0)(47.2,0)(47.3,0)(47.4,0)(47.5,0)(47.6,0)(47.7,0)(47.8,0)(47.9,0)(48,0)(48.1,0)(48.2,0)(48.3,0)(48.4,0)(48.5,0)(48.6,0)(48.7,0)(48.8,0)(48.9,0)(49,0)(49.1,0)(49.2,0)(49.3,0)(49.4,0)(49.5,0)(49.6,0)(49.7,0)(49.8,0)(49.9,0)(50,0)(50.1,0)(50.2,0)(50.3,0)(50.4,0)(50.5,0)(50.6,0)(50.7,0)(50.8,0)(50.9,0)(51,0)(51.1,0)(51.2,0)(51.3,0)(51.4,0)(51.5,0)(51.6,0)(51.7,0)(51.8,0)(51.9,0)(52,0)(52.1,0)(52.2,0)(52.3,0)(52.4,0)(52.5,0)(52.6,0)(52.7,0)(52.8,0)(52.9,0)(53,0)(53.1,0)(53.2,0)(53.3,0)(53.4,0)(53.5,0)(53.6,0)(53.7,0)(53.8,0)(53.9,0)(54,0)(54.1,0)(54.2,0)(54.3,0)(54.4,0)(54.5,0)(54.6,0)(54.7,0)(54.8,0)(54.9,0)(55,0)(55.1,0)(55.2,0)(55.3,0)(55.4,0)(55.5,0)(55.6,0)(55.7,0)(55.8,0)(55.9,0)(56,0)(56.1,0)(56.2,0)(56.3,0)(56.4,0)(56.5,0)(56.6,0)(56.7,0)(56.8,0)(56.9,0)(57,0)(57.1,0)(57.2,0)(57.3,0)(57.4,0)(57.5,0)(57.6,0)(57.7,0)(57.8,0)(57.9,0)(58,0)(58.1,0)(58.2,0)(58.3,0)(58.4,0)(58.5,0)(58.6,0)(58.7,0)(58.8,0)(58.9,0)(59,0)(59.1,0)(59.2,0)(59.3,0)(59.4,0)(59.5,0)(59.6,0)(59.7,0)(59.8,0)(59.9,0)(60,0) 
};
\addplot [
color=green!50!black,
solid,
line width=1.0pt,
forget plot
]
coordinates{
 (0,10)(0.1,9.95004165278025)(0.2,9.80066577841239)(0.3,9.55336489125601)(0.4,9.21060994002879)(0.5,8.77582561890366)(0.6,8.2533561490967)(0.7,7.6484218728448)(0.8,6.96706709347156)(0.9,6.21609968270655)(1,5.40302305868131)(1.1,4.53596121425569)(1.2,3.62357754476665)(1.3,2.67498828624581)(1.4,1.69967142900236)(1.5,0.707372016677002)(1.6,-0.291995223012888)(1.7,-1.28844494295522)(1.8,-2.2720209469308)(1.9,-3.23289566863493)(2,-4.16146836547127)(2.1,-5.04846104599838)(2.2,-5.88501117255322)(2.3,-6.66276021279796)(2.4,-7.37393715541213)(2.5,-8.01143615546895)(2.6,-8.56888753368904)(2.7,-9.04072142017013)(2.8,-9.42222340668606)(2.9,-9.70958165149534)(3,-9.89992496600385)(3.1,-9.99135150273214)(3.2,-9.98294775794686)(3.3,-9.87479769908797)(3.4,-9.66798192579391)(3.5,-9.36456687290726)(3.6,-8.96758416334077)(3.7,-8.4810003171034)(3.8,-7.9096771191435)(3.9,-7.25932304200076)(4,-6.53643620863553)(4.1,-5.74823946533215)(4.2,-4.90260821340652)(4.3,-4.00799172079936)(4.4,-3.07332869978387)(4.5,-2.10795799430756)(4.6,-1.1215252693504)(4.7,-0.123886634628858)(4.8,0.874989834394413)(4.9,1.8651236942256)(5,2.836621854632)(5.1,3.77977742712945)(5.2,4.6851667130033)(5.3,5.54374336179103)(5.4,6.34692875942565)(5.5,7.08669774291181)(5.6,7.75565878510162)(5.7,8.34712784839062)(5.8,8.85519516941212)(5.9,9.2747843074392)(6,9.60170286650243)(6.1,9.83268438442456)(6.2,9.96542097023083)(6.3,9.99858636383278)(6.4,9.93184918758053)(6.5,9.76587625727883)(6.6,9.5023259195839)(6.7,9.14383148235182)(6.8,8.69397490349692)(6.9,8.1572510012523)(7,7.53902254343185)(7.1,6.84546666442695)(7.2,6.08351314532153)(7.3,5.26077517381015)(7.4,4.38547327574314)(7.5,3.46635317834964)(7.6,2.51259842582209)(7.7,1.53373862037834)(7.8,0.539554205626344)(7.9,-0.460021256395345)(8,-1.45500033808592)(8.1,-2.43544153735752)(8.2,-3.39154860983779)(8.3,-4.31376844970544)(8.4,-5.19288654116592)(8.5,-6.02011902684714)(8.6,-6.78720047319884)(8.7,-7.48646645597257)(8.8,-8.110930140615)(8.9,-8.65435209240944)(9,-9.11130261884496)(9.1,-9.47721602130918)(9.2,-9.74843621403964)(9.3,-9.92225325452397)(9.4,-9.99693042034995)(9.5,-9.97172156196165)(9.6,-9.84687855793913)(9.7,-9.623648798311)(9.8,-9.30426272104547)(9.9,-8.89191152625161)(10,-8.39071529076262)(10.1,-7.80568180169002)(10.2,-7.14265652027032)(10.3,-6.4082641759484)(10.4,-5.60984257427093)(10.5,-4.75536927995876)(10.6,-3.85338190771733)(10.7,-2.91289281721271)(10.8,-1.94329906455286)(10.9,-0.954288510009244)(11,0.0442569798805476)(11.1,1.04236026865683)(11.2,2.0300486381871)(11.3,2.99745343276949)(11.4,3.93490866347797)(11.5,4.83304758752889)(11.6,5.68289629767835)(11.7,6.47596338653714)(11.8,7.20432478990654)(11.9,7.86070296140837)(12,8.43853958732272)(12.1,8.93206111509085)(12.2,9.33633644074385)(12.3,9.64732617886344)(12.4,9.86192302278588)(12.5,9.97798279178298)(12.6,9.99434585500718)(12.7,9.91084871813965)(12.8,9.7283256569715)(12.9,9.44860038159581)(13,9.07446781449924)(13.1,8.60966616462046)(13.2,8.05883957640203)(13.3,7.42749172703439)(13.4,6.72193083553255)(13.5,5.949206633097)(13.6,5.11703992452977)(13.7,4.23374544450518)(13.8,3.30814877948931)(13.9,2.34949818539735)(14,1.36737218207777)(14.1,0.371583847907989)(14.2,-0.627917229240763)(14.3,-1.62114436499678)(14.4,-2.59817356213686)(14.5,-3.54924266788606)(14.6,-4.46484891412136)(14.7,-5.33584386588951)(14.8,-6.15352482954523)(14.9,-6.90972180718901)(15,-7.59687912858572)(15.1,-8.20813094492394)(15.2,-8.73736983010784)(15.3,-9.17930780413977)(15.4,-9.52952916886849)(15.5,-9.7845346281854)(15.6,-9.94177625183462)(15.7,-9.9996829334898)(15.8,-9.95767608872927)(15.9,-9.81617543606023)(16,-9.57659480323029)(16.1,-9.24132800072785)(16.2,-8.81372490361905)(16.3,-8.29805798070338)(16.4,-7.69947960541778)(16.5,-7.02397057502446)(16.6,-6.27828035246148)(16.7,-5.46985962794027)(16.8,-4.60678587411184)(16.9,-3.69768263863027)(17,-2.75163338051491)(17.1,-1.77809071123051)(17.2,-0.786781947318125)(17.3,0.212388081736323)(17.4,1.20943599928422)(17.5,2.19439963211367)(17.6,3.15743754919105)(17.7,4.08892739398702)(17.8,4.97956202788202)(17.9,5.82044252401872)(18,6.60316708243795)(18.1,7.31991497808632)(18.2,7.96352470291584)(18.3,8.52756552130511)(18.4,9.00640172384385)(18.5,9.39524893747853)(18.6,9.69022192938634)(18.7,9.8883734269372)(18.8,9.98772356586779)(18.9,9.98727967243069)(19,9.8870461818624)(19.1,9.68802459406788)(19.2,9.39220346696462)(19.3,9.00253854746912)(19.4,8.52292323865092)(19.5,7.95814969813595)(19.6,7.31386095645178)(19.7,6.59649453373176)(19.8,5.81321811814183)(19.9,4.97185794870988)(20,4.0808206181321)(20.1,3.14900907687794)(20.2,2.18573367785167)(20.3,1.20061915042384)(20.4,0.203508433316855)(20.5,-0.795635672784825)(20.6,-1.78683005024625)(20.7,-2.76017101249318)(20.8,-3.70593325837446)(20.9,-4.6146670441567)(21,-5.47729260223985)(21.1,-6.28519086319366)(21.2,-7.03028957465033)(21.3,-7.70514395658184)(21.4,-8.3030110870811)(21.5,-8.81791727540879)(21.6,-9.24471774913656)(21.7,-9.57914805901232)(21.8,-9.8178666879278)(21.9,-9.95848843825277)(22,-9.99960826394131)(22.1,-9.94081530928759)(22.2,-9.78269701406012)(22.3,-9.52683324399787)(22.4,-9.17578050531395)(22.5,-8.73304640093077)(22.6,-8.20305458367074)(22.7,-7.59110055658004)(22.8,-6.90329876201225)(22.9,-6.14652148814162)(23,-5.3283302033314)(23.1,-4.45690000444124)(23.2,-3.54093793396194)(23.3,-2.58959598212501)(23.4,-1.61237964324124)(23.5,-0.619052939944165)(23.6,0.380459135697285)(23.7,1.37616978941788)(23.8,2.35813020950376)(23.9,3.31652897203057)(24,4.24179007336748)(24.1,5.12466861044071)(24.2,5.95634315274867)(24.3,6.7285038831796)(24.4,7.43343562695754)(24.5,8.06409493911793)(24.6,8.61418048028208)(24.7,9.07819597755673)(24.8,9.45150514147625)(24.9,9.73037799027438)(25,9.91202811862901)(25.1,9.99464053850317)(25.2,9.97738981390553)(25.3,9.86044830837392)(25.4,9.64498446277588)(25.5,9.3331511206338)(25.6,8.9280640176239)(25.7,8.43377065017549)(25.8,7.85520983422448)(25.9,7.19816235819701)(26,6.4691932232826)(26.1,5.67558604811205)(26.2,4.8252702932481)(26.3,3.92674203263639)(26.4,2.98897906364293)(26.5,2.02135120387065)(26.6,1.03352667103907)(26.7,0.0353754813490353)(26.8,-0.963129168457056)(26.9,-1.95201054998004)(27,-2.9213880873366)(27.1,-3.86157608060704)(27.2,-4.7631804821473)(27.3,-5.61719275880807)(27.4,-6.41507990221985)(27.5,-7.14886968779221)(27.6,-7.81123033054625)(27.7,-8.39554374188693)(27.8,-8.89597165535643)(27.9,-9.3075139606629)(28,-9.6260586631295)(28.1,-9.84842296938604)(28.2,-9.97238508878828)(28.3,-9.99670643281558)(28.4,-9.921143990638)(28.5,-9.74645275720014)(28.6,-9.47437818956129)(28.7,-9.10763876686538)(28.8,-8.64989882819606)(28.9,-8.10573195971176)(29,-7.48057529688482)(29.1,-6.78067519844083)(29.2,-6.0130248348073)(29.3,-5.18529431466613)(29.4,-4.30575404776313)(29.5,-3.38319210970798)(29.6,-2.42682643442723)(29.7,-1.4462127116159)(29.8,-0.451148909444447)(29.9,0.548422623500138)(30,1.54251449887511)(30.1,2.5211940792649)(30.2,3.47468272181056)(30.3,4.39345348317726)(30.4,5.2683263096227)(30.5,6.0905597610595)(30.6,6.85193835263525)(30.7,7.54485464114118)(30.8,8.16238523607009)(30.9,8.69835997584606)(31,9.1474235780389)(31.1,9.50508914757621)(31.2,9.76778300831567)(31.3,9.93288041003581)(31.4,9.9987317540726)(31.5,9.96467907556397)(31.6,9.83106261761728)(31.7,9.5992174317126)(31.8,9.27146003830968)(31.9,8.85106528094118)(32,8.34223360505871)(32.1,7.75004908857027)(32.2,7.08042864341444)(32.3,6.3400628957322)(32.4,5.53634933534201)(32.5,4.67731840246682)(32.6,3.77155325023014)(32.7,2.82810398462684)(32.8,1.85639723885605)(32.9,0.866141985518046)(33,-0.132767472230898)(33.1,-1.13035036128435)(33.2,-2.11663916317215)(33.3,-3.08177920620922)(33.4,-4.01612713011851)(33.5,-4.9103472392989)(33.6,-5.75550478200927)(33.7,-6.54315522345474)(33.8,-7.26542862078699)(33.9,-7.9151082569717)(34,-8.48570274783963)(34.1,-8.97151090185159)(34.2,-9.36767868451939)(34.3,-9.67024771831441)(34.4,-9.87619483346696)(34.5,-9.9834622744795)(34.6,-9.99097826053928)(34.7,-9.89866769439806)(34.8,-9.70745291271894)(34.9,-9.41924447039322)(35,-9.0369220509076)(35.1,-8.56430569349856)(35.2,-8.00611762458313)(35.3,-7.36793507483358)(35.4,-6.6561345533317)(35.5,-5.87782813559856)(35.6,-5.04079240208622)(35.7,-4.15339073715659)(35.8,-3.22448976490986)(35.9,-2.26337075680669)(36,-1.27963689627253)(36.1,-0.283117326862405)(36.2,0.716231057291603)(36.3,1.70842309747568)(36.4,2.68354513879927)(36.5,3.63185408415812)(36.6,4.54387474403944)(36.7,5.41049450948362)(36.8,6.22305440226066)(36.9,6.97343559251857)(37,7.65414051944741)(37.1,8.2583678044284)(37.2,8.78008020816102)(37.3,9.21406495276224)(37.4,9.55598580612044)(37.5,9.80242640809278)(37.6,9.95092440564683)(37.7,9.99999605587796)(37.8,9.94915105107795)(37.9,9.79889741772768)(38,9.5507364404643)(38.1,9.20714766174157)(38.2,8.77156410706099)(38.3,8.24833798331628)(38.4,7.64269719298041)(38.5,6.96069309863206)(38.6,6.2091400597415)(38.7,5.39554734584273)(38.8,4.52804410639494)(38.9,3.61529814700855)(39,2.66642932359605)(39.1,1.6909184197866)(39.2,0.698512418069775)(39.3,-0.300872888830909)(39.4,-1.29725197328163)(39.5,-2.28066934482985)(39.6,-3.2412990221737)(39.7,-4.16954271111884)(39.8,-5.05612570756197)(39.9,-5.89218956726824)(40,-6.66938061651728)(40.1,-7.37993341925011)(40.2,-8.01674836673946)(40.3,-8.5734626145331)(40.4,-9.04451365789224)(40.5,-9.42519491050038)(40.6,-9.71170273111803)(40.7,-9.90117442830848)(40.8,-9.9917168635044)(40.9,-9.98242536662266)(41,-9.87339277522874)(41.1,-9.66570850693443)(41.2,-9.36144767429733)(41.3,-8.96365035098186)(41.4,-8.47629119634816)(41.5,-7.90423974196997)(41.6,-7.25321173688436)(41.7,-6.52971203771677)(41.8,-5.7409696143042)(41.9,-4.89486532021775)(42,-3.99985314987908)(42.1,-3.0648757690425)(42.2,-2.09927516263499)(42.3,-1.11269929273043)(42.4,-0.115005699302341)(42.5,0.883836993057193)(42.6,1.87384867833988)(42.7,2.84513748704067)(42.8,3.78799862244842)(42.9,4.6930113277664)(43,5.55113301520072)(43.1,6.35378961650773)(43.2,7.09296125225013)(43.3,7.76126236378147)(43.4,8.35201550730604)(43.5,8.85931807269088)(43.6,9.27810126039459)(43.7,9.60418072723701)(43.8,9.83429839497288)(43.9,9.96615500393302)(44,9.99843308646674)(44.1,9.93081013064303)(44.2,9.7639618026833)(44.3,9.4995551959278)(44.4,9.14023217379006)(44.5,8.68958297313093)(44.6,8.15211033179845)(44.7,7.53318449876011)(44.8,6.83898957634974)(44.9,6.07646173076196)(45,5.25321988817156)(45.1,4.37748960894214)(45.2,3.45802090054598)(45.3,2.50400079038121)(45.4,1.52496153203159)(45.5,0.53068536213921)(45.6,-0.468893240470339)(45.7,-1.46378681681672)(45.8,-2.44405471915307)(45.9,-3.39990243463277)(46,-4.32177944884381)(46.1,-5.2004746713923)(46.2,-6.02720847007268)(46.3,-6.79372039405031)(46.4,-7.492351709556)(46.5,-8.11612192342188)(46.6,-8.65879852986226)(46.7,-9.11495928361041)(46.8,-9.48004637720171)(46.9,-9.75041198107873)(47,-9.92335469149854)(47.1,-9.99714652206582)(47.2,-9.97105016920191)(47.3,-9.8453263790383)(47.4,-9.62123134212757)(47.5,-9.30100414200251)(47.6,-8.88784438299373)(47.7,-8.38588022084081)(47.8,-7.80012711552481)(47.9,-7.13643771844962)(48,-6.40144339468448)(48.1,-5.60248796455541)(48.2,-4.74755432662085)(48.3,-3.84518469518751)(48.4,-2.90439524932881)(48.5,-1.93458604620439)(48.6,-0.945447098795235)(48.7,0.0531384435016174)(48.8,1.05119304403632)(48.9,2.03874447115322)(49,3.00592543743359)(49.1,3.94307219037017)(49.2,4.8408210693869)(49.3,5.69020206444095)(49.4,6.48272844139772)(49.5,7.21048153867318)(49.6,7.86618988788254)(49.7,8.44330186794888)(49.8,8.93605116673519)(49.9,9.33951439612931)(50,9.64966028491021)(50.1,9.86338995787795)(50.2,9.97856789878979)(50.3,9.9940432877329)(50.4,9.90966149973647)(50.5,9.72626564973325)(50.6,9.44568816843388)(50.7,9.0707324932849)(50.8,8.60514505744855)(50.9,8.05357785668097)(51,7.42154196812836)(51.1,6.71535248546576)(51.2,5.94206542056882)(51.3,5.10940720217515)(51.4,4.22569747596287)(51.5,3.29976597740079)(51.6,2.34086430795005)(51.7,1.35857349612093)(51.8,0.362708267003488)(51.9,-0.636781023222398)(52,-1.62990780795601)(52.1,-2.60674909264856)(52.2,-3.55754460208414)(52.3,-4.47279430182338)(52.4,-5.34335331940813)(52.5,-6.16052331690307)(52.6,-6.91613940181385)(52.7,-7.60265170799337)(52.8,-8.21320083140919)(52.9,-8.74168636704082)(53,-9.18282786211027)(53.1,-9.53221757662085)(53.2,-9.78636452403805)(53.3,-9.94272935207307)(53.4,-9.9997497150515)(53.5,-9.95685588435493)(53.6,-9.81447644096083)(53.7,-9.57403399320323)(53.8,-9.2379309625404)(53.9,-8.80952557935384)(54,-8.29309832862041)(54.1,-7.69380918072121)(54.2,-7.01764603472341)(54.3,-6.27136488927204)(54.4,-5.46242233888475)(54.5,-4.59890107012422)(54.6,-3.68942910206537)(54.7,-2.74309357798184)(54.8,-1.76934996961331)(54.9,-0.777927601217636)(55,0.221267562620619)(55.1,1.21825189411437)(55.2,2.20306385538277)(55.3,3.16586353084433)(55.4,4.09703094440101)(55.5,4.98726217905932)(55.6,5.82766233859441)(55.7,6.60983442241138)(55.8,7.32596322560035)(55.9,7.96889342588064)(56,8.53220107721541)(56.1,9.01025779575733)(56.2,9.39828699679926)(56.3,9.69241162082982)(56.4,9.88969287183038)(56.5,9.98815958075341)(56.6,9.98682790079218)(56.7,9.88571113765264)(56.8,9.68581961660727)(56.9,9.3891505876591)(57,8.9986682696796)(57.1,8.51827423291371)(57.2,7.95276841577966)(57.3,7.30780116547078)(57.4,6.5898167815545)(57.5,5.80598912666081)(57.6,4.96414994761838)(57.7,4.0727106232291)(57.8,3.14057812055164)(57.9,2.17706599943083)(58,1.19180135448593)(58.1,0.194628624364286)(58.2,-0.804488770636511)(58.3,-1.79556797976957)(58.4,-2.76870646718474)(58.5,-3.71418095479238)(58.6,-4.62254457404462)(58.7,-5.48472125592329)(58.8,-6.29209641602056)(58.9,-7.03660302861941)(59,-7.71080222974809)(59.1,-8.30795764384937)(59.2,-8.82210269141906)(59.3,-9.24810020509561)(59.4,-9.58169375853824)(59.5,-9.81955019523241)(59.6,-9.95929293228756)(59.7,-9.99952570646783)(59.8,-9.93984652519281)(59.9,-9.78085168311447)(60,-9.52412980413819) 
};
\addplot [
color=red,
solid,
line width=1.0pt,
forget plot
]
coordinates{
 (0,0)(0.1,0)(0.2,0)(0.3,0)(0.4,0)(0.5,0)(0.6,0)(0.7,0)(0.8,0)(0.9,0)(1,0)(1.1,0)(1.2,0)(1.3,0)(1.4,0)(1.5,0)(1.6,0)(1.7,0)(1.8,0)(1.9,0)(2,0)(2.1,0)(2.2,0)(2.3,0)(2.4,0)(2.5,0)(2.6,0)(2.7,0)(2.8,0)(2.9,0)(3,0)(3.1,0)(3.2,0)(3.3,0)(3.4,0)(3.5,0)(3.6,0)(3.7,0)(3.8,0)(3.9,0)(4,0)(4.1,0)(4.2,0)(4.3,0)(4.4,0)(4.5,0)(4.6,0)(4.7,0)(4.8,0)(4.9,0)(5,0)(5.1,0)(5.2,0)(5.3,0)(5.4,0)(5.5,0)(5.6,0)(5.7,0)(5.8,0)(5.9,0)(6,0)(6.1,0)(6.2,0)(6.3,0)(6.4,0)(6.5,0)(6.6,0)(6.7,0)(6.8,0)(6.9,0)(7,0)(7.1,0)(7.2,0)(7.3,0)(7.4,0)(7.5,0)(7.6,0)(7.7,0)(7.8,0)(7.9,0)(8,0)(8.1,0)(8.2,0)(8.3,0)(8.4,0)(8.5,0)(8.6,0)(8.7,0)(8.8,0)(8.9,0)(9,0)(9.1,0)(9.2,0)(9.3,0)(9.4,0)(9.5,0)(9.6,0)(9.7,0)(9.8,0)(9.9,0)(10,0)(10.1,0)(10.2,0)(10.3,0)(10.4,0)(10.5,0)(10.6,0)(10.7,0)(10.8,0)(10.9,0)(11,0)(11.1,0)(11.2,0)(11.3,0)(11.4,0)(11.5,0)(11.6,0)(11.7,0)(11.8,0)(11.9,0)(12,0)(12.1,0)(12.2,0)(12.3,0)(12.4,0)(12.5,0)(12.6,0)(12.7,0)(12.8,0)(12.9,0)(13,0)(13.1,0)(13.2,0)(13.3,0)(13.4,0)(13.5,0)(13.6,0)(13.7,0)(13.8,0)(13.9,0)(14,0)(14.1,0)(14.2,0)(14.3,0)(14.4,0)(14.5,0)(14.6,0)(14.7,0)(14.8,0)(14.9,0)(15,0)(15.1,0)(15.2,0)(15.3,0)(15.4,0)(15.5,0)(15.6,0)(15.7,0)(15.8,0)(15.9,0)(16,0)(16.1,0)(16.2,0)(16.3,0)(16.4,0)(16.5,0)(16.6,0)(16.7,0)(16.8,0)(16.9,0)(17,0)(17.1,0)(17.2,0)(17.3,0)(17.4,0)(17.5,0)(17.6,0)(17.7,0)(17.8,0)(17.9,0)(18,0)(18.1,0)(18.2,0)(18.3,0)(18.4,0)(18.5,0)(18.6,0)(18.7,0)(18.8,0)(18.9,0)(19,0)(19.1,0)(19.2,0)(19.3,0)(19.4,0)(19.5,0)(19.6,0)(19.7,0)(19.8,0)(19.9,0)(20,0)(20.1,0)(20.2,0)(20.3,0)(20.4,0)(20.5,0)(20.6,0)(20.7,0)(20.8,0)(20.9,0)(21,0)(21.1,0)(21.2,0)(21.3,0)(21.4,0)(21.5,0)(21.6,0)(21.7,0)(21.8,0)(21.9,0)(22,0)(22.1,0)(22.2,0)(22.3,0)(22.4,0)(22.5,0)(22.6,0)(22.7,0)(22.8,0)(22.9,0)(23,0)(23.1,0)(23.2,0)(23.3,0)(23.4,0)(23.5,0)(23.6,0)(23.7,0)(23.8,0)(23.9,0)(24,0)(24.1,0)(24.2,0)(24.3,0)(24.4,0)(24.5,0)(24.6,0)(24.7,0)(24.8,0)(24.9,0)(25,0)(25.1,0)(25.2,0)(25.3,0)(25.4,0)(25.5,0)(25.6,0)(25.7,0)(25.8,0)(25.9,0)(26,0)(26.1,0)(26.2,0)(26.3,0)(26.4,0)(26.5,0)(26.6,0)(26.7,0)(26.8,0)(26.9,0)(27,0)(27.1,0)(27.2,0)(27.3,0)(27.4,0)(27.5,0)(27.6,0)(27.7,0)(27.8,0)(27.9,0)(28,0)(28.1,0)(28.2,0)(28.3,0)(28.4,0)(28.5,0)(28.6,0)(28.7,0)(28.8,0)(28.9,0)(29,0)(29.1,0)(29.2,0)(29.3,0)(29.4,0)(29.5,0)(29.6,0)(29.7,0)(29.8,0)(29.9,0)(30,0)(30.1,0)(30.2,0)(30.3,0)(30.4,0)(30.5,0)(30.6,0)(30.7,0)(30.8,0)(30.9,0)(31,0)(31.1,0)(31.2,0)(31.3,0)(31.4,0)(31.5,0)(31.6,0)(31.7,0)(31.8,0)(31.9,0)(32,0)(32.1,0)(32.2,0)(32.3,0)(32.4,0)(32.5,0)(32.6,0)(32.7,0)(32.8,0)(32.9,0)(33,0)(33.1,0)(33.2,0)(33.3,0)(33.4,0)(33.5,0)(33.6,0)(33.7,0)(33.8,0)(33.9,0)(34,0)(34.1,0)(34.2,0)(34.3,0)(34.4,0)(34.5,0)(34.6,0)(34.7,0)(34.8,0)(34.9,0)(35,0)(35.1,0)(35.2,0)(35.3,0)(35.4,0)(35.5,0)(35.6,0)(35.7,0)(35.8,0)(35.9,0)(36,0)(36.1,0)(36.2,0)(36.3,0)(36.4,0)(36.5,0)(36.6,0)(36.7,0)(36.8,0)(36.9,0)(37,0)(37.1,0)(37.2,0)(37.3,0)(37.4,0)(37.5,0)(37.6,0)(37.7,0)(37.8,0)(37.9,0)(38,0)(38.1,0)(38.2,0)(38.3,0)(38.4,0)(38.5,0)(38.6,0)(38.7,0)(38.8,0)(38.9,0)(39,0)(39.1,0)(39.2,0)(39.3,0)(39.4,0)(39.5,0)(39.6,0)(39.7,0)(39.8,0)(39.9,0)(40,0)(40.1,0)(40.2,0)(40.3,0)(40.4,0)(40.5,0)(40.6,0)(40.7,0)(40.8,0)(40.9,0)(41,0)(41.1,0)(41.2,0)(41.3,0)(41.4,0)(41.5,0)(41.6,0)(41.7,0)(41.8,0)(41.9,0)(42,0)(42.1,0)(42.2,0)(42.3,0)(42.4,0)(42.5,0)(42.6,0)(42.7,0)(42.8,0)(42.9,0)(43,0)(43.1,0)(43.2,0)(43.3,0)(43.4,0)(43.5,0)(43.6,0)(43.7,0)(43.8,0)(43.9,0)(44,0)(44.1,0)(44.2,0)(44.3,0)(44.4,0)(44.5,0)(44.6,0)(44.7,0)(44.8,0)(44.9,0)(45,50)(45.1,50)(45.2,50)(45.3,50)(45.4,50)(45.5,50)(45.6,50)(45.7,50)(45.8,50)(45.9,50)(46,50)(46.1,50)(46.2,50)(46.3,50)(46.4,50)(46.5,50)(46.6,50)(46.7,50)(46.8,50)(46.9,50)(47,50)(47.1,50)(47.2,50)(47.3,50)(47.4,50)(47.5,50)(47.6,50)(47.7,50)(47.8,50)(47.9,50)(48,50)(48.1,50)(48.2,50)(48.3,50)(48.4,50)(48.5,50)(48.6,50)(48.7,50)(48.8,50)(48.9,50)(49,50)(49.1,50)(49.2,50)(49.3,50)(49.4,50)(49.5,50)(49.6,50)(49.7,50)(49.8,50)(49.9,50)(50,50)(50.1,50)(50.2,50)(50.3,50)(50.4,50)(50.5,50)(50.6,50)(50.7,50)(50.8,50)(50.9,50)(51,50)(51.1,50)(51.2,50)(51.3,50)(51.4,50)(51.5,50)(51.6,50)(51.7,50)(51.8,50)(51.9,50)(52,50)(52.1,50)(52.2,50)(52.3,50)(52.4,50)(52.5,50)(52.6,50)(52.7,50)(52.8,50)(52.9,50)(53,50)(53.1,50)(53.2,50)(53.3,50)(53.4,50)(53.5,50)(53.6,50)(53.7,50)(53.8,50)(53.9,50)(54,50)(54.1,50)(54.2,50)(54.3,50)(54.4,50)(54.5,50)(54.6,50)(54.7,50)(54.8,50)(54.9,50)(55,50)(55.1,50)(55.2,50)(55.3,50)(55.4,50)(55.5,50)(55.6,50)(55.7,50)(55.8,50)(55.9,50)(56,50)(56.1,50)(56.2,50)(56.3,50)(56.4,50)(56.5,50)(56.6,50)(56.7,50)(56.8,50)(56.9,50)(57,50)(57.1,50)(57.2,50)(57.3,50)(57.4,50)(57.5,50)(57.6,50)(57.7,50)(57.8,50)(57.9,50)(58,50)(58.1,50)(58.2,50)(58.3,50)(58.4,50)(58.5,50)(58.6,50)(58.7,50)(58.8,50)(58.9,50)(59,50)(59.1,50)(59.2,50)(59.3,50)(59.4,50)(59.5,50)(59.6,50)(59.7,50)(59.8,50)(59.9,50)(60,50) 
};
\addplot [
color=mycolor1,
solid,
line width=1.0pt,
forget plot
]
coordinates{
 (0,0)(0.1,0)(0.2,0)(0.3,0)(0.4,0)(0.5,0)(0.6,0)(0.7,0)(0.8,0)(0.9,0)(1,0)(1.1,0)(1.2,0)(1.3,0)(1.4,0)(1.5,0)(1.6,0)(1.7,0)(1.8,0)(1.9,0)(2,0)(2.1,0)(2.2,0)(2.3,0)(2.4,0)(2.5,0)(2.6,0)(2.7,0)(2.8,0)(2.9,0)(3,0)(3.1,0)(3.2,0)(3.3,0)(3.4,0)(3.5,0)(3.6,0)(3.7,0)(3.8,0)(3.9,0)(4,0)(4.1,0)(4.2,0)(4.3,0)(4.4,0)(4.5,0)(4.6,0)(4.7,0)(4.8,0)(4.9,0)(5,0)(5.1,0)(5.2,0)(5.3,0)(5.4,0)(5.5,0)(5.6,0)(5.7,0)(5.8,0)(5.9,0)(6,0)(6.1,0)(6.2,0)(6.3,0)(6.4,0)(6.5,0)(6.6,0)(6.7,0)(6.8,0)(6.9,0)(7,0)(7.1,0)(7.2,0)(7.3,0)(7.4,0)(7.5,0)(7.6,0)(7.7,0)(7.8,0)(7.9,0)(8,0)(8.1,0)(8.2,0)(8.3,0)(8.4,0)(8.5,0)(8.6,0)(8.7,0)(8.8,0)(8.9,0)(9,0)(9.1,0)(9.2,0)(9.3,0)(9.4,0)(9.5,0)(9.6,0)(9.7,0)(9.8,0)(9.9,0)(10,0)(10.1,0)(10.2,0)(10.3,0)(10.4,0)(10.5,0)(10.6,0)(10.7,0)(10.8,0)(10.9,0)(11,0)(11.1,0)(11.2,0)(11.3,0)(11.4,0)(11.5,0)(11.6,0)(11.7,0)(11.8,0)(11.9,0)(12,0)(12.1,0)(12.2,0)(12.3,0)(12.4,0)(12.5,0)(12.6,0)(12.7,0)(12.8,0)(12.9,0)(13,0)(13.1,0)(13.2,0)(13.3,0)(13.4,0)(13.5,0)(13.6,0)(13.7,0)(13.8,0)(13.9,0)(14,0)(14.1,0)(14.2,0)(14.3,0)(14.4,0)(14.5,0)(14.6,0)(14.7,0)(14.8,0)(14.9,0)(15,-50)(15.1,-50)(15.2,-50)(15.3,-50)(15.4,-50)(15.5,-50)(15.6,-50)(15.7,-50)(15.8,-50)(15.9,-50)(16,-50)(16.1,-50)(16.2,-50)(16.3,-50)(16.4,-50)(16.5,-50)(16.6,-50)(16.7,-50)(16.8,-50)(16.9,-50)(17,-50)(17.1,-50)(17.2,-50)(17.3,-50)(17.4,-50)(17.5,-50)(17.6,-50)(17.7,-50)(17.8,-50)(17.9,-50)(18,-50)(18.1,-50)(18.2,-50)(18.3,-50)(18.4,-50)(18.5,-50)(18.6,-50)(18.7,-50)(18.8,-50)(18.9,-50)(19,-50)(19.1,-50)(19.2,-50)(19.3,-50)(19.4,-50)(19.5,-50)(19.6,-50)(19.7,-50)(19.8,-50)(19.9,-50)(20,-50)(20.1,-50)(20.2,-50)(20.3,-50)(20.4,-50)(20.5,-50)(20.6,-50)(20.7,-50)(20.8,-50)(20.9,-50)(21,-50)(21.1,-50)(21.2,-50)(21.3,-50)(21.4,-50)(21.5,-50)(21.6,-50)(21.7,-50)(21.8,-50)(21.9,-50)(22,-50)(22.1,-50)(22.2,-50)(22.3,-50)(22.4,-50)(22.5,-50)(22.6,-50)(22.7,-50)(22.8,-50)(22.9,-50)(23,-50)(23.1,-50)(23.2,-50)(23.3,-50)(23.4,-50)(23.5,-50)(23.6,-50)(23.7,-50)(23.8,-50)(23.9,-50)(24,-50)(24.1,-50)(24.2,-50)(24.3,-50)(24.4,-50)(24.5,-50)(24.6,-50)(24.7,-50)(24.8,-50)(24.9,-50)(25,-50)(25.1,-50)(25.2,-50)(25.3,-50)(25.4,-50)(25.5,-50)(25.6,-50)(25.7,-50)(25.8,-50)(25.9,-50)(26,-50)(26.1,-50)(26.2,-50)(26.3,-50)(26.4,-50)(26.5,-50)(26.6,-50)(26.7,-50)(26.8,-50)(26.9,-50)(27,-50)(27.1,-50)(27.2,-50)(27.3,-50)(27.4,-50)(27.5,-50)(27.6,-50)(27.7,-50)(27.8,-50)(27.9,-50)(28,-50)(28.1,-50)(28.2,-50)(28.3,-50)(28.4,-50)(28.5,-50)(28.6,-50)(28.7,-50)(28.8,-50)(28.9,-50)(29,-50)(29.1,-50)(29.2,-50)(29.3,-50)(29.4,-50)(29.5,-50)(29.6,-50)(29.7,-50)(29.8,-50)(29.9,-50)(30,-50)(30.1,-50)(30.2,-50)(30.3,-50)(30.4,-50)(30.5,-50)(30.6,-50)(30.7,-50)(30.8,-50)(30.9,-50)(31,-50)(31.1,-50)(31.2,-50)(31.3,-50)(31.4,-50)(31.5,-50)(31.6,-50)(31.7,-50)(31.8,-50)(31.9,-50)(32,-50)(32.1,-50)(32.2,-50)(32.3,-50)(32.4,-50)(32.5,-50)(32.6,-50)(32.7,-50)(32.8,-50)(32.9,-50)(33,-50)(33.1,-50)(33.2,-50)(33.3,-50)(33.4,-50)(33.5,-50)(33.6,-50)(33.7,-50)(33.8,-50)(33.9,-50)(34,-50)(34.1,-50)(34.2,-50)(34.3,-50)(34.4,-50)(34.5,-50)(34.6,-50)(34.7,-50)(34.8,-50)(34.9,-50)(35,-50)(35.1,-50)(35.2,-50)(35.3,-50)(35.4,-50)(35.5,-50)(35.6,-50)(35.7,-50)(35.8,-50)(35.9,-50)(36,-50)(36.1,-50)(36.2,-50)(36.3,-50)(36.4,-50)(36.5,-50)(36.6,-50)(36.7,-50)(36.8,-50)(36.9,-50)(37,-50)(37.1,-50)(37.2,-50)(37.3,-50)(37.4,-50)(37.5,-50)(37.6,-50)(37.7,-50)(37.8,-50)(37.9,-50)(38,-50)(38.1,-50)(38.2,-50)(38.3,-50)(38.4,-50)(38.5,-50)(38.6,-50)(38.7,-50)(38.8,-50)(38.9,-50)(39,-50)(39.1,-50)(39.2,-50)(39.3,-50)(39.4,-50)(39.5,-50)(39.6,-50)(39.7,-50)(39.8,-50)(39.9,-50)(40,-50)(40.1,-50)(40.2,-50)(40.3,-50)(40.4,-50)(40.5,-50)(40.6,-50)(40.7,-50)(40.8,-50)(40.9,-50)(41,-50)(41.1,-50)(41.2,-50)(41.3,-50)(41.4,-50)(41.5,-50)(41.6,-50)(41.7,-50)(41.8,-50)(41.9,-50)(42,-50)(42.1,-50)(42.2,-50)(42.3,-50)(42.4,-50)(42.5,-50)(42.6,-50)(42.7,-50)(42.8,-50)(42.9,-50)(43,-50)(43.1,-50)(43.2,-50)(43.3,-50)(43.4,-50)(43.5,-50)(43.6,-50)(43.7,-50)(43.8,-50)(43.9,-50)(44,-50)(44.1,-50)(44.2,-50)(44.3,-50)(44.4,-50)(44.5,-50)(44.6,-50)(44.7,-50)(44.8,-50)(44.9,-50)(45,-50)(45.1,-50)(45.2,-50)(45.3,-50)(45.4,-50)(45.5,-50)(45.6,-50)(45.7,-50)(45.8,-50)(45.9,-50)(46,-50)(46.1,-50)(46.2,-50)(46.3,-50)(46.4,-50)(46.5,-50)(46.6,-50)(46.7,-50)(46.8,-50)(46.9,-50)(47,-50)(47.1,-50)(47.2,-50)(47.3,-50)(47.4,-50)(47.5,-50)(47.6,-50)(47.7,-50)(47.8,-50)(47.9,-50)(48,-50)(48.1,-50)(48.2,-50)(48.3,-50)(48.4,-50)(48.5,-50)(48.6,-50)(48.7,-50)(48.8,-50)(48.9,-50)(49,-50)(49.1,-50)(49.2,-50)(49.3,-50)(49.4,-50)(49.5,-50)(49.6,-50)(49.7,-50)(49.8,-50)(49.9,-50)(50,-50)(50.1,-50)(50.2,-50)(50.3,-50)(50.4,-50)(50.5,-50)(50.6,-50)(50.7,-50)(50.8,-50)(50.9,-50)(51,-50)(51.1,-50)(51.2,-50)(51.3,-50)(51.4,-50)(51.5,-50)(51.6,-50)(51.7,-50)(51.8,-50)(51.9,-50)(52,-50)(52.1,-50)(52.2,-50)(52.3,-50)(52.4,-50)(52.5,-50)(52.6,-50)(52.7,-50)(52.8,-50)(52.9,-50)(53,-50)(53.1,-50)(53.2,-50)(53.3,-50)(53.4,-50)(53.5,-50)(53.6,-50)(53.7,-50)(53.8,-50)(53.9,-50)(54,-50)(54.1,-50)(54.2,-50)(54.3,-50)(54.4,-50)(54.5,-50)(54.6,-50)(54.7,-50)(54.8,-50)(54.9,-50)(55,-50)(55.1,-50)(55.2,-50)(55.3,-50)(55.4,-50)(55.5,-50)(55.6,-50)(55.7,-50)(55.8,-50)(55.9,-50)(56,-50)(56.1,-50)(56.2,-50)(56.3,-50)(56.4,-50)(56.5,-50)(56.6,-50)(56.7,-50)(56.8,-50)(56.9,-50)(57,-50)(57.1,-50)(57.2,-50)(57.3,-50)(57.4,-50)(57.5,-50)(57.6,-50)(57.7,-50)(57.8,-50)(57.9,-50)(58,-50)(58.1,-50)(58.2,-50)(58.3,-50)(58.4,-50)(58.5,-50)(58.6,-50)(58.7,-50)(58.8,-50)(58.9,-50)(59,-50)(59.1,-50)(59.2,-50)(59.3,-50)(59.4,-50)(59.5,-50)(59.6,-50)(59.7,-50)(59.8,-50)(59.9,-50)(60,-50) 
};
\addplot [
color=mycolor2,
solid,
line width=1.0pt,
forget plot
]
coordinates{
 (0,0)(0.1,0)(0.2,0)(0.3,0)(0.4,0)(0.5,0)(0.6,0)(0.7,0)(0.8,0)(0.9,0)(1,0)(1.1,0)(1.2,0)(1.3,0)(1.4,0)(1.5,0)(1.6,0)(1.7,0)(1.8,0)(1.9,0)(2,0)(2.1,0)(2.2,0)(2.3,0)(2.4,0)(2.5,0)(2.6,0)(2.7,0)(2.8,0)(2.9,0)(3,0)(3.1,0)(3.2,0)(3.3,0)(3.4,0)(3.5,0)(3.6,0)(3.7,0)(3.8,0)(3.9,0)(4,0)(4.1,0)(4.2,0)(4.3,0)(4.4,0)(4.5,0)(4.6,0)(4.7,0)(4.8,0)(4.9,0)(5,0)(5.1,0)(5.2,0)(5.3,0)(5.4,0)(5.5,0)(5.6,0)(5.7,0)(5.8,0)(5.9,0)(6,0)(6.1,0)(6.2,0)(6.3,0)(6.4,0)(6.5,0)(6.6,0)(6.7,0)(6.8,0)(6.9,0)(7,0)(7.1,0)(7.2,0)(7.3,0)(7.4,0)(7.5,0)(7.6,0)(7.7,0)(7.8,0)(7.9,0)(8,0)(8.1,0)(8.2,0)(8.3,0)(8.4,0)(8.5,0)(8.6,0)(8.7,0)(8.8,0)(8.9,0)(9,0)(9.1,0)(9.2,0)(9.3,0)(9.4,0)(9.5,0)(9.6,0)(9.7,0)(9.8,0)(9.9,0)(10,0)(10.1,0)(10.2,0)(10.3,0)(10.4,0)(10.5,0)(10.6,0)(10.7,0)(10.8,0)(10.9,0)(11,0)(11.1,0)(11.2,0)(11.3,0)(11.4,0)(11.5,0)(11.6,0)(11.7,0)(11.8,0)(11.9,0)(12,0)(12.1,0)(12.2,0)(12.3,0)(12.4,0)(12.5,0)(12.6,0)(12.7,0)(12.8,0)(12.9,0)(13,0)(13.1,0)(13.2,0)(13.3,0)(13.4,0)(13.5,0)(13.6,0)(13.7,0)(13.8,0)(13.9,0)(14,0)(14.1,0)(14.2,0)(14.3,0)(14.4,0)(14.5,0)(14.6,0)(14.7,0)(14.8,0)(14.9,0)(15,0)(15.1,0)(15.2,0)(15.3,0)(15.4,0)(15.5,0)(15.6,0)(15.7,0)(15.8,0)(15.9,0)(16,0)(16.1,0)(16.2,0)(16.3,0)(16.4,0)(16.5,0)(16.6,0)(16.7,0)(16.8,0)(16.9,0)(17,0)(17.1,0)(17.2,0)(17.3,0)(17.4,0)(17.5,0)(17.6,0)(17.7,0)(17.8,0)(17.9,0)(18,0)(18.1,0)(18.2,0)(18.3,0)(18.4,0)(18.5,0)(18.6,0)(18.7,0)(18.8,0)(18.9,0)(19,0)(19.1,0)(19.2,0)(19.3,0)(19.4,0)(19.5,0)(19.6,0)(19.7,0)(19.8,0)(19.9,0)(20,0)(20.1,0)(20.2,0)(20.3,0)(20.4,0)(20.5,0)(20.6,0)(20.7,0)(20.8,0)(20.9,0)(21,0)(21.1,0)(21.2,0)(21.3,0)(21.4,0)(21.5,0)(21.6,0)(21.7,0)(21.8,0)(21.9,0)(22,0)(22.1,0)(22.2,0)(22.3,0)(22.4,0)(22.5,0)(22.6,0)(22.7,0)(22.8,0)(22.9,0)(23,0)(23.1,0)(23.2,0)(23.3,0)(23.4,0)(23.5,0)(23.6,0)(23.7,0)(23.8,0)(23.9,0)(24,0)(24.1,0)(24.2,0)(24.3,0)(24.4,0)(24.5,0)(24.6,0)(24.7,0)(24.8,0)(24.9,0)(25,0)(25.1,0)(25.2,0)(25.3,0)(25.4,0)(25.5,0)(25.6,0)(25.7,0)(25.8,0)(25.9,0)(26,0)(26.1,0)(26.2,0)(26.3,0)(26.4,0)(26.5,0)(26.6,0)(26.7,0)(26.8,0)(26.9,0)(27,0)(27.1,0)(27.2,0)(27.3,0)(27.4,0)(27.5,0)(27.6,0)(27.7,0)(27.8,0)(27.9,0)(28,0)(28.1,0)(28.2,0)(28.3,0)(28.4,0)(28.5,0)(28.6,0)(28.7,0)(28.8,0)(28.9,0)(29,0)(29.1,0)(29.2,0)(29.3,0)(29.4,0)(29.5,0)(29.6,0)(29.7,0)(29.8,0)(29.9,0)(30,0)(30.1,0)(30.2,0)(30.3,0)(30.4,0)(30.5,0)(30.6,0)(30.7,0)(30.8,0)(30.9,0)(31,0)(31.1,0)(31.2,0)(31.3,0)(31.4,0)(31.5,0)(31.6,0)(31.7,0)(31.8,0)(31.9,0)(32,0)(32.1,0)(32.2,0)(32.3,0)(32.4,0)(32.5,0)(32.6,0)(32.7,0)(32.8,0)(32.9,0)(33,0)(33.1,0)(33.2,0)(33.3,0)(33.4,0)(33.5,0)(33.6,0)(33.7,0)(33.8,0)(33.9,0)(34,0)(34.1,0)(34.2,0)(34.3,0)(34.4,0)(34.5,0)(34.6,0)(34.7,0)(34.8,0)(34.9,0)(35,0)(35.1,0)(35.2,0)(35.3,0)(35.4,0)(35.5,0)(35.6,0)(35.7,0)(35.8,0)(35.9,0)(36,0)(36.1,0)(36.2,0)(36.3,0)(36.4,0)(36.5,0)(36.6,0)(36.7,0)(36.8,0)(36.9,0)(37,0)(37.1,0)(37.2,0)(37.3,0)(37.4,0)(37.5,0)(37.6,0)(37.7,0)(37.8,0)(37.9,0)(38,0)(38.1,0)(38.2,0)(38.3,0)(38.4,0)(38.5,0)(38.6,0)(38.7,0)(38.8,0)(38.9,0)(39,0)(39.1,0)(39.2,0)(39.3,0)(39.4,0)(39.5,0)(39.6,0)(39.7,0)(39.8,0)(39.9,0)(40,0)(40.1,0)(40.2,0)(40.3,0)(40.4,0)(40.5,0)(40.6,0)(40.7,0)(40.8,0)(40.9,0)(41,0)(41.1,0)(41.2,0)(41.3,0)(41.4,0)(41.5,0)(41.6,0)(41.7,0)(41.8,0)(41.9,0)(42,0)(42.1,0)(42.2,0)(42.3,0)(42.4,0)(42.5,0)(42.6,0)(42.7,0)(42.8,0)(42.9,0)(43,0)(43.1,0)(43.2,0)(43.3,0)(43.4,0)(43.5,0)(43.6,0)(43.7,0)(43.8,0)(43.9,0)(44,0)(44.1,0)(44.2,0)(44.3,0)(44.4,0)(44.5,0)(44.6,0)(44.7,0)(44.8,0)(44.9,0)(45,0)(45.1,0)(45.2,0)(45.3,0)(45.4,0)(45.5,0)(45.6,0)(45.7,0)(45.8,0)(45.9,0)(46,0)(46.1,0)(46.2,0)(46.3,0)(46.4,0)(46.5,0)(46.6,0)(46.7,0)(46.8,0)(46.9,0)(47,0)(47.1,0)(47.2,0)(47.3,0)(47.4,0)(47.5,0)(47.6,0)(47.7,0)(47.8,0)(47.9,0)(48,0)(48.1,0)(48.2,0)(48.3,0)(48.4,0)(48.5,0)(48.6,0)(48.7,0)(48.8,0)(48.9,0)(49,0)(49.1,0)(49.2,0)(49.3,0)(49.4,0)(49.5,0)(49.6,0)(49.7,0)(49.8,0)(49.9,0)(50,0)(50.1,0)(50.2,0)(50.3,0)(50.4,0)(50.5,0)(50.6,0)(50.7,0)(50.8,0)(50.9,0)(51,0)(51.1,0)(51.2,0)(51.3,0)(51.4,0)(51.5,0)(51.6,0)(51.7,0)(51.8,0)(51.9,0)(52,0)(52.1,0)(52.2,0)(52.3,0)(52.4,0)(52.5,0)(52.6,0)(52.7,0)(52.8,0)(52.9,0)(53,0)(53.1,0)(53.2,0)(53.3,0)(53.4,0)(53.5,0)(53.6,0)(53.7,0)(53.8,0)(53.9,0)(54,0)(54.1,0)(54.2,0)(54.3,0)(54.4,0)(54.5,0)(54.6,0)(54.7,0)(54.8,0)(54.9,0)(55,0)(55.1,0)(55.2,0)(55.3,0)(55.4,0)(55.5,0)(55.6,0)(55.7,0)(55.8,0)(55.9,0)(56,0)(56.1,0)(56.2,0)(56.3,0)(56.4,0)(56.5,0)(56.6,0)(56.7,0)(56.8,0)(56.9,0)(57,0)(57.1,0)(57.2,0)(57.3,0)(57.4,0)(57.5,0)(57.6,0)(57.7,0)(57.8,0)(57.9,0)(58,0)(58.1,0)(58.2,0)(58.3,0)(58.4,0)(58.5,0)(58.6,0)(58.7,0)(58.8,0)(58.9,0)(59,0)(59.1,0)(59.2,0)(59.3,0)(59.4,0)(59.5,0)(59.6,0)(59.7,0)(59.8,0)(59.9,0)(60,0) 
};
\end{axis}
\end{tikzpicture}%

%% file: figures/sim_svd_Delta.tikz
%
%
%

\definecolor{mycolor1}{rgb}{0,0.75,0.75}

\begin{tikzpicture}

\begin{semilogxaxis}[%
width=\figurewidth,
height=\figureheight,
scale only axis,
every outer x axis line/.append style={gray!0!black},
every x tick label/.append style={font=\color{gray!0!black}},
xmin=0.001, xmax=1000,
xminorticks=true,
every outer y axis line/.append style={gray!0!black},
every y tick label/.append style={font=\color{gray!0!black}},
ymin=-100, ymax=0,
name=plot1,
unbounded coords=jump,
xlabel={Frequency [rad/s]},
ylabel={Singular Values [dB]}]
\addplot [
color=blue,
solid,
line width=1.0pt,
forget plot
]
coordinates{
 (6.33574562021173e-05,-72.0863153416949)(0.0107933934669818,-72.0827701568504)(0.0192412101845034,-72.0750455820217)(0.025690323372429,-70.5482182080827)(0.0343009981519521,-68.0151768642378)(0.0457977292525215,-65.4649116568224)(0.06114784168658,-62.8844257576693)(0.0816428806395725,-60.2515471407197)(0.109007280964921,-57.5296767204642)(0.145543459641299,-54.6616781438467)(0.1943255391461,-51.5661404667167)(0.259458001462178,-48.1424711519076)(0.346421035642335,-44.2898413219777)(0.46253163617697,-39.9332377616499)(0.617559248582775,-35.0435542112486)(0.824547762099878,-29.6802457907474)(1.10091301125222,-24.2227190452113)(1.46990812910317,-20.0907462478452)(1.9625800457622,-19.4571790550219)(2.62038175023494,-21.593952614035)(3.49866010906969,-24.3196679047943)(4.67131270384478,-26.9918933900082)(6.23700550977608,-29.660606277651)(8.32747456554565,-32.3928790334611)(11.1186101296709,-35.1692809414098)(14.8452559347469,-37.932512842356)(19.8209687360141,-40.6425938161422)(26.4644007055813,-43.2914905269545)(35.3345244641364,-45.8901731766148)(47.1776645538518,-48.4540208071687)(62.9902925399451,-50.9958412457244)(84.1028692664254,-53.5244421370132)(112.291788680928,-56.0453400850853)(149.928842085251,-58.5618236748972)(200.180778604356,-61.0758011038815)(267.27575271915,-63.5883631867271)(356.859077528004,-66.1001283186852)(476.468216508795,-68.611445449404)(4764.68216508795,-88.6121711643443) 
};
\addplot [
color=blue,
solid,
line width=1.0pt,
forget plot
]
coordinates{
 (6.33574562021173e-05,-122.736278913253)(0.000200353868353052,-112.736277342203)(0.000633574562021173,-102.736261631727)(0.00200353868353052,-92.7361045295061)(0.00633574562021173,-82.7345337604109)(0.00808389794443477,-80.6169762374886)(0.0107933934669818,-78.1040098995209)(0.014411035781728,-75.5893062560528)(0.0192412101845034,-73.0715105057539)(0.025690323372429,-72.0662187051685)(0.0343009981519521,-72.0504700673853)(0.0457977292525215,-72.0223546122336)(0.06114784168658,-71.9721103847143)(0.0816428806395725,-71.8821811782696)(0.109007280964921,-71.7208966821083)(0.145543459641299,-71.4312035331313)(0.1943255391461,-70.912256378184)(0.259458001462178,-69.9975920475726)(0.346421035642335,-68.4580436546346)(0.46253163617697,-66.0889999512313)(0.617559248582775,-62.8720624377962)(0.824547762099878,-59.040183944698)(1.10091301125222,-54.9836519677133)(1.46990812910317,-51.1443279204653)(1.9625800457622,-47.9431238858152)(2.62038175023494,-45.6980624749549)(3.49866010906969,-44.5535444920384)(4.67131270384478,-44.4661926927223)(6.23700550977608,-45.2555309149156)(8.32747456554565,-46.6886843080346)(11.1186101296709,-48.5521196665449)(14.8452559347469,-50.6845405825497)(19.8209687360141,-52.9779169794787)(26.4644007055813,-55.3650390371996)(35.3345244641364,-57.8058924860247)(47.1776645538518,-60.2772574430338)(62.9902925399451,-62.765856501254)(84.1028692664254,-65.2641607339316)(112.291788680928,-67.7679210193109)(149.928842085251,-70.2747456486688)(200.180778604356,-72.7832904129824)(267.27575271915,-75.2928004636161)(356.859077528004,-77.8028521109882)(476.468216508795,-80.3132076050781)(4764.68216508795,-100.312716498721) 
};
\addplot [
color=green!65!black,
solid,
line width=1.0pt,
forget plot
]
coordinates{
 (3.448150933975e-05,-93.027825747259)(0.000571445316922507,-93.0143281693466)(0.000762835828043288,-93.00376350516)(0.00101832753426064,-92.9850005038013)(0.00135938943729649,-91.4416647143613)(0.00181468101378105,-88.9324220005409)(0.0024224604748486,-86.4230896528911)(0.00323379960865767,-83.9135975900729)(0.00431687534947631,-81.4038209565423)(0.00576269869444741,-78.8935373550958)(0.00769276236966505,-76.3823507801932)(0.0102692498799487,-73.8695565276826)(0.013708663810113,-71.3539019233269)(0.0183000185656831,-68.8331645494672)(0.0244291262914551,-66.3034151183752)(0.0326110167168337,-63.7577489517829)(0.0435332151718253,-61.1841633945969)(0.058113515431065,-58.5621858757054)(0.0775771020455743,-55.858027377346)(0.10355950276194,-53.0189592942182)(0.138244022134264,-49.969978422967)(0.184545204893379,-46.6189807042304)(0.246353745524438,-42.8753311548155)(0.328863423836903,-38.6740188740447)(0.439007538965966,-33.9849362592692)(0.586041515411986,-28.8066293844969)(0.782320637580219,-23.2077789823136)(1.04433826595658,-17.6071738731864)(1.39411177636147,-13.5460298238264)(1.86103268293967,-13.2852115193791)(2.4843364109648,-16.1497950396564)(3.31639925479244,-19.8117362617831)(4.42713956477276,-23.3965793380773)(5.9098929954388,-26.9260242786792)(7.8892555128493,-30.5516393412866)(10.5315532100258,-34.3137285613729)(14.058818710455,-38.1291179325981)(18.7674485986815,-41.8590905010365)(25.0531096643436,-45.385018932166)(33.443986834615,-48.6542308698591)(44.6451666231997,-51.6817321205157)(59.5978856429365,-54.5204679332265)(79.5586228423394,-57.2292416782801)(106.204681597118,-59.8555806448936)(141.775133733747,-62.4322258172633)(189.258969029923,-64.9798122733356)(252.646260419243,-67.510705312783)(337.263450345308,-70.0321055852674)(450.220932421762,-72.5481389076883)(601.01053874408,-75.0611478113433)(802.303139790633,-77.5724555126577)(1071.01337933777,-80.0828072976591) 
};
\addplot [
color=green!65!black,
solid,
line width=1.0pt,
forget plot
]
coordinates{
 (3.448150933975e-05,-123.356963668301)(0.000109040106673979,-113.356962819488)(0.000179948843765257,-109.005705337913)(0.000240217954673898,-106.49657524194)(0.000320672611950687,-103.98744357513)(0.000428073431042535,-101.478309109062)(0.000571445316922507,-98.9691696546721)(0.000762835828043288,-96.4600213110309)(0.00101832753426064,-93.9508571266973)(0.00135938943729649,-92.9517640126964)(0.00181468101378105,-92.8931589691992)(0.0024224604748486,-92.7906439778708)(0.00323379960865767,-92.6137496121957)(0.00431687534947631,-92.3153393959213)(0.00576269869444741,-91.8296125167633)(0.00769276236966505,-91.079346009375)(0.0102692498799487,-89.997902479091)(0.013708663810113,-88.5587725036223)(0.0183000185656831,-86.7897707622356)(0.0244291262914551,-84.757951119838)(0.0326110167168337,-82.5398811555457)(0.0435332151718253,-80.1998823520221)(0.058113515431065,-77.7828569684322)(0.0775771020455743,-75.3158948235546)(0.10355950276194,-72.8120792006776)(0.138244022134264,-70.2731511134126)(0.184545204893379,-67.6900027139712)(0.246353745524438,-65.041074594644)(0.328863423836903,-62.2898869495311)(0.439007538965966,-59.3857282990869)(0.586041515411986,-56.2769161180278)(0.782320637580219,-52.9499658905271)(1.04433826595658,-49.4951689271723)(1.39411177636147,-46.1580164257133)(1.86103268293967,-43.306700748816)(2.4843364109648,-41.298842852431)(3.31639925479244,-40.3322452361396)(4.42713956477276,-40.3878107092701)(5.9098929954388,-41.2871190008788)(7.8892555128493,-42.7974072711089)(10.5315532100258,-44.7104042184071)(14.058818710455,-46.8725584744207)(18.7674485986815,-49.182808566585)(25.0531096643436,-51.5789730996587)(33.443986834615,-54.0242730077455)(44.6451666231997,-56.4974438579705)(59.5978856429365,-58.986349980446)(79.5586228423394,-61.4841163336304)(106.204681597118,-63.986864256754)(141.775133733747,-66.4924106579371)(189.258969029923,-68.9995284062621)(252.646260419243,-71.50752823258)(337.263450345308,-74.0160231408007)(450.220932421762,-76.5247958992462)(601.01053874408,-79.0337245856689)(802.303139790633,-81.5427407756841)(1071.01337933777,-84.0518060701993) 
};
\addplot [
color=red,
solid,
line width=1.0pt,
forget plot
]
coordinates{
 (6.56263992411199e-05,-60.597415252401)(0.0102938925253544,-60.5978776834159)(0.0244413700896683,-60.6000081203166)(0.0435000235571196,-60.6055095816276)(0.0580325246634107,-60.6115794912656)(0.0774200481612874,-60.6218546162988)(0.103284561408634,-60.6384532141685)(0.137789899111792,-60.6625604511349)(0.183822790534218,-60.6878575722357)(0.245234364330084,-60.5561178037304)(0.327162335386188,-55.577257586596)(0.436460827942021,-50.2927529507456)(0.58227379414861,-44.6750403442465)(0.776800000473936,-38.748024820312)(1.0363135809995,-32.8919722288416)(1.32352206890392,-29.1075751402279)(1.69178527959376,-28.4655756666573)(2.03467002228675,-30.9771234224253)(2.33747529820029,-34.708851413181)(2.59443728089181,-38.4804931074718)(2.80604897372162,-37.9955047491799)(2.97642134626885,-37.6859360197288)(3.15713806618821,-37.4233872097222)(3.41464567124924,-37.1461524325658)(3.79002252445013,-36.9017179217951)(4.35406426274872,-35.8054512962908)(5.2365297992498,-33.843880854886)(6.69356729190137,-32.7500884260138)(7.79408247692148,-32.626503358426)(10.3979319224201,-33.1910629087578)(13.8716761829787,-34.4947163852724)(18.5059299831067,-36.2775145698557)(24.6883967029075,-38.3586379703258)(32.9363037856804,-40.6187616070768)(43.9396741763661,-42.9835283584049)(58.6190538953128,-45.4084704814427)(78.2025252574552,-47.8676769888055)(104.328448691183,-50.3462821933129)(139.182528575333,-52.8358338248714)(185.680670072693,-55.3315508361884)(247.712924830095,-57.8307367169927)(330.468934132225,-60.3318731478694)(440.872096203279,-62.8341060075689)(588.15878025286,-65.336955067377)(784.651044526594,-67.8401503988762)(1046.78750423822,-70.3435403050394) 
};
\addplot [
color=red,
solid,
line width=1.0pt,
forget plot
]
coordinates{
 (6.56263992411199e-05,-140.484073888805)(0.000207528896237484,-130.484058356785)(0.000656263992411199,-120.48390303954)(0.00207528896237485,-110.482350162819)(0.00656263992411199,-100.466850891349)(0.0102938925253544,-96.531793412949)(0.0137328792724838,-93.9954740064396)(0.0183207637585202,-91.4342539318349)(0.0244413700896683,-88.8299258793506)(0.0326067504463239,-86.1524812332486)(0.0435000235571196,-83.3551968357025)(0.0580325246634107,-80.3719916739723)(0.0774200481612874,-77.1231869880218)(0.103284561408634,-73.5352986648686)(0.137789899111792,-69.5686145176844)(0.183822790534218,-65.2301351992239)(0.245234364330084,-60.6758131217706)(0.327162335386188,-60.4760708248052)(0.436460827942021,-59.6510171540612)(0.58227379414861,-57.4830574982695)(0.776800000473936,-53.8178915407698)(1.0363135809995,-49.4404809061129)(1.32352206890392,-45.788347505992)(1.69178527959376,-42.5388573490437)(2.03467002228675,-40.5065016115303)(2.33747529820029,-39.2545105598146)(2.59443728089181,-38.7907225375193)(2.80604897372162,-42.5182819843751)(2.97642134626885,-44.5792683153097)(3.15713806618821,-44.222784680958)(3.41464567124924,-41.5846389359875)(3.79002252445013,-38.4612535738235)(4.35406426274872,-36.7862303809185)(5.2365297992498,-36.9713888009679)(6.69356729190137,-37.7262473774238)(7.79408247692148,-38.4295120873303)(10.3979319224201,-40.1246895535799)(13.8716761829787,-42.1486641241523)(18.5059299831067,-44.3739158389494)(24.6883967029075,-46.7181610878263)(32.9363037856804,-49.1312686950426)(43.9396741763661,-51.5837276347419)(58.6190538953128,-54.0585102230698)(78.2025252574552,-56.5459040703486)(104.328448691183,-59.0404055267757)(139.182528575333,-61.538907423556)(185.680670072693,-64.0396592280869)(247.712924830095,-66.5416758778273)(330.468934132225,-69.044403424152)(440.872096203279,-71.5475304708794)(588.15878025286,-74.050882006715)(784.651044526594,-76.5543596831612)(1046.78750423822,-79.0579082364307) 
};
\addplot [
color=mycolor1,
solid,
line width=1.0pt,
forget plot
]
coordinates{
 (5.40628243117286e-05,-56.3442023256462)(0.0170961661566587,-56.3475905869552)(0.0345496394036994,-56.3580379419436)(0.0460772902101354,-56.3688065896413)(0.0614511963005233,-56.3879503427295)(0.0819546789653621,-56.421966961476)(0.109299245721248,-56.482352029868)(0.14576745667301,-55.9097897219432)(0.194403458914145,-52.6698083609643)(0.25679760502612,-49.1919763044806)(0.326983416449176,-45.8448904159105)(0.398205138954853,-42.8714085514521)(0.467631600691363,-40.2819687812676)(0.533123523540709,-38.0629984528239)(0.607787606595147,-35.752701775391)(0.713754453542311,-32.8102751592848)(0.869220507996734,-29.0993279961644)(1.10678871527463,-24.7001612111791)(1.45884068993612,-21.1559777089122)(1.75066400088983,-20.7841407778168)(2.20424636753704,-22.8268393327583)(2.65368526787738,-25.7539658010639)(3.08147408325115,-28.4174820292067)(3.47566833396817,-30.4245540659706)(3.92028945932384,-31.9028744702736)(4.55226191062657,-32.5810940632877)(5.480453793941,-32.1749671841246)(6.90039342878408,-31.4805282940659)(8.2085857051785,-31.3374242122865)(10.9474191997436,-31.8832702933579)(14.6000774602754,-33.1773320016064)(19.4714624476091,-34.9603720994092)(25.96820810576,-37.0438310250329)(34.6326237198926,-39.3053443161015)(46.187962636422,-41.6701014985155)(61.5987951059602,-44.0939461969795)(82.1515248112287,-46.5513500320857)(109.561770115805,-49.0277305054048)(146.117573575069,-51.5148148774968)(194.87039397865,-54.0079283117502)(259.889823792381,-56.5044349529802)(346.603294280987,-59.0028504595363)(462.249124853789,-61.5023395353854)(4622.49124853789,-81.5005858340806) 
};
\addplot [
color=mycolor1,
solid,
line width=1.0pt,
forget plot
]
coordinates{
 (5.40628243117286e-05,-125.663134838671)(0.000170961661566587,-115.66313325831)(0.000540628243117286,-105.663117454731)(0.00170961661566587,-95.6629594216902)(0.00540628243117286,-85.6613793659511)(0.0170961661566587,-75.6456062077163)(0.0259059848676407,-72.0129494911213)(0.0345496394036994,-69.481034488394)(0.0460772902101354,-66.9254164076155)(0.0614511963005233,-64.3286204252117)(0.0819546789653621,-61.6615245063083)(0.109299245721248,-58.8778647503951)(0.14576745667301,-56.5893043910855)(0.194403458914145,-56.7776497583194)(0.25679760502612,-57.090030835455)(0.326983416449176,-57.5160082291595)(0.398205138954853,-57.9760407733845)(0.467631600691363,-58.3487870927782)(0.533123523540709,-58.4892174828523)(0.607787606595147,-58.2134037685244)(0.713754453542311,-56.9043852254118)(0.869220507996734,-53.9020373627147)(1.10678871527463,-49.4706760358658)(1.45884068993612,-44.812397916075)(1.75066400088983,-42.2703207625901)(2.20424636753704,-39.7968803863519)(2.65368526787738,-38.4403088773984)(3.08147408325115,-37.7549774571034)(3.47566833396817,-37.4521220634521)(3.92028945932384,-37.353692031276)(4.55226191062657,-37.4858390617117)(5.480453793941,-37.9836628160589)(6.90039342878408,-39.0072784356924)(8.2085857051785,-40.0014119934536)(10.9474191997436,-41.941449341972)(14.6000774602754,-44.1178217626178)(19.4714624476091,-46.4331891943468)(25.96820810576,-48.8287660119989)(34.6326237198926,-51.2701154325924)(46.187962636422,-53.7374179948571)(61.5987951059602,-56.2193820594508)(82.1515248112287,-58.7096114903305)(109.561770115805,-61.2044950082252)(146.117573575069,-63.7019974246585)(194.87039397865,-66.200972969321)(259.889823792381,-68.7007769731427)(346.603294280987,-71.2010468314524)(462.249124853789,-73.7015786290085)(4622.49124853789,-93.7011508599108) 
};
\end{semilogxaxis}

\begin{axis}[%
width=\figurewidth,
height=\figureheight,
scale only axis,
xmin=0, xmax=1,
xtick={\empty},
xlabel={Frequency (rad/s)},
ymin=0, ymax=1,
ytick={\empty},
ylabel={Singular Values (dB)},
hide x axis,
hide y axis]
\end{axis}
\end{tikzpicture}%